\documentclass[draftclsnofoot,onecolumn]{IEEEtran}
\usepackage{graphicx}
\usepackage{subfigure}  
\usepackage{bm}
\usepackage{amsthm}
\usepackage{amsfonts}
\usepackage{amsmath}

\usepackage{multirow}
\usepackage{makecell}
\usepackage{xcolor}
\usepackage[citebordercolor=white]{hyperref}
\hypersetup{ 
	colorlinks = true, 
	citecolor = blue, 
	linkcolor = red, 
	urlcolor = blue}

\usepackage{cuted} 
\usepackage{enumitem}
\usepackage{tablefootnote}

\usepackage[ruled,linesnumbered]{algorithm2e}

\begin{document}

\title{Maneuver Detection via a Confidence Dominance Maneuver Indicator}

\author{Xingyu~Zhou,~Roberto~Armellin,~Laura~Pirovano,~Dong~Qiao~and~Xiangyu~Li 

\thanks{This work was supported by the National Natural Science Foundation of China (No. 124B2049 and No. 62394353) and the Changjiang Scholars Program (No. T2023191). Xingyu Zhou is grateful for the financial support provided by the China Scholarship Council (No. 202406030186) and the China Association for Science and Technology (CAST) Young Talent Support Program (No. 156-O-590-0000198-7). Xingyu Zhou also thanks Zeno Pavanello for his helpful discussion.}

\thanks{Authors’ address: Xingyu Zhou, Dong Qiao, and Xiangyu Li are with School of Aerospace Engineering, Beijing Institute of Technology, E-mail: \texttt{zhouxingyu@bit.edu.cn}, \texttt{qiaodong@bit.edu.cn}, and \texttt{lixiangy@bit.edu.cn}; Roberto Armellin is with Te Punaha Atea - Space Institute, The University of Auckland, E-mail: \texttt{roberto.armellin@auckland.ac.nz}; Laura Pirovano is an independent researcher, E-mail: \texttt{laura0pirovano@gmail.com}; (Corresponding author: Dong Qiao.)}
}

\markboth{Journal of \LaTeX\ Class Files}
{Shell \MakeLowercase{\textit{et al.}}: Bare Demo of IEEEtran.cls for IEEE Journals}

\maketitle

\begin{abstract}
Accurate and efficient maneuver detection is critical for ensuring the safety and predictability of spacecraft trajectories. 
This paper presents a novel maneuver detection approach based on comparing the confidence levels associated with the orbital state estimation and the observation likelihood. 
First, a confidence-dominance maneuver indicator (CDMI) is proposed by setting a confidence level for the state estimation and computing the maximum likelihood of the observation and its confidence level. 
The CDMI then flag a maneuver when the observation's confidence level exceeds that of the state estimation, indicating that the observation is unlikely under the no-maneuver hypothesis while maintaining consistency with the prior state estimation confidence.
To efficiently compute the maximum likelihood of the observation and obtain the CDMI, a recursive polynomial optimization method is developed, taking advantage of convex optimization and polynomial approximation. 
In addition, an integrated CDMI approach is developed to eliminate the need to manually select the state confidence level. The integrated CDMI approach maintains high detection accuracy while simultaneously providing an indication of maneuver likelihood, thereby enhancing robustness and practical applicability.
The performance of the proposed CDMI-based maneuver detection approaches is evaluated against an optimal control distance metric and two mixture-based approaches. 
The simulation results demonstrate that the proposed integrated CDMI approach can achieve up to 99.33\% detection accuracy, at least 10\% higher than the competing methods, while substantially reducing computational costs.
\end{abstract}


\section{Introduction}
\IEEEPARstart{S}{ince} space activities grow increasingly complex, the need for reliable maneuver detection techniques has become more critical \cite{Wu2002, Li2003, Hall2021, Zhou2022Aero, Zhou2023AA}. Spacecraft trajectories are influenced not only by natural disturbances such as gravitational forces, solar radiation pressure, and atmospheric drag but also by intentional maneuvers for orbit adjustments, station-keeping, or evasive actions. 
If a spacecraft does not transmit its maneuver intentions, its trajectory becomes unpredictable \cite{Goff2015, Lin2023ieee, Zhou2024JGCD, Zhang2024SST, Li2024SST, Ren2025}. 
While estimation is central to maintaining catalog accuracy and predicting conjunctions, it is accurate maneuver detection that enables these capabilities in the first place. Maneuver detection serves as a prerequisite for maintaining up-to-date orbital catalogs \cite{Pirovano2020CMDA, Pirovano2021AA, Pirovano2024}, correlating new tracking data with known objects \cite{Serra2021, Serra2022, Zhou2022AS, Cipollone2025}, and forecasting future trajectories to assess potential conjunctions \cite{Shu2022, Khatri2023, Parigini2024, Mao2024Astro}. These functions are essential to ensuring timely collision avoidance and safeguarding space operations \cite{Armellin2021, Cui2024ieee, Pavanello2024, Pavanello2024IEEE}.

The maneuver detection problems can be broadly divided into two types: maneuver detection of two uncorrelated tracks (UCTs) \cite{Pirovano2024} and maneuver detection of one UCT and a certain number of measurements after that UCT \cite{Montilla2023, Montilla2025}. 
In this paper, a track refers to a known orbital state obtained by an orbit determination (OD) process based on a set of measurements, typically represented by a mean state vector and an associated covariance matrix.
For the second type, the measurements given (\emph{e.g.}, only one angle measurement) are inadequate for an independent OD due to the limited observation capacity. Thus, the second track is not available. 

Conducting a two-line element (TLE) analysis is one of the most widely used ways to detect unknown maneuvers between two UCTs \cite{Lemmens2014, Clark2020, Zollo2024}. An alternative approach can be the optimal control distance metric (OCDM) proposed by Holzinger and Scheeres \cite{Holzinger2010, Holzinger2011, Holzinger2012}. The OCDM formulates the maneuver detection problem as an uncertainty two-point boundary-value problem (UTPBVP) by identifying the energy-optimal control to connect two UCTs (the UTCs serve as boundary conditions), which can further estimate the maneuver, enhancing its applicability to object correlation and maneuver characterization \cite{Holzinger2012, Lubey2015}. Subsequent studies applied OCDM to detect anomalies caused by unknown dynamics \cite{Lubey2014} (\emph{e.g.}, solar radiation pressure and atmospheric drag) and maneuver detection in cislunar periodic orbits \cite{Greaves2023}. More recent advancements include combining OCDM with binary hypothesis testing \cite{Jaunzemis2016}, solving for the OCDM using sequential convex optimization \cite{Pirovano2024}, and simplifying OCDM by approximating continuous maneuvers with impulsive ones \cite{Pastor2022}.

The second type of problem is more intractable since limited measurements are available, resulting in much larger uncertainty levels. A basic implementation for this type of maneuver detection involves propagating the last estimated state (within the UCT) forward to obtain predicted observations and then comparing them with sensors’ observed measurements \cite{Moose1975, Sung1994, Bogler1987, Lee1999}. The residual between the predicted and observed ones is then evaluated—if it exceeds a pre-defined threshold, a maneuver is detected. This approach is simple and easy to implement, but it heavily depends on the OD quality of the UCT and the measurement noise level. To improve maneuver detection accuracy, Kelecy and Jah utilized the difference between the solutions obtained by the extended Kalman filter (EKF) and the batch least squares (BLS) as a metric for the detection of continuous maneuvers, leveraging the fact that BLS may converge to incorrect solutions under unknown maneuvers \cite{Kelecy2010}. Li \emph{et al.} employed the Mahalanobis Distance (MD) as a metric based on measurement residual and covariance from the EKF process \cite{Li2019JSR}. The OCDM can also be adopted here by replacing the boundary condition in the UTPBVP with the measurement constraints, which finally becomes a measurement residual boundary-value problem (MRBVP) \cite{Holzinger2012}. Recently, Montilla \emph{et al.} made significant progress in this field by proposing two mixture-based approaches (MBAs). The MBAs employ Gaussian mixture representation to approximate the uncertainties \cite{Montilla2025}, which is particularly useful when crucial uncertainties are quantified, for example, after a long-term propagation under highly nonlinear dynamics. The MBAs have promising maneuver detection accuracy, but at the cost of not being efficient.

A key limitation of most of the aforementioned methods is their reliance on manually defined thresholds, which can be difficult to tune and may not generalize well across scenarios. To overcome this, recent years have seen the emergence of data-driven maneuver detection approaches, which leverage artificial intelligence techniques to learn decision boundaries directly from data. These methods typically frame maneuver detection as a classification problem—similar to facial or pattern recognition—by identifying latent patterns in measurement sequences. For example, Gaussian binary classification has been applied to detect maneuver-induced trajectory deviations \cite{Wang2021AA}, and convolutional neural networks have been used to identify anomalies in TLE data \cite{Li2024ASR}. While these methods often achieve high detection accuracy, many are restricted to fixed-length input sequences. The adoption of long short-term memory (LSTM) networks has improved flexibility by accommodating variable-length data and has shown promising preliminary results in maneuver detection tasks \cite{Zhou2023AA, Cipollone2025NCA}. Nevertheless, such methods remain in early stages of development and face notable challenges, including limited availability of real-world labeled data and low interpretability, which hinder their practical reliability in operational systems \cite{Lu2023JGCD}.

This paper aims to handle the second type of maneuver detection problem and develop accurate and efficient approaches. 
First, a confidence-dominance maneuver indicator (CDMI) is proposed. Specifically, given a user-defined state confidence level, a confidence-constrained state region is first defined based on the state estimate obtained from the first UCT. The CDMI then computes the maximum likelihood of the observed measurement, under the no-maneuver hypothesis, within this region, yielding an associated measurement confidence level. The CDMI flags a maneuver when the measurement confidence exceeds that of the state estimate (\emph{i.e.}, when the observation statistically dominates the assumed state), indicating an inconsistency with the no-maneuver hypothesis.
Second, a recursive polynomial optimization (RPO) method is developed to obtain the maximum likelihood and compute the maneuver indicator efficiently, leveraging the fast computational speed of convex optimization and polynomial approximation. 
Third, an integrated CDMI approach is further developed to eliminate the need to manually select the state confidence level, which is often difficult to tune and may significantly impact detection performance. Instead of relying on a fixed state confidence level like the originally developed CDMI, the integrated CDMI integrates the measurement confidence over a range of state confidence levels, effectively balancing maneuver detection outcomes across varying assumptions of state uncertainty.
Finally, the proposed approaches (\emph{i.e.}, the CDMI and its integrated version) are applied to several angles-only maneuver detection scenarios in the cislunar space. It will be shown that they outperform the competitive methods in terms of maneuver detection accuracy and computational efficiency.

The remainder of this paper is organized as follows. Section~\ref{sec:Preliminary} presents the maneuver detection problem and briefly introduces the Taylor polynomial and differential algebra (DA) techniques. The CDMI and the associated RPO algorithm are given in Sec.~\ref{sec:Polynomial Optimization-Based Indicator}. The integrated CDMI approach is proposed in Sec.~\ref{sec:Maneuver Detection Approaches}. Simulation examples are given in Sec.~\ref{sec:Numerical Results}, and conclusions are provided in Sec.~\ref{sec:Conclusion}.

\section{Preliminary} \label{sec:Preliminary}

\subsection{Problem Statement} \label{sec:Problem Statement}
Maneuver detection is a sub-problem in orbit estimation. An orbit estimation system can be generally modeled as a combination of an orbital state equation and a measurement equation, expressed as
\begin{equation} \label{eq1}
    \left\{ {\begin{array}{*{20}{l}}
        {\boldsymbol{\dot x} = \boldsymbol{f}(t,\boldsymbol{x})}\\
        {\boldsymbol{z} = \boldsymbol{h}(\boldsymbol{x}) + \boldsymbol{\varepsilon }}
    \end{array}} \right. \,,
\end{equation}
where $\boldsymbol{x} = [\boldsymbol{r};\boldsymbol{v}] \in {\mathbb{R}^6}$ is the orbital state vector of a spacecraft at a given epoch $t$, with $\boldsymbol{r} = {[x,y,z]^T} \in {\mathbb{R}^3}$ and $\boldsymbol{v} = {[\dot x,\dot y,z]^T} \in {\mathbb{R}^3}$ being the time-dependent position and velocity vectors, respectively, $\boldsymbol{f}(t,\boldsymbol{x}):{\mathbb{R}^6} \mapsto {\mathbb{R}^6}$ is a collection of ordinary differential equations (ODEs) representing the nonlinear orbit dynamics, $\boldsymbol{z} = \boldsymbol{h}(\boldsymbol{x}) + \boldsymbol{\varepsilon} \in {\mathbb{R}^m}$ denotes the $m$-dimensional measurement vector, $\boldsymbol{h}(\boldsymbol{x}):{\mathbb{R}^6} \mapsto {\mathbb{R}^m}$ is the measurement model (usually nonlinear), and $\boldsymbol{\varepsilon} \in {\mathbb{R}^m}$ is a zero-mean, Gaussian-distributed measurement noise satisfying $\mathbb{E}\{\boldsymbol{\varepsilon }\} = {\mathbf{0}_{m \times 1}}$ and $\mathbb{E}\{\boldsymbol{\varepsilon }{\boldsymbol{\varepsilon }^T}\} = \boldsymbol{R}$ ($\boldsymbol{R} \in {\mathbb{R}^{m \times m}}$ is the covariance matrix of the measurement noise $\boldsymbol{\varepsilon}$, which is a positive definite matrix).

The discrete form of the orbit estimation system in Eq.~\eqref{eq1} can be written as
\begin{equation} \label{eq2}
    \left\{ {\begin{array}{*{20}{l}}
        {{\boldsymbol{x}_k} = \boldsymbol{F}({\boldsymbol{x}_{k - 1}},{t_{k - 1}},{t_k})}\\
        {{\boldsymbol{z}_k} = \boldsymbol{h}({\boldsymbol{x}_k}) + {\boldsymbol{\varepsilon }_k}}
    \end{array}} \right. \,,
\end{equation}
where ${\boldsymbol{x}_k} \in {\mathbb{R}^6}$, ${\boldsymbol{z}_k} \in {\mathbb{R}^m}$, and ${\boldsymbol{\varepsilon }_k} \in {\mathbb{R}^m}$ (satisfying $\mathbb{E} \{ {\boldsymbol{\varepsilon }_k}\boldsymbol{\varepsilon }_k^T\}  = {\boldsymbol{R}_k}$) represent the orbital state, measurement vector, and measurement noise at the epoch ${t_k}$, respectively, and $\boldsymbol{F}({\boldsymbol{x}_{k - 1}},{t_{k - 1}},{t_k}):{\mathbb{R}^6} \mapsto {\mathbb{R}^6}$ is a nonlinear orbit propagation from ${t_{k - 1}}$ to ${t_k}$, formulated as
\begin{equation} \label{eq3}
    \boldsymbol{F}({\boldsymbol{x}_{k - 1}},{t_{k - 1}},{t_k}) = \int\limits_{{t_{k - 1}}}^{{t_k}} {\boldsymbol{f}(\tau ,\boldsymbol{x})d\tau }  + {\boldsymbol{x}_{k - 1}} \,.
\end{equation}

Let ${\boldsymbol{x}_0} = [{\boldsymbol{r}_0};{\boldsymbol{v}_0}] \in {\mathbb{R}^6}$ and ${\boldsymbol{\hat x}_0} = [{\boldsymbol{\hat r}_0};{\boldsymbol{\hat v}_0}] \in {\mathbb{R}^6}$ be the true and estimated orbital state at the initial epoch $t_0$. 
Assume that the estimated orbital state ${\boldsymbol{\hat x}_0}$ is Gaussian-distributed, with ${\boldsymbol{x}_0}$ being the mean and ${\boldsymbol{P}_0} = \mathbb{E} \{ ({\boldsymbol{\hat x}_0} - {\boldsymbol{x}_0}){({\boldsymbol{\hat x}_0} - {\boldsymbol{x}_0})^T}\} $ being the covariance (\emph{i.e.}, ${\boldsymbol{\hat x}_0} \sim {\cal N}({\boldsymbol{\hat x}_0};{\boldsymbol{x}_0},{\boldsymbol{P}_0})$). 
Note that the mean ${\boldsymbol{\hat{x}}_0}$ and covariance ${\boldsymbol{P}_0}$ can be obtained via an OD process based on the data collected in the first UCT.
In addition, we assume that a measurement ${\boldsymbol{z}_k} = \boldsymbol{h}({\boldsymbol{x}_k}) + {\boldsymbol{\varepsilon }_k}$ is available at the epoch $t_k$ (${t_k} > {t_0}$), whose predicted orbital state can be propagated based on the initial estimated orbital state as
\begin{equation} \label{eq4}
    {\boldsymbol{\hat x}_k} = \boldsymbol{F}({\boldsymbol{\hat x}_0},{t_0},{t_k}) \,.
\end{equation}

Based on the above definitions and discussions, the maneuver detection problem investigated in this work can be generalized as follows: given the initial estimation ${\boldsymbol{\hat x}_0}$ and associated covariance matrix ${\boldsymbol{P}_0}$, the measurement ${\boldsymbol{z}_k}$ at the given epochs $t_k$ ($k \in \{ 1,2, \cdots ,N\} $), and its associated noise covariance ${\boldsymbol{R}_k}$, identify whether a spacecraft has executed a maneuver.

\subsection{Taylor Polynomial Approximation} \label{sec:Taylor Polynomial Approximation}
The Taylor polynomial is an efficient technique to approximate the nonlinear orbit propagation. Let $\delta {\boldsymbol{x}_0} = {\boldsymbol{x}_0} - {\boldsymbol{\hat x}_0} \in {\mathbb{R}^6}$ represents an orbital state deviation at the initial epoch (which defines a neighboring orbit) and ${\boldsymbol{\tilde x}_k} = \boldsymbol{F}({\boldsymbol{\hat x}_0} + \delta {\boldsymbol{x}_0},{t_0},{t_k})$ be a propagated state of the neighboring orbit at the given epoch $t_k$. Then, the propagated state ${\boldsymbol{\tilde x}_k}$ can be approximated using a Taylor polynomial up to the \emph{n}-th order as
\begin{equation} \label{eq5}
    {\boldsymbol{\tilde x}_k} \approx {\cal T}_{{{\boldsymbol{\tilde x}}_k}}^n(\delta {\boldsymbol{x}_0}) = \boldsymbol{F}({\boldsymbol{\hat x}_0},{t_0},{t_k}) + \sum\limits_{k = 1}^n {\frac{1}{{k!}}\frac{{{\partial ^k}\boldsymbol{F}}}{{\partial \boldsymbol{\hat x}_0^k}}{{(\delta {\boldsymbol{x}_0})}^k}}  \,,
\end{equation}
where ${\cal T}_{{{\boldsymbol{\tilde x}}_k}}^n(\delta {\boldsymbol{x}_0})$ is an \emph{n}-th order Taylor polynomial expanded at the point ${\boldsymbol{\hat x}_0}$, and $\boldsymbol{F}({\boldsymbol{\hat x}_0},{t_0},{t_k})$ and $\frac{{\partial \boldsymbol{F}}}{{\partial {{\boldsymbol{\hat x}}_0}}}\delta {\boldsymbol{x}_0}$ are its constant and linear (first-order) terms, respectively.

For nonlinear orbital propagation problems (\emph{e.g.}, Eq.~\eqref{eq5}), high-order Taylor polynomial approximations can be derived using a variational approach, leading to the so-called state transition tensor (STT) \cite{Park2006}, or by leveraging the algebra of Taylor polynomials, as in DA \cite{Armellin2010, Valli2013} and jet transport (JT) \cite{Perez-Palau2015}. The variational approach necessitates computing intricate partial derivatives of the dynamical equations and integrating an expanded system of ODEs. In contrast, DA and JT integrate only the original set of ODEs but require specialized numerical schemes to accommodate the Taylor polynomial data structure effectively. In this work, the DA is employed to generate Taylor polynomials, which will be briefly introduced in the next subsection.

\subsection{Differential Algebra} \label{sec:Differential Algebra}
The DA technique was originally developed as an algebraic framework for solving analytical problems \cite{Griewank2008}. In this work, DA is utilized to obtain high-order Taylor polynomial expansions of the flow in nonlinear orbital dynamics with respect to the initial state deviation. Unlike conventional numerical methods that evaluate functions at discrete points, DA exploits the richer structural information embedded in functions beyond their numerical values \cite{Armellin2009CMDA, Armellin2010, Armellin2012CMDA, Valli2013, Fu2024Astro, Fossa2024Astro, Liu2024IEEE}. Its fundamental principle lies in manipulating functions and their operations in a manner analogous to the arithmetic of real numbers within a computational environment. 

The Differential Algebra Core Engine (DACE) \footnote{Available at \url{https://github.com/dacelib/dace/}.} is a robust and versatile C++ computational toolbox for DA. Its Python interface, DACEyPy \footnote{Available at \url{https://github.com/giovannipurpura/daceypy/}.}, serves as a wrapper for DACE, providing a user-friendly environment for DA-based computations. In this work, DACEyPy is utilized to implement the DA framework efficiently.

\section{confidence-Dominance Maneuver Indicator} \label{sec:Polynomial Optimization-Based Indicator}

\subsection{Definition of Confidence-Constrained Region} \label{sec:Definition of Confidence-Constrained Region}
Two concepts of confidence-constrained regions (CCR) are here proposed: one is the confidence-constrained (initial) state region (CCSR) and the other is the confidence-constrained measurement region (CCMR). According to the problem statement in Sec.~\ref{sec:Problem Statement}, the initial orbital deviation $\delta {\boldsymbol{x}_0}$ satisfies a Gaussian distribution, that is $\delta {\boldsymbol{x}_0} \sim {\cal N}(\delta {\boldsymbol{x}_0};{\mathbf{0}_{6 \times 1}},{\boldsymbol{P}_0})$. In this work, the CCSR is defined by bounding the initial orbital deviation $\delta {\boldsymbol{x}_0}$ with a $\chi^2$ distribution of 6 degrees of freedom (DoF) and a given confidence level $\alpha_x$ as
\begin{equation} \label{eq6}
    \frac{1}{2}\delta \boldsymbol{x}_0^T\boldsymbol{P}_0^{ - 1}\delta {\boldsymbol{x}_0} \le {q_{{\chi ^2}}}({\alpha _x},6) = {{\cal M}_x}  \,,
\end{equation}
where ${{\cal M}_x} = {q_{{\chi ^2}}}({\alpha _x},6)$ refers to the quantile (inverse cumulative distribution function [CDF]) of the $\chi^2$ distribution with 6 DoF. Specifically, it represents the value corresponding to a cumulative probability of the given confidence level $\alpha_x$ for the $\chi^2$ distribution with 6 DoF. In other words, ${q_{{\chi ^2}}}({\alpha _x},6)$ is the value such that:
\begin{equation} \label{eq7}
    P\{ \chi _6^2 \le {q_{{\chi ^2}}}({\alpha _x},6)\}  = {\alpha _x} \,,
\end{equation}
where $\chi _6^2$ is a random variable following the $\chi^2$ distribution with 6 DoF.

Based on the CCSR defined in Eq.~\eqref{eq6}, the corresponding CCMR at an (arbitrary) epoch $t_k$, labeled as ${\cal Z}({\alpha _x},{t_k})$, is then defined as
\begin{equation} \label{eq8}
    {\cal Z}({\alpha _x},{t_k}) = \left\{ {{{\boldsymbol{\tilde z}}_k}\left| {{{\boldsymbol{\tilde z}}_k} = \boldsymbol{h}({{\boldsymbol{\tilde x}}_k}),{\rm{ }}\frac{1}{2}\delta \boldsymbol{x}_0^T\boldsymbol{P}_0^{ - 1}\delta {\boldsymbol{x}_0} \le {{\cal M}_x}} \right.} \right\} \,,
\end{equation}
which is a time-dependent measurement set containing all possible observations that satisfy the confidence constraint of the initial state. With the definitions of two CCRs, the confidence-constrained maximum likelihood maneuver indicator under two conditions is introduced in the following two subsections.

\subsection{Indicator with Measurement at Single Epoch} \label{sec:Indicator with Measurement at Single Epoch}
The proposed maneuver indicator is based on the maximum likelihood estimation within the CCSR.  
The goal of the maximum likelihood estimation is to find the observation with the highest likelihood without any maneuver (\emph{i.e.}, a ballistic motion), given the confidence level constraint on the state estimate. 
The highest likelihood obtained is called the confidence-constrained maximum likelihood throughout this paper. 
Then, based on this confidence-constrained maximum likelihood, the measurement confidence level can be obtained. 
Finally, we derive the CDMI by comparing the confidence levels associated with the state estimate and the observation, which can assess whether the observed measurement is consistent with the ballistic motion hypothesis, thus enabling effective maneuver detection.

This section details the indicator with measurements collected at a single epoch (\emph{i.e.}, $N=1$ and the corresponding CCMR is written as ${\cal Z}({\alpha _x},{t_1})$). 
Note that more than one measurement may be available at a given epoch. For example, an optical sensor can provide two independent measurements (\emph{i.e.}, right ascension and declination) at one epoch, whereas a radar usually has three (\emph{i.e.}, two angles and one range). 
Given the assumption that the measurement noises satisfy a Gaussian distribution (see Sec.~\ref{sec:Problem Statement}), we maximize logarithmic likelihood, which is equivalent to minimizing the squared Mahalanobis, to find the point with the maximum likelihood in the CCMR ${\cal Z}({\alpha _x},{t_1})$ (this point is also called the closest point as it is closest to the observed measurement $\boldsymbol{z}_1$). Therefore, a nonlinear programming (NLP) optimization problem is formulated as
\begin{equation} \label{eq9}
    \begin{array}{*{20}{l}}
        {\mathop {\min }\limits_{\delta {\boldsymbol{x}_0}} }&{J = \delta \boldsymbol{\tilde z}_1^T\boldsymbol{R}_1^{ - 1}\delta {{\boldsymbol{\tilde z}}_1}}\\
        {{\rm{s}}{\rm{.t}}{\rm{.}}}&{\delta {{\boldsymbol{\tilde z}}_1} = {{\boldsymbol{\tilde z}}_1} - {\boldsymbol{z}_1}}\\
        {}&{{{\boldsymbol{\tilde z}}_1} \in {\cal Z}({\alpha _x},{t_1})}
    \end{array} \,.
\end{equation}

Let $\delta \boldsymbol{x}_0^ * $ be the optimal solution of the NLP optimization problem in Eq.~\eqref{eq9}, and $\delta \boldsymbol{\tilde z}_1^ *  = \boldsymbol{h}[\boldsymbol{F}({\boldsymbol{\hat x}_0} + \delta \boldsymbol{x}_0^ * ,{t_0},{t_1})] - {\boldsymbol{z}_1}$ be the corresponding measurement residual. 
Then, by computing $\mathcal{M}_z = \frac{1}{2} \delta \tilde{\boldsymbol{z}}_1^\top R_1^{-1} \delta \tilde{\boldsymbol{z}}_1={q_{\chi^2}}({\alpha_z}, m)$, one can evaluate the corresponding confidence level $\alpha_z$:
\begin{equation} \label{eq10}
    \begin{aligned}
        \alpha_z &= F{\chi_{m}^2(\mathcal{M}_z)} \\
        &= \int_0^{\mathcal{M}_z} \frac{1}{2^{m/2} \Gamma(m/2)} x^{(m/2)-1} e^{-x/2}dx
    \end{aligned} \,,
\end{equation}
where $F{\chi_m^2}$ represents the CDF of the $\chi_m^2$ distribution. For the $\chi_m^2$ distribution in Eq.~\eqref{eq10}, the DoF equals the measurement dimension $m$. 

Two key situations are considered here. First, if the confidence-constrained maximum likelihood of the observation is low, $\delta \boldsymbol{\tilde z}_1^* \to \infty$, then ${q{_{\chi^2}}}({\alpha_z}, m) = {\mathcal{M}_z} \to +\infty$. In this case, ${\alpha_z} \to 1$. This suggests that the measurement is either an outlier or that an unmodeled maneuver may have occurred.

Second, when the confidence-constrained maximum likelihood is highest (\emph{i.e.}, when ${q_{\chi^2}}({\alpha_z}, m) = {\mathcal{M}_z} \to 0$ and ${\alpha_z} \to 0$), the measurement ${\boldsymbol{z}_1}$ is consistent with the state estimation. This indicates that the measurement aligns with the ballistic hypothesis, and the likelihood that a maneuver has occurred is low. 

These outcomes reveal that larger values of $\alpha_z$ are associated with a higher likelihood of a maneuver. However, since $\alpha_z$ depends on the confidence level of the state estimate—denoted by $\alpha_x$—it cannot be used directly as a maneuver indicator. To address this, we compare $\alpha_z$ against $\alpha_x$ to assess whether the observation is statistically consistent with the estimated state. A maneuver is flagged when the measurement confidence exceeds that of the state estimate, \emph{i.e.}, when $\alpha_z > \alpha_x$, indicating that the observation statistically dominates the estimated state and is inconsistent with the no-maneuver hypothesis.
Therefore, we adopt $(\alpha_z > \alpha_x)$ as the proposed CDMI, where a true outcome indicates the presence of a maneuver.

Three issues are worth noting for the NLP optimization problem in Eq.~\eqref{eq9}. First, when different measurement variables are independent and have the same standard deviations (STDs), weights are no longer required, and the objective function of the NLP in Eq.~\eqref{eq9} can be simplified as
\begin{equation} \label{eq12}
    J = \delta \boldsymbol{\tilde z}_1^T\delta {\boldsymbol{\tilde z}_1} \,.
\end{equation}

Second, when ${\alpha _x} = 0$, one doesn’t need to solve for the NLP optimization problem since the CCMR only contains one point:
\begin{equation} \label{eq13}
    {\cal Z}({\alpha _x} = 0,{t_1}) = \left\{ {{{\boldsymbol{\tilde z}}_1}\left| {{{\boldsymbol{\tilde z}}_1} = \boldsymbol{h}({{\boldsymbol{\hat x}}_1})} \right.} \right\} \,.
\end{equation}

Third, when ${\alpha _x} = 1$, one has ${{\cal M}_x} = {q_{{\chi ^2}}}({\alpha _x} = 1,6) =  + \infty $, which means that the CCSR is a set of real numbers with 6 DoF (\emph{i.e.}, $ \mathrm{CCSR} \equiv {\mathbb{R}^6}$). In this case, the CCMR can be expressed as
\begin{equation} \label{eq14}
    {\cal Z}({\alpha _x} = 1,{t_1}) = \left\{ {{{\boldsymbol{\tilde z}}_1}\left| {{{\boldsymbol{\tilde z}}_1} \in {\mathbb{R}^m}} \right.} \right\} \,,
\end{equation}
and the maneuver indicator can be assigned directly as $(\alpha_z > \alpha_x)=(1>1)=\mathrm{false}$. This extreme case corresponds to a situation where the state estimate is completely untrusted, that is, the estimation uncertainty is infinitely large. Under such conditions, any measurement can be explained by the uncertainty in the state estimate, and thus, no maneuver should be detected. While this scenario is unrealistic in practice, it serves as a theoretical limit illustrating that the CDMI will never falsely indicate a maneuver when the prior information is fully uninformative.

\subsection{Indicator with Measurement at Multiple Epochs} \label{sec:Indicator with Measurement at Multiple Epochs}
When measurements are available at more than one epoch (\emph{i.e.}, $N > 1$), the NLP optimization problem in Eq.~\eqref{eq9} can be rewritten as
\begin{equation} \label{eq16}
    \begin{array}{*{20}{l}}
        {\mathop {\min }\limits_{\delta {\boldsymbol{x}_0}} }&{J = \delta {{\boldsymbol{\tilde Z}}^T}{{\boldsymbol{\bar R}}^{ - 1}}\delta \boldsymbol{\tilde Z}}\\
        {{\rm{s}}{\rm{.t}}{\rm{.}}}&{\delta \boldsymbol{\tilde Z} = [\delta {{\boldsymbol{\tilde z}}_1};\delta {{\boldsymbol{\tilde z}}_2}; \cdots ;\delta {{\boldsymbol{\tilde z}}_N}] \in {\mathbb{R}^{Nm}}}\\
        {}&{\delta {{\boldsymbol{\tilde z}}_k} = {{\boldsymbol{\tilde z}}_k} - {\boldsymbol{z}_k}{\rm{, } \,}k \in \{ 1, \cdots ,N\} }\\
        {}&{{{\boldsymbol{\tilde z}}_k} \in {\cal Z}({\alpha _x},{t_k}){\rm{, } \, }k \in \{ 1, \cdots ,N\} }
    \end{array} \,,
\end{equation}
where
\begin{equation} \label{eq17}
    \boldsymbol{\bar R} = \left[ {\begin{array}{*{20}{c}}
        {{\boldsymbol{R}_1}}&{{\mathbf{0}_{m \times m}}}& \cdots &{{\mathbf{0}_{m \times m}}}\\
        {{\mathbf{0}_{m \times m}}}&{{\boldsymbol{R}_2}}& \cdots &{{\mathbf{0}_{m \times m}}}\\
         \vdots & \vdots & \ddots & \vdots \\
        {{\mathbf{0}_{m \times m}}}&{{\mathbf{0}_{m \times m}}}& \cdots &{{\boldsymbol{R}_N}}
    \end{array}} \right] \in {\mathbb{R}^{Nm \times Nm}} \,.
\end{equation}

Similar to Eq.~\eqref{eq10}, once the optimal solution $\delta {\boldsymbol{\tilde Z}^ * }$ is solved, the associated confidence level can be computed through
\begin{equation} \label{eq18}
    \alpha_z = F{\chi_{Nm}^2(\mathcal{M}_z)}
    = F{\chi_{Nm}^2 \left [ \frac{1}{2}{(\delta {\boldsymbol{\tilde Z}^ * })^T}{\boldsymbol{\bar R}^{ - 1}}\delta {\boldsymbol{\tilde Z}^ * } \right]}\,,
\end{equation}
which can further yield the proposed CMDI. Additionally, if the STDs of the measurements at different epochs are the same, the objective function in Eq.~\eqref{eq16} can be simplified as
\begin{equation} \label{eq19}
    J = \delta {\boldsymbol{\tilde Z}^T}\delta \boldsymbol{\tilde Z} \,.
\end{equation}

\subsection{Recursive Polynomial Optimization Method} \label{sec:Recursive Polynomial Optimization Method}
\begin{figure*}[!h]
	\centering
	\includegraphics[width=0.8\textwidth]{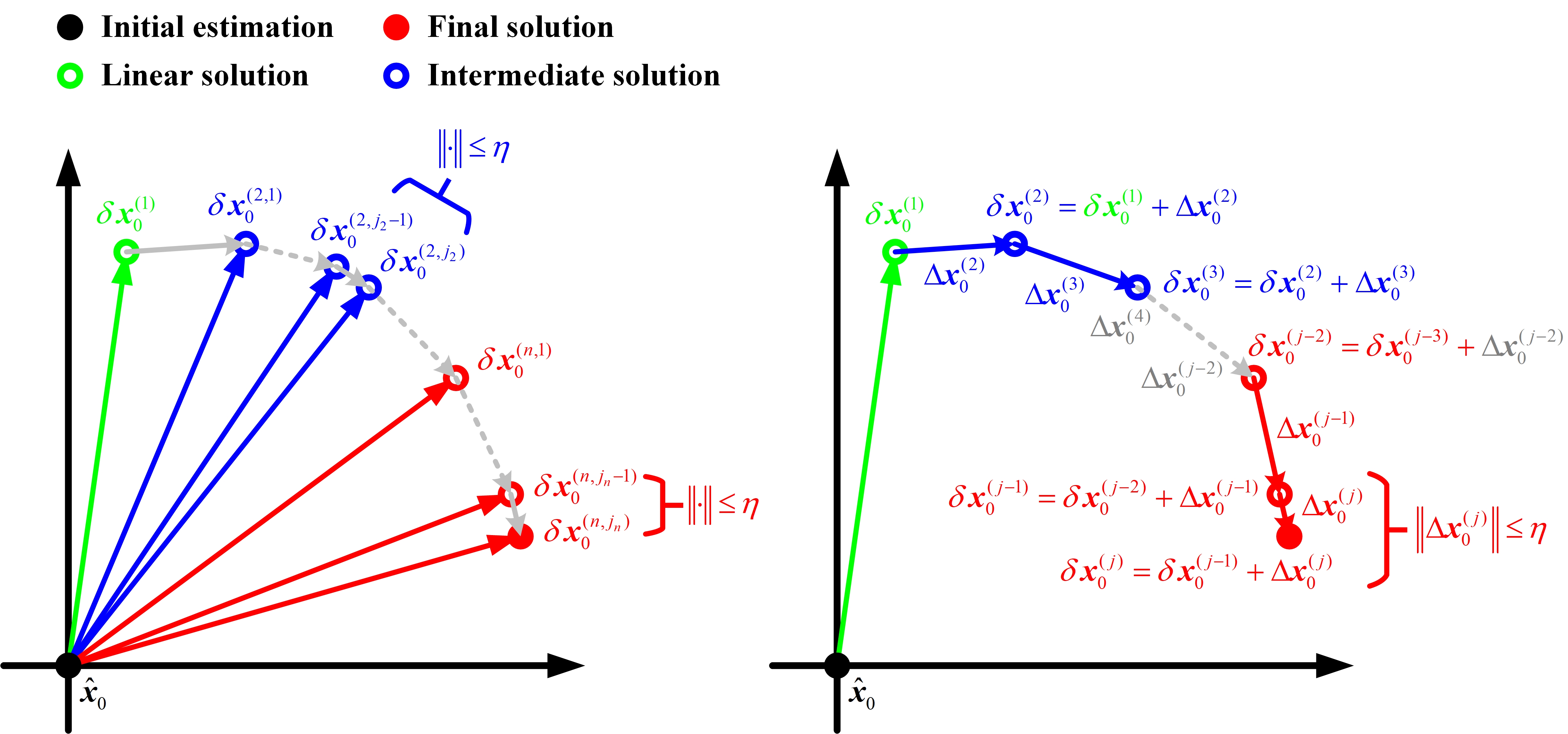}
	\caption{A comparison between two RPO approaches (left: Ref.~\cite{Pavanello2024IEEE}, right: this work).}
	\label{fig1}
\end{figure*}

The key to the CMDI proposed above lies in estimating the confidence-constrained maximum likelihood, that is, solving the NLP problems in Eqs.~\eqref{eq9} and \eqref{eq16} to find the closest point within the CCMR (defined by Eq.~\eqref{eq8}). Since the NLP problems contain nonlinear objectives and constraints, it’s inefficient to solve them using conventional numerical optimization methods. Motivated by Ref.~\cite{Pavanello2024IEEE}, a novel RPO approach using both the polynomial approximation and convex optimization is proposed in this work to resolve the NLP problems. It’s worth noting that the RPO approach was first introduced in Ref.~\cite{Pavanello2024IEEE} for solving the collision avoidance optimization problems, and the RPO proposed in this paper is an improved version of the one in Ref.~\cite{Pavanello2024IEEE}. The difference between the two RPOs is pictorially shown in Fig.~\ref{fig1} and will be detailed later in this subsection. Moreover, the advantages of the proposed RPO over the previous one in handling the maneuver detection problem will be shown in Sec.~\ref{sec:Maneuver Detection Results Using Single-Epoch Angle}.

Take the NLP problem in Eq.~\eqref{eq9} as an example to show the process of the proposed RPO approach (\emph{i.e.}, consider measurements at a single epoch $t_1$). First, a new nonlinear function is defined for lighter notation:
\begin{equation} \label{eq20}
    {\boldsymbol{\tilde z}_1} = \boldsymbol{g}(\delta {\boldsymbol{x}_0}) = \boldsymbol{h}({\boldsymbol{\tilde x}_1}) = \boldsymbol{h}[\boldsymbol{F}({\boldsymbol{\hat x}_0} + \delta {\boldsymbol{x}_0},{t_0},{t_1})] \,.
\end{equation}
Like Ref.~\cite{Pavanello2024IEEE}, an \emph{n}-th order Taylor polynomial ($n \in {\mathbb{N}^ +}$ is a user-defined parameter representing the highest order of the polynomial) is derived in the DA framework to approximate the nonlinear function ${\boldsymbol{\tilde z}_1} = \boldsymbol{g}(\delta {\boldsymbol{x}_0})$ as
\begin{equation} \label{eq21}
    {\boldsymbol{\tilde z}_1} \approx {\cal T}_{\boldsymbol{g}(\delta {\boldsymbol{x}_0})}^n(\delta {\boldsymbol{x}_0}) = \boldsymbol{g}(\mathbf{0}) + \sum\limits_{i = 1}^n {{{\cal G}^{(i)}}{{(\delta {\boldsymbol{x}_0})}^i}} \,,
\end{equation}
where $\delta {\boldsymbol{x}_0} = {\mathbf{0}_{6 \times 1}}$ is the reference point on which to base the \emph{n}-th order Taylor polynomial is expanded, $\boldsymbol{g}(\mathbf{0})$ is the constant term of the polynomial, and ${{\cal G}^{(i)}} \in {\mathbb{R}^{m \times 6 \times  \cdots  \times 6}}$ is an (\emph{i}+1)-th order tensor, defined as
\begin{equation} \label{eq22}
    {{\cal G}^{(i)}} = {\left. {\frac{{{\partial ^i}\boldsymbol{g}(\delta {\boldsymbol{x}_0})}}{{\partial \delta {\boldsymbol{x}_0} \cdots \partial \delta {\boldsymbol{x}_0}}}} \right|_{\delta {\boldsymbol{x}_0} = {\mathbf{0}_{6 \times 1}}}} \,.
\end{equation}

The derivation process of Eq.~\eqref{eq21} includes two steps. In the first step, one should derive the high-order Taylor polynomial for ${\boldsymbol{\tilde x}_1}$ with respect to $\delta {\boldsymbol{x}_0}$ as
\begin{equation} \label{eq23}
    {\boldsymbol{\tilde x}_1} = \boldsymbol{F}({\boldsymbol{\hat x}_0} + \delta {\boldsymbol{x}_0},{t_0},{t_1}) \approx {\cal T}_{\boldsymbol{F}(\delta {\boldsymbol{x}_0})}^n(\delta {\boldsymbol{x}_0}) \,,
\end{equation}
which can be obtained by implementing an 8$^{\mathrm{th}}$-order Runge-Kutta (RK78) with a relative tolerance of $10^{-12}$ and an absolute tolerance of $10^{-12}$ in the DA framework. Then, in the second step, one should derive a Taylor polynomial for ${\boldsymbol{\tilde z}_1}$ with respect to ${\boldsymbol{\tilde x}_1}$ as ${\boldsymbol{\tilde z}_1} \approx {\cal T}_{\boldsymbol{h}({{\boldsymbol{\tilde x}}_1})}^n({\boldsymbol{\tilde x}_1})$, which, after replacing ${\boldsymbol{\tilde x}_1}$ with Eq.~\eqref{eq23}, becomes Eq.~\eqref{eq21}.

By substituting the high-order Taylor polynomial approximation in Eq.~\eqref{eq21} into Eq.~\eqref{eq9}, the NLP optimization problem can be rewritten as a polynomial optimization problem (POP):
\begin{equation} \label{eq24}
    \begin{array}{*{20}{l}}
        {\mathop {\min }\limits_{\delta {\boldsymbol{x}_0}} }&{J = \delta \boldsymbol{\tilde z}_1^T\boldsymbol{R}_1^{ - 1}\delta {{\boldsymbol{\tilde z}}_1}}\\
        {{\rm{s}}{\rm{.t}}{\rm{.}}}&{\delta {{\boldsymbol{\tilde z}}_1} = {{\boldsymbol{\tilde z}}_1} - {\boldsymbol{z}_1}}\\
        {}&{{{\boldsymbol{\tilde z}}_1} \approx \boldsymbol{g}(\mathbf{0}) + \sum\limits_{i = 1}^n {{{\cal G}^{(i)}}{{(\delta {\boldsymbol{x}_0})}^i}} }\\
        {}&{\delta \boldsymbol{x}_0^T\boldsymbol{P}_0^{ - 1}\delta {\boldsymbol{x}_0} \le 2{{\cal M}_x}}
    \end{array} \,.
\end{equation}

Recall that both ${\boldsymbol{R}_1}$ and ${\boldsymbol{P}_0}$ are assumed to be positive definite matrices; therefore, the third constraint becomes a second-order cone constraint. Moreover, the nonlinear objective in Eq.~\eqref{eq24} can be easily convexified by adding a slack variable $s$ with a second-order cone constraint \cite{Han2019ASR} as
\begin{equation} \label{eq25}
    \delta \boldsymbol{\tilde z}_1^T\boldsymbol{R}_1^{ - 1}\delta {\boldsymbol{\tilde z}_1} \le s \,,
\end{equation}
and rewriting the objective function as $J = s$. Hence, the second constraint (\emph{i.e.}, ${\boldsymbol{\tilde z}_1} \approx \boldsymbol{g}(\mathbf{0}) + \sum\limits_{i = 1}^n {{{\cal G}^{(i)}}{{(\delta {\boldsymbol{x}_0})}^i}} $) becomes the only challenge in this POP as it is still nonlinear. Note that the POP in Eq.~\eqref{eq22} can be solved using general numerical optimization methods (\emph{e.g.}, sequential least squares programming [SLSQP]). However, these methods suffer from drawbacks related to requiring an initial guess or high computational cost, which will be shown in Sec.~\ref{sec:Maneuver Detection Results Using Single-Epoch Angle} and Sec.~\ref{sec:Maneuver Detection Results Using Multiple-Epoch Angle}.

Both RPOs (in this work and Ref.~\cite{Pavanello2024IEEE}) linearize this constraint. As shown in Fig.~\ref{fig1}, they adopt the same linearization approach in the first iteration; however, they differ in the second and subsequent iterations. In the first iteration, the nonlinear constraint ${\boldsymbol{\tilde z}_k} \approx \boldsymbol{g}(\mathbf{0}) + \sum\limits_{i = 1}^n {{{\cal G}^{(i)}}{{(\delta {\boldsymbol{x}_0})}^i}} $ is linearized as 
\begin{equation} \label{eq26}
    {\boldsymbol{\tilde z}_1} \approx \boldsymbol{g}(\mathbf{0}) + {{\cal G}^{(1)}}\delta {\boldsymbol{x}_0} \,.
\end{equation}
Then, the POP is reformulated as
\begin{equation} \label{eq27}
    \begin{array}{*{20}{l}}
        {\mathop {\min }\limits_{\delta {\boldsymbol{x}_0}} }&{J = s}\\
        {{\rm{s}}{\rm{.t}}{\rm{.}}}&{\delta {{\boldsymbol{\tilde z}}_1} = {{\boldsymbol{\tilde z}}_1} - {\boldsymbol{z}_1}}\\
        {}&{{{\boldsymbol{\tilde z}}_1} \approx \boldsymbol{g}(\mathbf{0}) + {{\cal G}^{(1)}}\delta {\boldsymbol{x}_0}}\\
        {}&{\delta \boldsymbol{\tilde z}_1^T\boldsymbol{R}_1^{ - 1}\delta {{\boldsymbol{\tilde z}}_1} \le s}\\
        {}&{\delta \boldsymbol{x}_0^T\boldsymbol{P}_0^{ - 1}\delta {\boldsymbol{x}_0} \le 2{{\cal M}_x}}
    \end{array} \,,
\end{equation}
becoming a second-order cone problem (SOCP).

In this work, the SOCP in Eq.~\eqref{eq27} (also the SOCPs in the second and subsequent iterations) is solved using an interior-point optimizer, MOSEK \footnote{Available at \url{https://www.mosek.com/}}, to obtain the optimal solution $\delta \boldsymbol{x}_0^{(1)}$. This solution is called the linear solution in Fig.~\ref{fig1} (represented by green markers), as only the linear term of the polynomial ${\cal T}_{\boldsymbol{g}(\delta {\boldsymbol{x}_0})}^n(\delta {\boldsymbol{x}_0})$ is considered. Since only the linear term (\emph{i.e.}, the first-order term) of the polynomial is considered in the first iteration, the accuracy of the solution cannot be guaranteed. Therefore, higher-order terms need to be incorporated in subsequent iterations. 
In this process, the RPO method proposed in this work differs from the one in Ref.~\cite{Pavanello2024IEEE}. The following section first introduces the proposed approach, as illustrated in the right subplot of Fig.~\ref{fig1}.

As shown in Fig.~\ref{fig1}, we directly define $\delta \boldsymbol{x}_0^{(j - 1)}$ as the optimal solution of the (\emph{j}-1)-th iteration. In the \emph{j}-th iteration, the proposed RPO approach linearizes the polynomial up to the highest order (\emph{i.e.}, \emph{n}) based on $\delta \boldsymbol{x}_0^{(j - 1)}$ as
\begin{equation} \label{eq31}
    {\boldsymbol{\tilde z}_1} \approx \boldsymbol{g}[\delta \boldsymbol{x}_0^{(j - 1)}] + \boldsymbol{\cal A}_1^{(j)}\Delta \boldsymbol{x}_0^{(j)} \,,
\end{equation}
where
\begin{equation} \label{eq32}
    \Delta \boldsymbol{x}_0^{(j)} = \delta \boldsymbol{x}_0^{(j)} - \delta \boldsymbol{x}_0^{(j - 1)} \,,
\end{equation}
\begin{equation} \label{eq33}
    \boldsymbol{\cal A}_1^{(j)} = \sum\limits_{i = 1}^n {i{\boldsymbol{\cal G}^{(i)}}{{[\delta \boldsymbol{x}_0^{(j - 1)}]}^{i - 1}}}  \in {\mathbb{R}^{m \times 6}} \,.
\end{equation}
Therefore, the SOCP is rewritten as
\begin{equation} \label{eq34}
    \begin{array}{*{20}{l}}
        {\mathop {\min }\limits_{\Delta \boldsymbol{x}_0^{(j)}} }&{J = s}\\
        {{\rm{s}}{\rm{.t}}{\rm{.}}}&{\delta {{\boldsymbol{\tilde z}}_1} = {{\boldsymbol{\tilde z}}_1} - {\boldsymbol{z}_1}}\\
        {}&{{{\boldsymbol{\tilde z}}_1} \approx \boldsymbol{g}[\delta \boldsymbol{x}_0^{(j - 1)}] + \boldsymbol{\cal A}_1^{(j)}\Delta \boldsymbol{x}_0^{(j)}}\\
        {}&{\delta \boldsymbol{\tilde z}_1^T\boldsymbol{R}_1^{ - 1}\delta {{\boldsymbol{\tilde z}}_1} \le s}\\
        {}&{{{(\delta \boldsymbol{x}_0^{(j - 1)} + \Delta \boldsymbol{x}_0^{(j)})}^T}\boldsymbol{P}_0^{ - 1}(\delta \boldsymbol{x}_0^{(j - 1)} + \Delta \boldsymbol{x}_0^{(j)}) \le 2{{\cal M}_x}}
    \end{array} \,.
\end{equation}

As plotted in Fig.~\ref{fig1}, once the optimal solution $\Delta \boldsymbol{x}_0^{(j)}$ is solved (using MOSEK), the initial state deviation is updated as $\delta \boldsymbol{x}_0^{(j)} = \Delta \boldsymbol{x}_0^{(j)} + \delta \boldsymbol{x}_0^{(j - 1)}$. Similarly, the approach converges when a pre-defined threshold is satisfied as
\begin{equation} \label{eq35}
    \left\| {\delta \boldsymbol{x}_0^{(j - 1)} - \delta \boldsymbol{x}_0^{(j)}} \right\| = \left\| {\Delta \boldsymbol{x}_0^{(j)}} \right\| \le \eta \,.
\end{equation}
According to the above discussions, the main process of the proposed RPO method is presented in Algorithm~\ref{alg:RPO}.

\begin{algorithm}
	\caption{Pseudocode of the proposed RPO for determining the indicator with measurement at a single epoch} \label{alg:RPO}
	
	\SetKwInOut{Input}{Input}
	\SetKwInOut{Output}{Output}
	\SetKwInOut{Return}{Return}
	
	\Input{The initial estimation (mean) ${\boldsymbol{\hat x}_0}$ and its associated covariance ${\boldsymbol{P}_0}$, orbital dynamics $\boldsymbol{f}(t,\boldsymbol{x})$, measurement model $\boldsymbol{h}(\boldsymbol{x})$, observed measurement ${\boldsymbol{z}_1} = \boldsymbol{h}({\boldsymbol{x}_1}) + {\boldsymbol{\varepsilon }_1}$ at a given epoch $t_1$, measurement noise covariance ${\boldsymbol{R}_1}$, confidence level ${\alpha _x}$, maximal order of the polynomial \emph{n}, threshold $\eta$.}
	\Output{The optimal solution $\delta \boldsymbol{x}_0^ * $ and the assoicated closest point $\delta \boldsymbol{\tilde z}_1^ * $.}
	
	Derive the polynomial ${\boldsymbol{\tilde z}_1} \approx {\cal T}_{\boldsymbol{g}(\delta {\boldsymbol{x}_0})}^n(\delta {\boldsymbol{x}_0})$ in the DA scheme\;

    Determine the inverse CDF of the $\chi^2$ distribution to obtain ${{\cal M}_x} = {q_{{\chi ^2}}}({\alpha _x},6)$\;

    Linearize the polynomial ${\cal T}_{\boldsymbol{g}(\delta {\boldsymbol{x}_0})}^n(\delta {\boldsymbol{x}_0})$ at zero point to obtain ${\boldsymbol{\tilde z}_1} \approx \boldsymbol{g}(\mathbf{0}) + {\boldsymbol{\cal G}^{(1)}}\delta {\boldsymbol{x}_0}$\;

    Solve the SOCP in Eq.~\eqref{eq27} to obtain the linear optimal solution $\delta \boldsymbol{x}_0^{(1)}$\;

    Let $j \gets 2$ and  ${\rm{tol}} \gets  + \infty $\;
	
	\While{$\rm{tol} > \eta $}{
		Linearize the polynomial ${\cal T}_{\boldsymbol{g}(\delta {\boldsymbol{x}_0})}^n(\delta {\boldsymbol{x}_0})$ at $\delta \boldsymbol{x}_0^{(j - 1)}$ to obtain ${\boldsymbol{\tilde z}_1} \approx \boldsymbol{g}[\delta \boldsymbol{x}_0^{(j - 1)}] + \boldsymbol{\cal A}_1^{(j)}\Delta \boldsymbol{x}_0^{(j)}$\;

        Solve the SOCP in Eq.~\eqref{eq34} to obtain the optimal solution $\Delta \boldsymbol{x}_0^{(j)}$\;

        Update the results as $\delta \boldsymbol{x}_0^{(j)} \leftarrow \Delta \boldsymbol{x}_0^{(j)} + \delta \boldsymbol{x}_0^{(j - 1)}$ and ${\rm{tol}} = \left\| {\delta \boldsymbol{x}_0^{(j - 1)} - \delta \boldsymbol{x}_0^{(j)}} \right\| = \left\| {\Delta \boldsymbol{x}_0^{(j)}} \right\|$\;

        $j \leftarrow j + 1$\;
	}
	
	\Return{$\delta \boldsymbol{x}_0^ *  \leftarrow \delta \boldsymbol{x}_0^{(j)}$ and $\delta \boldsymbol{\tilde z}_1^ *  = \boldsymbol{h}[\boldsymbol{F}({\boldsymbol{\hat x}_0} + \delta \boldsymbol{x}_0^ * ,{t_0},{t_1})] - {\boldsymbol{z}_1}$.}
\end{algorithm}

Compared with the RPO method proposed in Ref.~\cite{Pavanello2024IEEE} (shown by the subplot on the left of Fig.~\ref{fig1}), the main improvements of the proposed RPO method lie in the use of the highest-order polynomial throughout the second and subsequent iterations, without the need to gradually increase the polynomial order step by step. This modification significantly accelerates the convergence (will be illustrated in Fig.~\ref{fig5}), as all higher-order terms are utilized in each iteration based on the solution from the previous step.

When measurements at multiple epochs are available, one can stack the linear matrices in Eq.~\eqref{eq33} at different epochs to obtain
\begin{equation} \label{eq36}
    {\boldsymbol{\cal A}^{(j)}} = [\boldsymbol{\cal A}_1^{(j)};\boldsymbol{\cal A}_2^{(j)}; \cdots ;\boldsymbol{\cal A}_N^{(j)}] \in {\mathbb{R}^{Nm \times 6}} \,.
\end{equation}
Then, the same RPO process in Algorithm~\ref{alg:RPO} can be implemented.

\section{Integrated Indicator Approach} \label{sec:Maneuver Detection Approaches}
The original CDMI approach (see Sec.~\ref{sec:Indicator with Measurement at Single Epoch} and Sec.~\ref{sec:Indicator with Measurement at Multiple Epochs}) offers a straightforward implementation, as it requires executing the RPO only once. 
However, it has two critical limitations. First, it depends on a pre-defined threshold (\emph{i.e.}, confidence level ${\alpha _x}$), which significantly impacts maneuver detection accuracy, as will be demonstrated in Fig.~\ref{fig17} and Fig.~\ref{fig20}. For example, if one selects $\alpha_x=0$ (an entirely unreasonable choice, but useful to illustrate an extreme case), the algorithm should certainly detect a maneuver, whether it exists, leading to 100\% accuracy in maneuvering cases while 0\% accuracy in non-maneuvering cases. 
Second, this method is inherently limited to binary maneuver detection—it can only determine whether a spacecraft has performed a maneuver without providing probabilistic insight. In many practical scenarios, however, estimating maneuver probability is also essential, as it facilitates tasks such as intention inference and sensor tasking (something related to decision) in space operations.

To overcome the limitations of the original CDMI approach in Sec.~\ref{sec:Indicator with Measurement at Single Epoch} and Sec.~\ref{sec:Indicator with Measurement at Multiple Epochs}, this section presents an integrated-indicator approach for maneuver detection. This method aims to enhance detection accuracy by eliminating the need to manually select the confidence level ${\alpha _x}$ while providing an indication of the likelihood of the potential maneuver. To be convenient, we define
\begin{equation} \label{eq37}
    P=\int\limits_0^1 {{\alpha _z}d{\alpha _x}}  \,,
\end{equation}
which represents the integration of the measurement confidence level (\emph{i.e.}, ${\alpha _z}$) over the interval ${\alpha _x} \in [0,1]$. By integrating the CDMI (\emph{i.e.}, ${\alpha _z} \ge {\alpha _x}$) over the interval ${\alpha _x} \in [0,1]$, one has
\begin{equation} \label{eq39}
    \int\limits_0^1 {{\alpha _z}d{\alpha _x}} \ge \int\limits_0^1 {{\alpha _x}d{\alpha _x}}  = 0.5 \,,
\end{equation}
which finally becomes the integrated CDMI:
\begin{equation} \label{eq40}
    P=\int\limits_0^1 {{\alpha _z}d{\alpha _x}}  \ge 0.5 \,.
\end{equation}
This integration strategy offers two key advantages. First, by evaluating the measurement confidence level (\emph{i.e.}, $\alpha_z$) across the entire range of possible state confidence levels ($\alpha_x \in [0, 1]$), it effectively balances the contributions from different levels of prior uncertainty. This eliminates the need to manually select a specific $\alpha_x$, which is often difficult to tune and can significantly influence detection outcomes. Second, the integral form allows the resulting indicator to reflect the overall tendency of $\alpha_z$ to dominate $\alpha_x$ across the full interval. In other words, the integrated CDMI captures maneuver indications that are supported under most plausible prior confidence levels, thereby maintaining reliable detection performance without parameter tuning.

Motivated by Eq.~\eqref{eq40}, the integration $P$ in Eq.~\eqref{eq37} can be considered as a probabilistic indication of the maneuver. Note that $P$ is not a rigorous maneuver probability measure. It will be shown in Fig.~\ref{fig12} and Fig.~\ref{fig19} that $P$ tends to be larger than 0.5 in maneuver cases, while in non-maneuver cases, it tends to be smaller than 0.5. Thus, it’s reasonable that $P \le 0.5$ indicates no maneuver.

\begin{figure}[!h]
	\centering
	\includegraphics[width=0.48\textwidth]{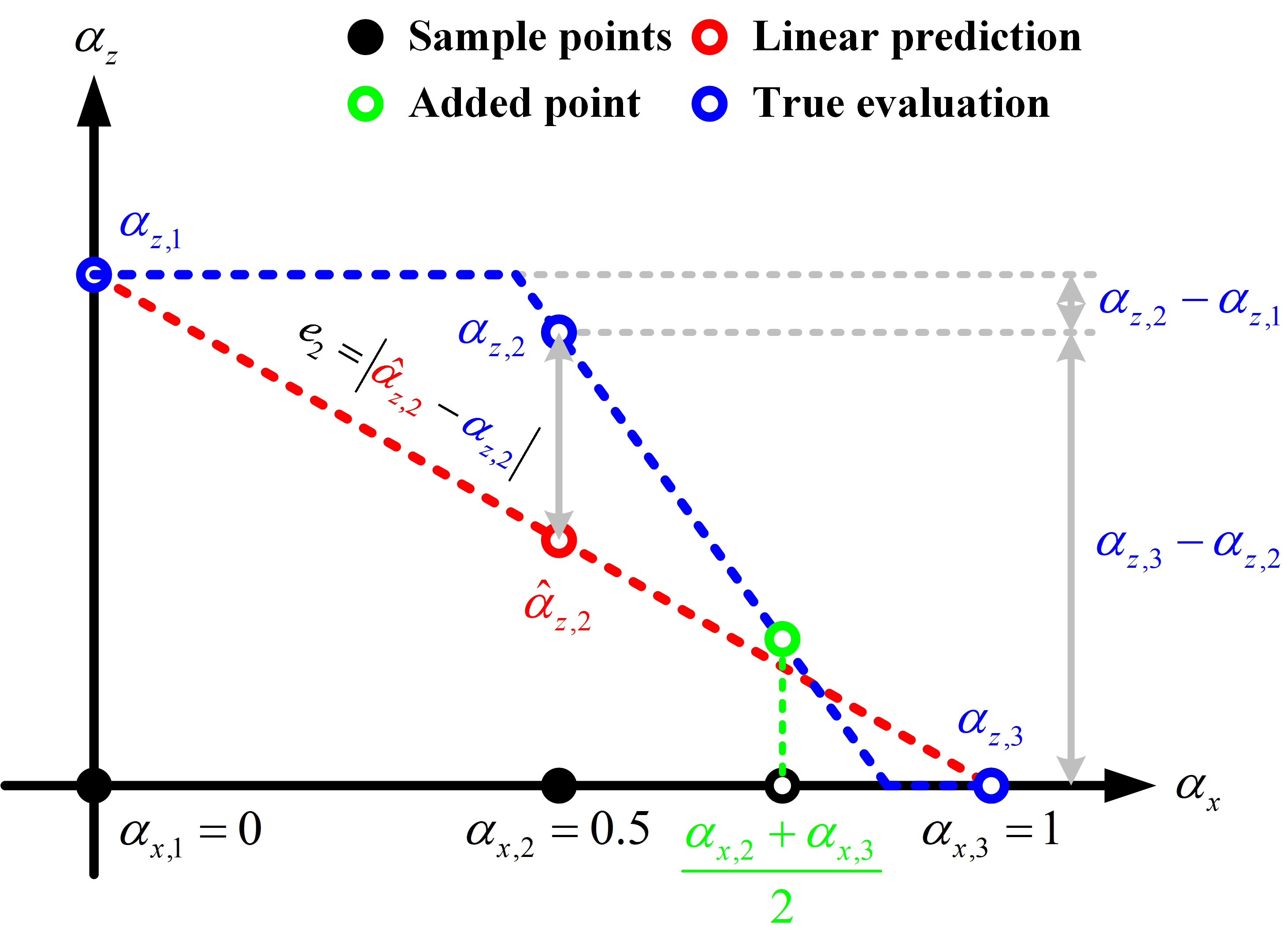}
	\caption{An illustration of the proposed adaptive sample method.}
	\label{fig2}
\end{figure}

Since the integrated CDMI (or the integration $P$) is obtained through integration, using numerical methods such as Runge-Kutta would result in a high computational cost, which would require repeated evaluations of $\alpha_z$ (under different $\alpha _x$). This, in turn, necessitates repeatedly solving for the nearest point using the RPO method. To improve computational efficiency, this work employs the trapezoidal rule to obtain the integration, expressed as
\begin{equation} \label{eq42}
    \begin{aligned}
        P &= \int\limits_0^1 {{\alpha _z}d{\alpha _x}} \\
         & \approx {\alpha _{z,1}}\frac{{{\alpha _{x,2}} - {\alpha _{x,1}}}}{2} + {\alpha _{z,{i_{\max }}}}\frac{{{\alpha _{x,{i_{\max }}}} - {\alpha _{x,{i_{\max }} - 1}}}}{2}\\
         & + \sum\limits_{i = 2}^{{i_{\max }} - 1} {\frac{{({\alpha _{z,i}} + {\alpha _{z,i + 1}})({\alpha _{x,i + 1}} - {\alpha _{x,i}})}}{2}} 
    \end{aligned} \,,
\end{equation}
where ${\alpha _{x,i}} \in [0,1]$ and ${\alpha _{z,i}} \in [0,1]$ represent the \emph{i}-th sample point of $\alpha_x$ and its associated $\alpha_z$. To further improve efficiency, an adaptive sampling method is employed in this work to select points over the interval ${\alpha _x} \in [0,1]$, as shown in Fig.~\ref{fig2}. 

First, choose three sample points as ${\alpha _{x,1}} = 0$, ${\alpha _{x,2}} = 0.5$, and ${\alpha _{x,3}} = 1$, and compute the corresponding values of $\alpha_z$. Re-note that the implementation of RPO is required only for the point ${\alpha _{x,2}} = 0.5$, and the values of $\alpha_z$ of the other two points (\emph{i.e.}, ${\alpha _{x,1}} = 0$ and ${\alpha _{x,3}} = 1$) can be directly obtained using Eqs.~\eqref{eq12}-\eqref{eq13}. Then, we have a point set as
\begin{equation} \label{eq43}
    {\cal S} = \left\{ {({\alpha _{x,1}},{\alpha _{z,1}}),({\alpha _{x,2}},{\alpha _{z,2}}),({\alpha _{x,3}},{\alpha _{z,3}})} \right\} \,.
\end{equation}
For any three consecutive points in the point set ${\cal S}$, say, $({\alpha _{x,i}},{\alpha _{z,i}})$, $({\alpha _{x,i + 1}},{\alpha _{z,i + 1}})$, and $({\alpha _{x,i + 2}},{\alpha _{z,i + 2}})$, the first point $({\alpha _{x,i}},{\alpha _{z,i}})$ and the third point $({\alpha _{x,i + 2}},{\alpha _{z,i + 2}})$ are used to construct a linear interpolation to predict the second point, $({\alpha _{x,2}},{\hat \alpha _{z,2}})$, and the interpolation error is then computed as
\begin{equation} \label{eq44}
    {e_{i + 1}} = \left| {{{\hat \alpha }_{z,i + 1}} - {\alpha _{z,i + 1}}} \right| \,.
\end{equation}

Next, the three consecutive points with the largest interpolation error are focused on. If this largest interpolation error is smaller than a pre-defined threshold (\emph{i.e.}, $e({\alpha _{z,i + 1}}) \le {\varepsilon _1}$), stop adding new points; else, one needs to add new points using the following steps:

\begin{enumerate}[label=(\arabic*)]
  \item \label{item:1} Compute the distance between ${\alpha _{x,i}}$ and ${\alpha _{x,i + 1}}$, and the distance between ${\alpha _{x,i + 1}}$ and ${\alpha _{x,i + 2}}$. If these two distances are smaller than a pre-defined threshold, say, ${\alpha _{x,i + 2}} - {\alpha _{x,i + 1}} \le {\varepsilon _2}$ and ${\alpha _{x,i + 1}} - {\alpha _{x,i}} \le {\varepsilon _2}$, stop adding new samples; otherwise, go to step~\ref{item:2}.
  \item \label{item:2} Add new sample points in the interval where the curve changes more rapidly. That is to say, compute ${\alpha _{z,i + 2}} - {\alpha _{z,i + 1}}$ and ${\alpha _{z,i + 1}} - {\alpha _{z,i}}$, and choose the larger one. A new sample is added at the midpoint of the selected interval. Obtain the value of $\alpha _z$ for this newly added sample using the RPO method. 
  \item \label{item:3} Update the point set ${\cal S}$, and sort the points in ${\cal S}$ according to the values of $\alpha _x$. Go back to step~\ref{item:1} until no new point is added.
\end{enumerate}

It should be noted that the above proposed adaptive sampling method is developed based on our empirical observations. While it may not be the optimal way, and there could exist more effective approaches, it has demonstrated advantages in the simulations. It will be shown in the numerical section that this adaptive sampling method can significantly reduce the computational cost while maintaining the same level of accuracy. 

\section{Numerical Results} \label{sec:Numerical Results}
The proposed methods are applied to maneuver detection in cislunar space, where the dynamics exhibit strong nonlinearity, resulting in non-Gaussian distributed uncertainties. All simulations are conducted on a personal laptop equipped with a 2.5 GHz processor (12th Gen Intel(R) Core(TM) i5-12500H) and 16 GB of RAM, using Python 3.10 within the PyCharm development environment.

\subsection{Scenario Design} \label{sec:Scenario Design}
The Circular Restricted Three-Body Problem (CRTBP) dynamics is employed in the numerical simulation, as it is a classical simplified dynamics model for describing the motions in the Earth-Moon system. The mathematical model is expressed as
\begin{equation} \label{eq45}
    \left\{ \begin{array}{l}
        \ddot x = 2\dot y + x - \frac{{(1 - \mu )(x + \mu )}}{{r_1^3}} - \frac{{\mu (x + \mu  - 1)}}{{r_2^3}}\\
        \ddot y =  - 2\dot x + y - \frac{{(1 - \mu )y}}{{r_1^3}} - \frac{{\mu y}}{{r_2^3}}\\
        \ddot z =  - \frac{{(1 - \mu )z}}{{r_1^3}} - \frac{{\mu z}}{{r_2^3}}
    \end{array} \right. \,,
\end{equation}
where $\boldsymbol{x} = [\boldsymbol{r};\boldsymbol{v}] = [x,y,z,\dot x,\dot y,\dot z] \in {\mathbb{R}^6}$ is the non-dimensional orbital state in the Earth-Moon rotating coordinate, $\mu=0.0121505839$ is the non-dimensional gravitational constant of the Earth-Moon three-body system, and
\begin{equation} \label{eq46}
    {r_1} = \sqrt {{{(x + \mu )}^2} + {y^2} + {z^2}} \,,
\end{equation}
\begin{equation} \label{eq47}
    {r_2} = \sqrt {{{(x + \mu  - 1)}^2} + {y^2} + {z^2}} \,.
\end{equation}

Angles-only measurements are considered, with one target and one observer included in the simulation. Assume that the target and the observer move on a linearly stable NRHO and a 9:2 NRHO (for the Gateway mission), respectively. The orbital parameters of the two NRHOs are given in Table~\ref{tab1} and Table~\ref{tab2}. These two NRHOs are located in the vicinity of the Earth-Moon L2 Lagrange point. As non-dimensional coordinates are used, the units for length (LU), velocity (VU), and time (TU) are provided in Table~\ref{tab1}.

\begin{table}[!h]
	\centering
	\caption{Orbital state of the linearly stable NRHO (target) at its apolune}
	\label{tab1}
	\begin{tabular}{|l|r|l|}
        \hline
        \multicolumn{2}{|l|}{Parameter}                     & Value                \\ \hline
        \multirow{3}{*}{Position vector (nd)} & $x$       & 1.07523949148639     \\ \cline{2-3}
                                              & $y$       & 0                    \\ \cline{2-3}
                                              & $z$       & -0.202146176080457   \\ \hline
        \multirow{3}{*}{Velocity vector (nd)} & $\dot{x}$ & 0                    \\ \cline{2-3}
                                              & $\dot{y}$ & -0.192431661980241   \\ \cline{2-3}
                                              & $\dot{z}$ & 0                    \\ \hline
        \multicolumn{2}{|l|}{Length unit (km)}              & 384400               \\ \hline
        \multicolumn{2}{|l|}{Velocity unit (km/s)}          & 1.02454629434750     \\ \hline
        \multicolumn{2}{|l|}{Time unit (s)}                 & 375190.464423878     \\ \hline
        \multicolumn{2}{|l|}{Period (nd)}                   & 2.26679784217712     \\
        \hline
    \end{tabular}
\end{table}

\begin{table}[!h]
	\centering
	\caption{Orbital state of the 9:2 NRHO (observer) at its apolune}
	\label{tab2}
	\begin{tabular}{|l|r|l|}
        \hline
        \multicolumn{2}{|l|}{Parameter}                     & Value                \\ \hline
        \multirow{3}{*}{Position vector (nd)} & $x$       & 1.02202815472411     \\ \cline{2-3}
                                              & $y$       & 0                    \\ \cline{2-3}
                                              & $z$       & -0.182101352652963   \\ \hline
        \multirow{3}{*}{Velocity vector (nd)} & $\dot{x}$ & 0                    \\ \cline{2-3}
                                              & $\dot{y}$ & -0.103270818092086   \\ \cline{2-3}
                                              & $\dot{z}$ & 0                    \\ \hline
        \multicolumn{2}{|l|}{Period (nd)}                 & 1.51119865689808     \\
        \hline
    \end{tabular}
\end{table}

Assume that the target is at its apolune at the initial epoch $t_0$. According to the assumption in Sec.~\ref{sec:Problem Statement}, the initial estimated errors follow a zero-mean Gaussian distribution. The covariance matrix of the initial estimated errors is set as
\begin{equation} \label{eq48}
    {\boldsymbol{P}_0} = \left[ {\begin{array}{*{20}{c}}
        {\sigma _R^2{\boldsymbol{I}_3}}&{{\mathbf{0}_{3 \times 3}}}\\
        {{\mathbf{0}_{3 \times 3}}}&{\sigma _V^2{\boldsymbol{I}_3}}
    \end{array}} \right] \in {\mathbb{R}^{6 \times 6}} \,.
\end{equation}
where ${\sigma _R} = 1{\,\rm{ km}}$ and ${\sigma _V} = 0.1{\,\rm{ m/s}}$ are the position and velocity STDs per axis, respectively. Note that the two values here are assigned for the standard test case. In the robustness analysis, we will examine the impact of varying initial estimation errors on maneuver detection.

As the angles-only sensor is simulated, the right ascension $\alpha$ and declination $\beta$ are employed to represent the measurement; thus, the measurement equation can be written as
\begin{equation} \label{eq49}
    \left\{ {\begin{array}{*{20}{l}}
        {\alpha  = {{\tan }^{ - 1}}\frac{{{x_k} - {x_{O,k}}}}{{{y_k} - {y_{O,k}}}}}\\
        {\beta  = {{\sin }^{ - 1}}\frac{{{x_k} - {x_{O,k}}}}{{\left\| {{\boldsymbol{r}_k} - {\boldsymbol{r}_{O,k}}} \right\|}}}
    \end{array}} \right. \,,
\end{equation}
where ${\boldsymbol{r}_k} = {[{x_k},{y_k},{z_k}]^T}$ and ${\boldsymbol{r}_{O,k}} = {[{x_{O,k}},{y_{O,k}},{z_{O,k}}]^T}$ denote the position vectors of the target and observer at the epoch $t_k$, respectively. In the standard test case, the STD of the measurement noises is set as $5"$. The impacts of measurement noise level on maneuver detection performance will be investigated in Sec.~\ref{sec:Robustness Analysis}. It should be noted that, in the simulations, the simplified objective function (\emph{i.e.}, Eqs.~\eqref{eq12} and \eqref{eq19}) can be employed since the STDs of the measurement noises are the same.

\begin{figure}[!h]
	\centering
	\subfigure[]
	{
		\label{fig3a}
		\centering
		\includegraphics[width=0.4\textwidth]{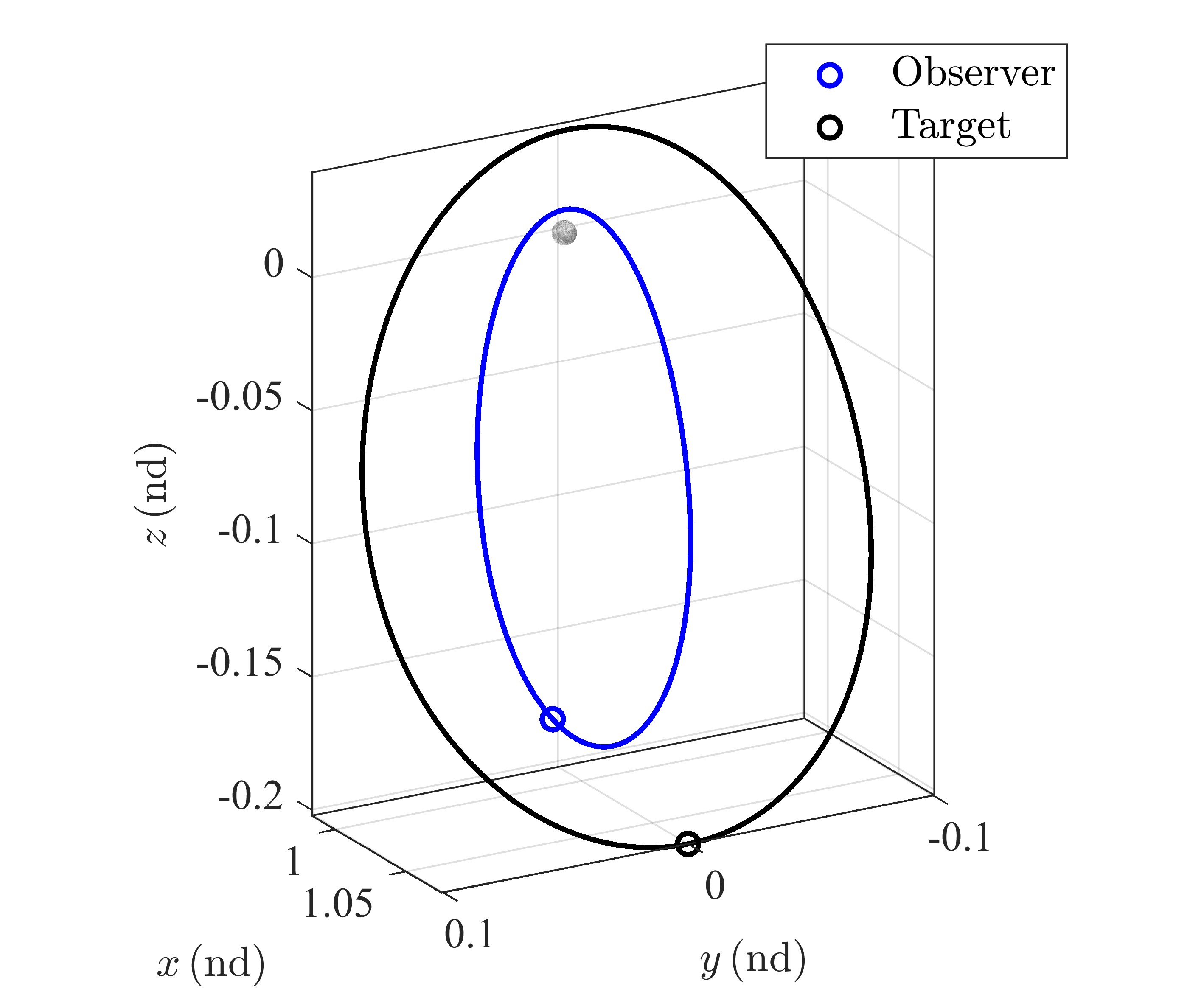}
	}
	\subfigure[]
	{
		\label{fig3b}
		\centering
		\includegraphics[width=0.4\textwidth]{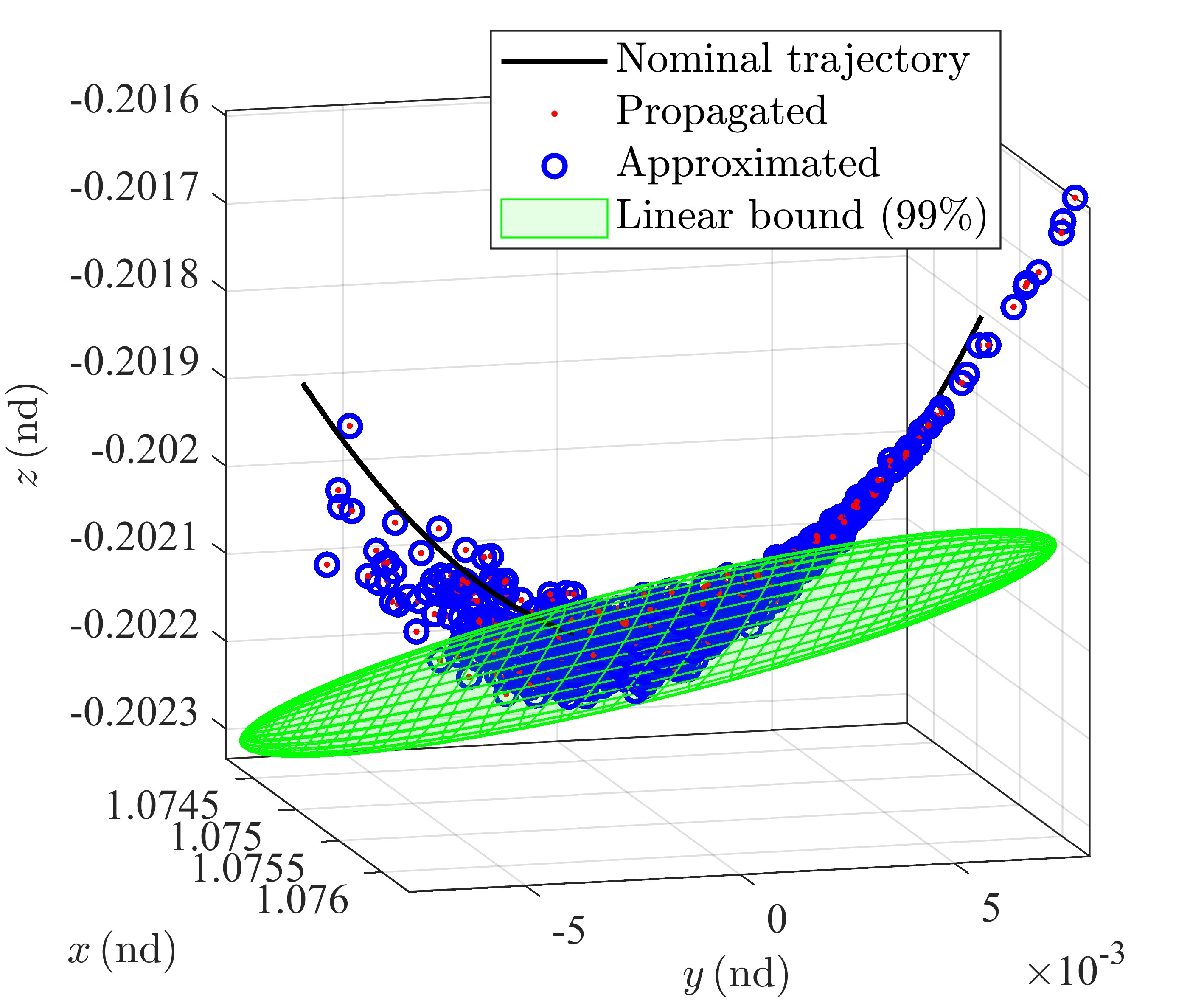}
	}
	\caption{Orbits of the target and observer. \subref{fig3a} Nominal orbits. \subref{fig3b} State distribution at $t_1$.}
	\label{fig3}
\end{figure}

Assume that the first measurement occurs after three orbital revolutions (of the target); that is to say, ${t_1} = 3{T} \approx {\rm{6.8004}} \, {\rm{(nd)}}$ (${T}$ is the period of the target). Figure~\ref{fig3}\subref{fig3a} shows the nominal orbits of two spacecraft. In Fig.~\ref{fig3}\subref{fig3a}, the blue circle represents the location of the observer at the epoch $t_1$, which is obtained by propagating the 9:2 NRHO from its apolune for 0.85 period (of the observer). It is worth noting that this location (0.85 periods from the apolune) is selected because it can provide a unique observational configuration. At this position, the angle measurement distribution is more non-Gaussian than in other locations. This characteristic is significant for testing the algorithm, as it allows us to evaluate its performance under challenging conditions, including strong nonlinear dynamics (CRTBP dynamics) and non-Gaussian measurement distributions (benefits from the selected observation configuration).

Additionally, Fig.~\ref{fig3}\subref{fig3b} shows the impacts of initial estimated uncertainties on the state distribution at the observation epoch (\emph{i.e.}, $t_1$). Using the covariance in Eq.~\eqref{eq48}, one thousand neighboring trajectories (for the target) are generated around the nominal orbit, with their final positions (at the epoch $t_1$) marked by red points in Fig.~\ref{fig3}\subref{fig3b}. A fifth-order Taylor polynomial (\emph{i.e.}, Eq.~\eqref{eq23} with $n = 5$) is derived within the DA framework to approximate these final states (\emph{i.e.}, the red points), with the approximated states indicated by blue circles. Moreover, linear propagation is performed by retaining only the constant and first-order terms of the polynomial, with the corresponding 99\% confidence region represented by the green ellipsoid in Fig.~\ref{fig3}\subref{fig3b}. As shown in Fig.~\ref{fig3}\subref{fig3b}, the highly nonlinear nature of the three-body dynamics leads to a final state distribution that is significantly non-Gaussian, deviating from the linear propagation result. However, the fifth-order Taylor polynomial provides an accurate approximation of the final states, thereby ensuring the reliability of maneuver detection.

\begin{figure*}[!h]
	\centering
	\subfigure[]
	{
		\label{fig4a}
		\centering
		\includegraphics[width=0.4\textwidth]{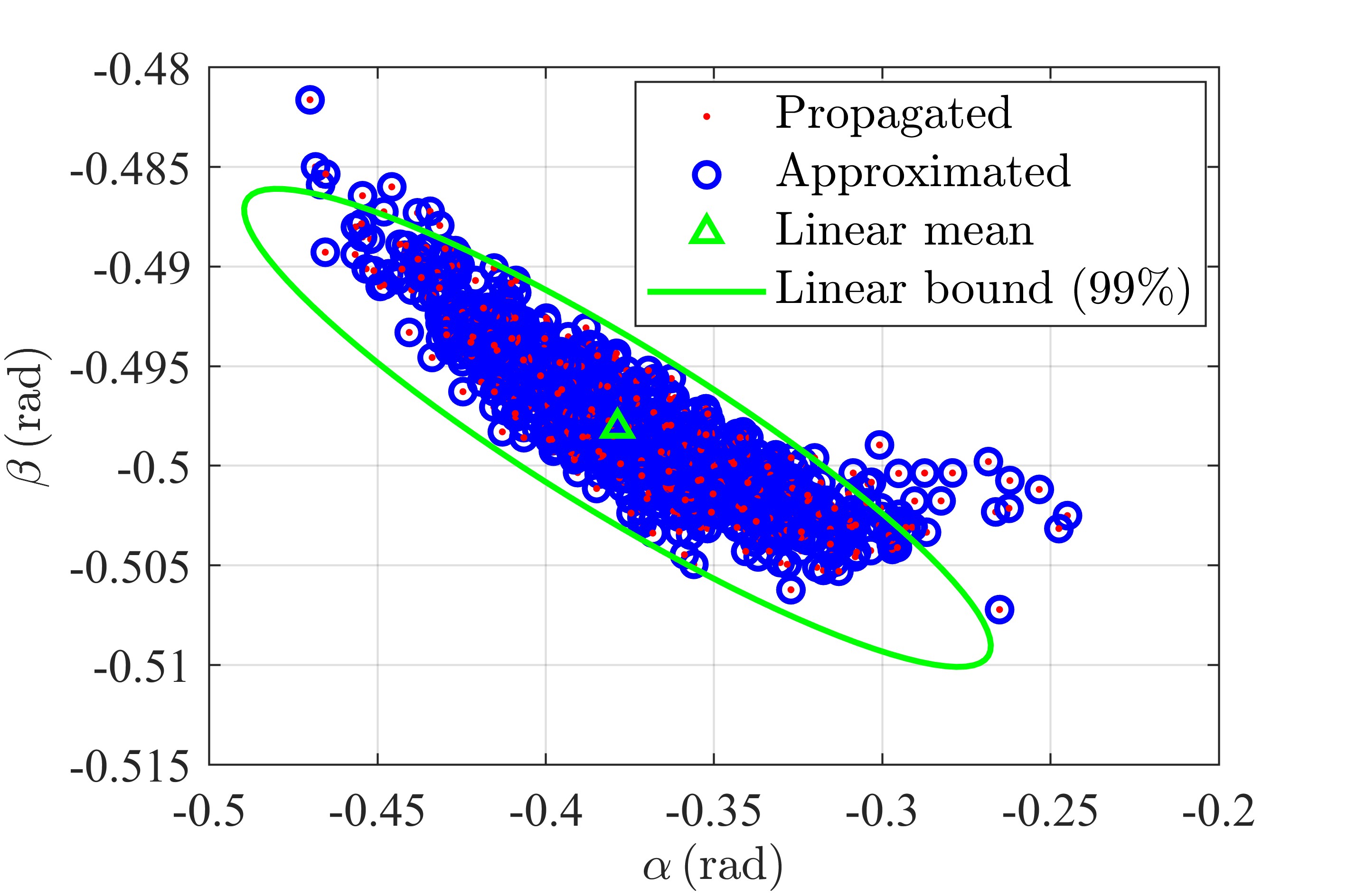}
	}
	\subfigure[]
	{
		\label{fig4b}
		\centering
		\includegraphics[width=0.4\textwidth]{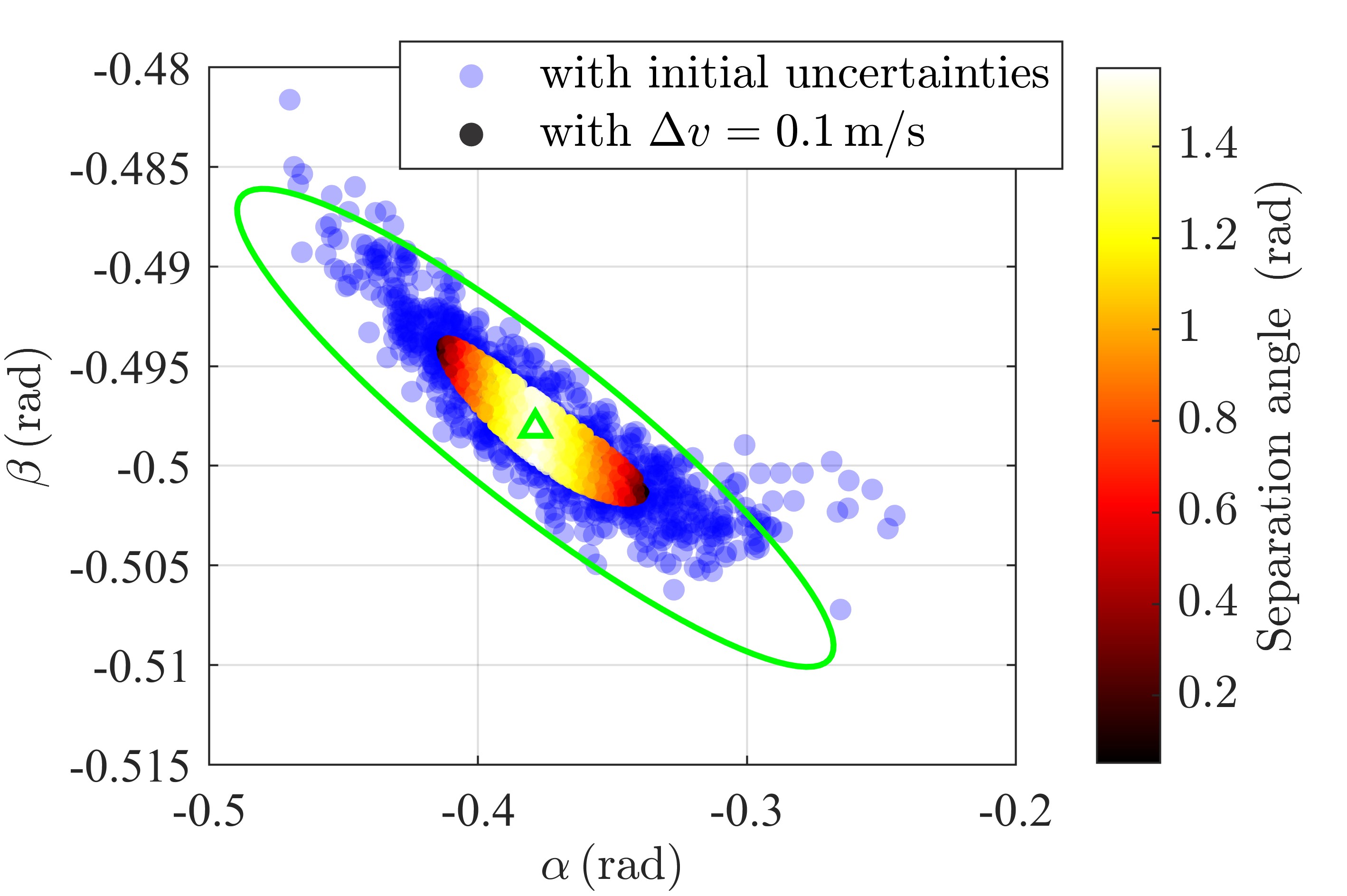}
	}
    \subfigure[]
	{
		\label{fig4c}
		\centering
		\includegraphics[width=0.4\textwidth]{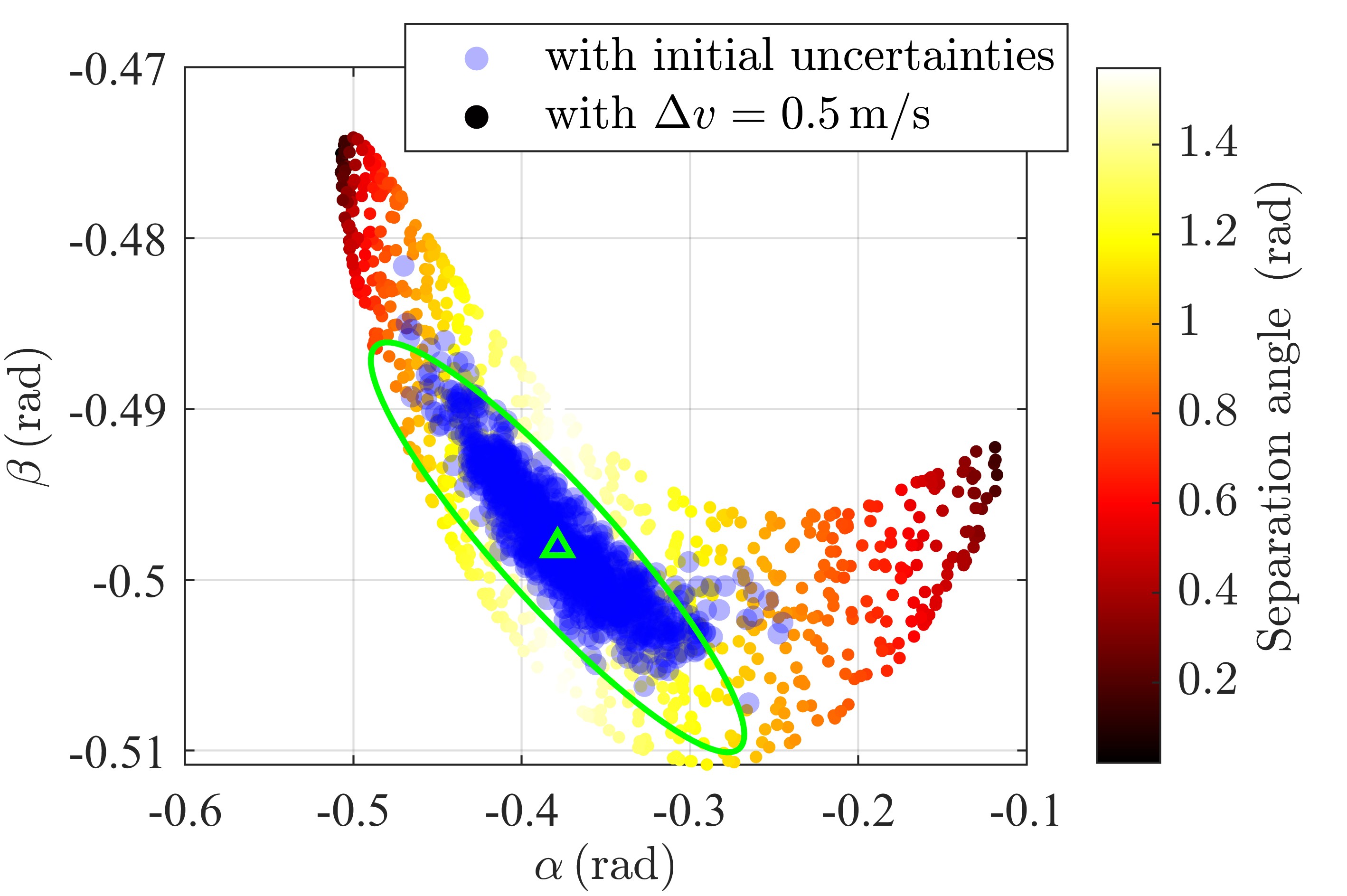}
	}
    \subfigure[]
	{
		\label{fig4d}
		\centering
		\includegraphics[width=0.4\textwidth]{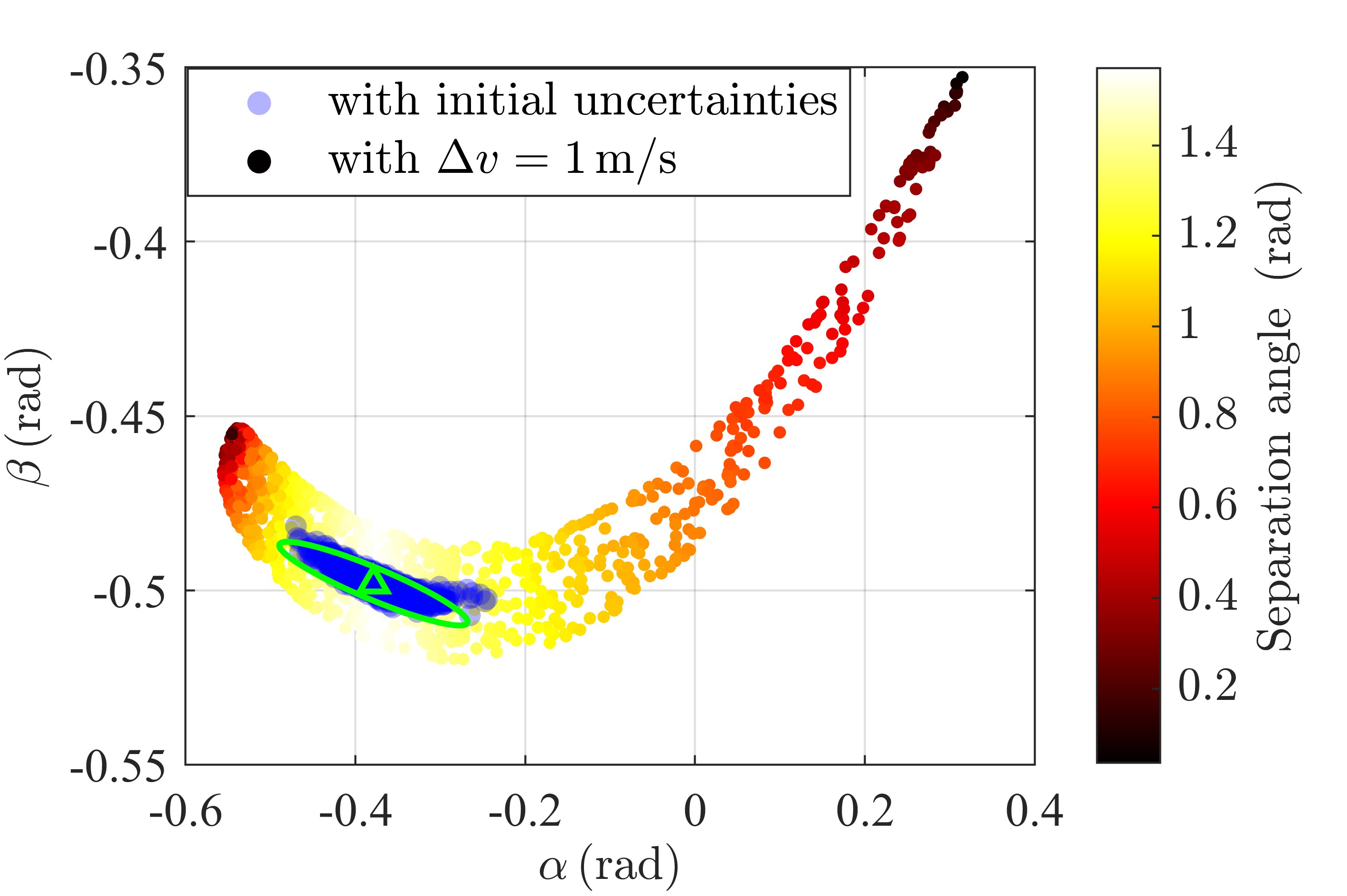}
	}
	\caption{Angle measurement distributions. \subref{fig4a} $\boldsymbol{P}_{0}$. \subref{fig4b} $\boldsymbol{P}_{0}$ vs $\Delta v=0.1\,\rm{m/s}$. \subref{fig4c} $\boldsymbol{P}_{0}$ vs $\Delta v=0.5\,\rm{m/s}$. \subref{fig4d} $\boldsymbol{P}_{0}$ vs $\Delta v=1\,\rm{m/s}$.}
	\label{fig4}
\end{figure*}

Then, we project the 1000 states (the red points in Fig.~\ref{fig3}\subref{fig3b}) on the observation space (using the measurement model in Eq.~\eqref{eq49}), with the angle measurements shown by the red points (labeled as “propagated” since they are obtained by numerically propagating the neighboring orbits and then projecting the states on the observation space) in Fig.~\ref{fig4}\subref{fig4a}. These angle measurements are approximated using a fifth-order Taylor polynomial (\emph{i.e.}, Eq.~\eqref{eq21}), as shown by blue circles in Fig.~\ref{fig4}\subref{fig4a}, which match the red points well. Additionally, linear propagation is performed, with the results colored green. Many red points are outside the 99\% confidence bound, indicating that linear propagation fails to capture the nonlinearity.

It is important to note that the cloud points in Fig.~\ref{fig4}\subref{fig4a} represent the effects of initial estimation uncertainties (in the absence of maneuvers) on the angle measurements. The subsequent analysis focuses on the impact of unknown maneuvers. Impulsive maneuvers with arbitrary directions are added to the nominal state immediately after the initial epoch $t_0$, while the initial estimation errors are disregarded to facilitate a separate analysis. Figures~\ref{fig4}\subref{fig4b}–\ref{fig4}\subref{fig4d} illustrate the distribution of angle cloud points under different magnitudes of impulsive maneuvers. These cloud points are color-coded, where darker shades indicate a smaller separation angle between the maneuver direction and the most sensitive direction (as determined by the Cauchy-Green tensor [CGT] approach \cite{Boone2022}), whereas lighter shades correspond to less sensitive directions. The stretching of the angle cloud points is primarily attributed to impulsive maneuvers along or near the sensitive direction. In other words, maneuvers aligned with the sensitive direction exert a more significant impact on the measurements and are, therefore, more likely to be detected.

Furthermore, the cloud points generated by initial estimation uncertainties (\emph{i.e.}, the red points in Fig.~\ref{fig4}\subref{fig4a}, however, are shown by blue points in Figs.~\ref{fig4}\subref{fig4b}–\ref{fig4}\subref{fig4d}), along with the linear mean and bound (represented by the green triangle and ellipse in Fig.~\ref{fig4}\subref{fig4a}, respectively), are incorporated into Figs.~\ref{fig4}\subref{fig4b}–\ref{fig4}\subref{fig4d} as a reference. This facilitates a direct comparison between the effects of initial estimation errors and those of maneuvers on the angle measurements. As shown in Fig.~\ref{fig4}\subref{fig4b}, when the impulsive maneuver is small (\emph{i.e.}, its magnitude is less than the velocity STD set in Eq.~\eqref{eq48}), the colored points (which represent angles affected by the maneuver) are largely obscured by the blue points. This suggests that under such conditions, the maneuver is difficult to detect. However, as the maneuver magnitude increases (as shown in Fig.~\ref{fig4}\subref{fig4d}), only a small fraction of the colored points remain obscured, indicating that the maneuver has a more pronounced effect on the angle measurements and is thus more likely to be detected. The aforementioned preliminary analysis suggests that the maneuver’s direction and magnitude play a crucial role in maneuver detection.

Finally, since many maneuver detection methods are compared in the simulations, abbreviations are employed in the following three subsections for convenience, which are introduced in Table~\ref{tab3} \footnote{The SLSQP is available at \url{https://docs.scipy.org/doc/scipy/reference/optimize.minimize-slsqp.html}.}.
The proposed methods are denoted as CDMI-X, where CDMI stands for the "confidence-dominance maneuver indicator", and X indicates the detection strategy—either S for the original CDMI approach (as it only employs a single $\alpha_x$-$\alpha_z$ pair) or I/AI for the integrated-CDMI approaches.

\begin{table*}[!h]
	\centering
	\caption{Abbreviations of proposed and competitive methods}
	\label{tab3}
	\begin{tabular}{|ll|l|l|}
        \hline
        \multicolumn{2}{|l|}{Methods} & Description & Additional statement \\ \hline
        \multicolumn{1}{|l|}{\multirow{3}{*}[-2.8em]{\makecell[l]{CDMI-based\\ methods}}} & CDMI-S$_c$ & \makecell[l]{The original CDMI approach. The subscript \emph{c} \\represents the given confidence level $\alpha_x$.} & \multirow{3}{*}{\makecell[l]{Two methods are employed to compute \\the indicators for maneuver detection: \\(1) the proposed RPO method, and \\(2) the SLSQP (with zero point as an \\initial guess)}} \\ \cline{2-3}
        \multicolumn{1}{|l|}{} & CDMI-AI$_{0.5}$ & \makecell[l]{The integrated CDMI approach. It is integrated \\using the proposed adaptive sampling method. The \\threshold for maneuver detection is 0.5.} & \\ \cline{2-3}
        \multicolumn{1}{|l|}{} & CDMI-I$_{0.5}$ & \makecell[l]{The integrated CDMI approach. It is integrated \\using an equally-spaced (non-adaptive) sampling \\method. The threshold for maneuver detection is \\0.5.} & \\ \hline
        \multicolumn{2}{|l|}{OCDM} & Optimal control distance metric with MRBVP. & Ref.~\cite{Holzinger2012} \\ \hline
        \multicolumn{2}{|l|}{MBA$_d$} & \makecell[l]{Mixture-based approach. The subscript \emph{d} denotes \\the number of directions to be split. A seven-\\component univariate library is employed for each\\ splitting direction \cite{Vittaldev2016}.} & Ref.~\cite{Montilla2025}  \\ \hline
    \end{tabular}
\end{table*}

As shown in Table~\ref{tab3}, one needs to set a value for the confidence level $\alpha_x$ for the CDMI-S$_c$ approach. It should also be noted that in Ref.~\cite{Montilla2025}, two mixture-based maneuver detection methods, a mixture-based cost metric, and a mixture-based cost variation metric, are proposed. The first one can be employed when only one measurement (the same as the scenario discussed in Sec.~\ref{sec:Indicator with Measurement at Single Epoch}) is available. In contrast, the second one requires measurements at more than one epoch (at least two epochs, as mentioned in Sec.~\ref{sec:Indicator with Measurement at Multiple Epochs}) to generate the cost variation. In our simulations, the mixture-based cost metric is compared in Sec.~\ref{sec:Maneuver Detection Results Using Single-Epoch Angle}, which focuses on the single-angle case, whereas the performances of the mixture-based cost variation metric are investigated in the multiple-angle case (will be shown in Sec.~\ref{sec:Maneuver Detection Results Using Multiple-Epoch Angle}). The final issue worth noting is that in Ref.~\cite{Montilla2025}, a least square-based strategy is employed to improve the quality of measurements (\emph{i.e.}, reduce the measurement noises). This method employs multiple observed measurements to fit a polynomial, which is inappropriate for cases when measurements at only a single epoch are available. Therefore, the simulation does not adopt this strategy for a fair comparison. 

\subsection{Maneuver Detection Results Using Single-Epoch Angle} \label{sec:Maneuver Detection Results Using Single-Epoch Angle}
In this subsection’s simulations, only the angle measurement at the epoch $t_1$ is employed for maneuver detection. To begin with, one-run simulations are implemented for both the non-maneuver and maneuver cases. Simulated parameters (\emph{i.e.}, the initial estimated errors, measurement noises, and impulsive maneuvers) are randomly generated, with their values provided in Table~\ref{tab4} \footnote{The parameters for the one-run simulation are generated with a fixed random seed of 42 to ensure reproducibility. The purpose of conducting a single-run simulation here is to demonstrate the workflow of the proposed method. A more realistic Monte Carlo (MC) simulation with multiple runs will be presented later.}. The magnitude of the added maneuver in the maneuver case is 1 m/s.

\begin{table*}[!h]
	\centering
	\caption{Parameters for the one-run simulations}
	\label{tab4}
	\begin{tabular}{|l|ll|l|}
        \hline
        Case & \multicolumn{2}{l|}{Parameter} & Value \\ \hline
        \multirow{4}{*}{Non-maneuver case} & \multicolumn{1}{l|}{\multirow{2}{*}{Initial estimated error (nd)}} & Position & $[-6.0909\times 10^{-7}, 4.1082\times 10^{-6},1.9964\times 10^{-6}]$ \\ \cline{3-4} 
        & \multicolumn{1}{l|}{} & Velocity & $[6.3217\times 10^{-5}, 1.4865\times 10^{-4}, -2.2854\times 10^{-5}]$ \\ \cline{2-4} 
        & \multicolumn{2}{l|}{Added maneuver (nd)} & $[0, 0, 0]$ \\ \cline{2-4} 
        & \multicolumn{2}{l|}{Measurement noises (rad)} & $[-1.1380\times 10^{-5}, 1.3152\times 10^{-5}]$ \\ \hline
        \multirow{4}{*}{Maneuver case} & \multicolumn{1}{l|}{\multirow{2}{*}{Initial estimated error (nd)}} & Position & $[-6.0909\times 10^{-7}, 4.1082\times 10^{-6},1.9964\times 10^{-6}]$ \\ \cline{3-4} 
        & \multicolumn{1}{l|}{} & Velocity & $[6.3217\times 10^{-5}, 1.4865\times 10^{-4},-2.2854\times 10^{-5}]$ \\ \cline{2-4} 
         & \multicolumn{2}{l|}{Added maneuver (nd)} & $[-8.5834\times 10^{-4}, 2.7464\times 10^{-4}, -3.7482\times 10^{-4}]$ \\ \cline{2-4} 
        & \multicolumn{2}{l|}{Measurement noises (rad)} & $[-1.1380\times 10^{-5}, 1.3152\times 10^{-5}]$ \\ \hline
    \end{tabular}
\end{table*}

\begin{figure*}[!h]
	\centering
	\subfigure[]
	{
		\label{fig5a}
		\centering
		\includegraphics[width=0.4\textwidth]{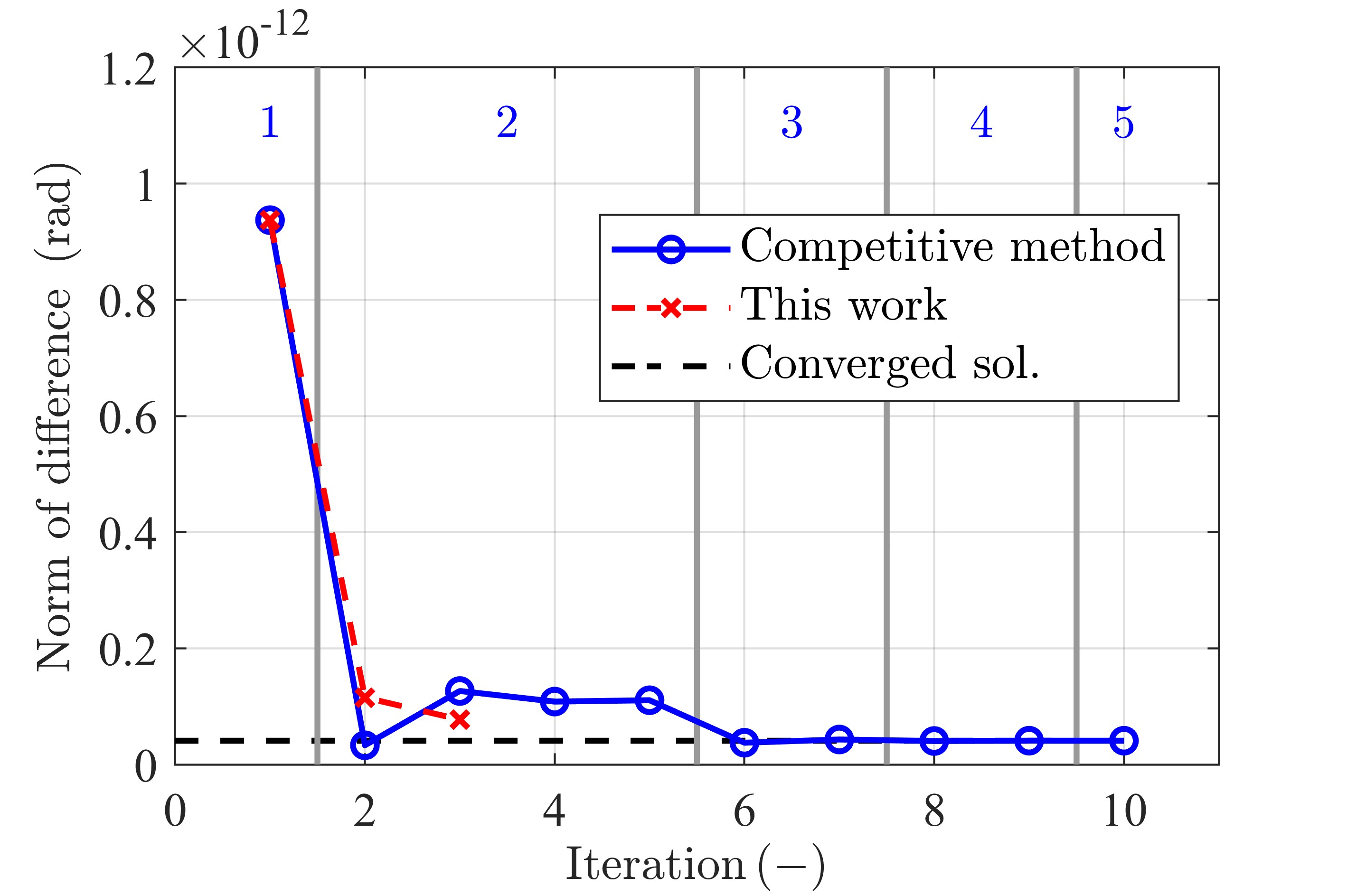}
	}
	\subfigure[]
	{
		\label{fig5b}
		\centering
		\includegraphics[width=0.4\textwidth]{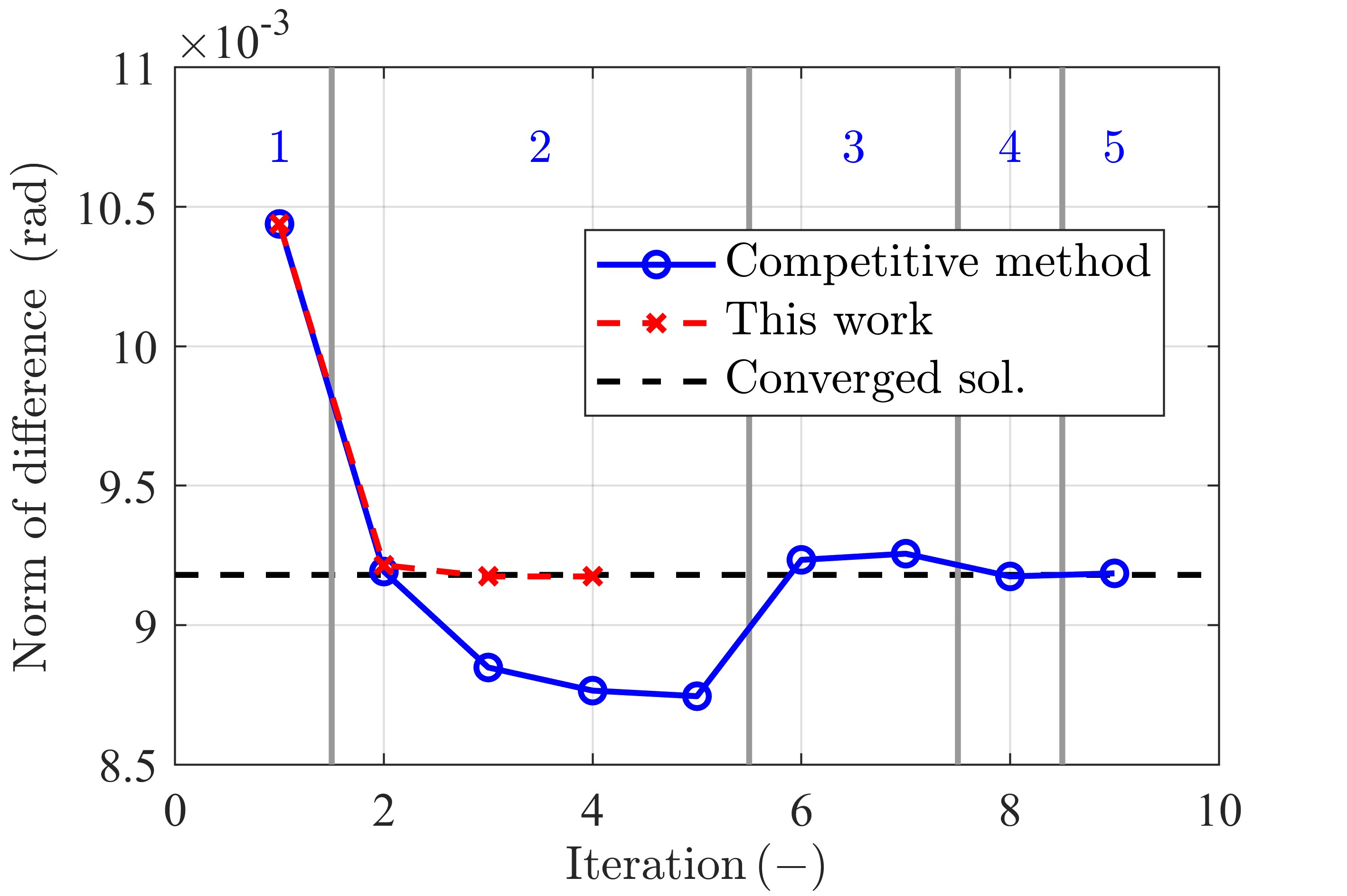}
	}
	\caption{Comparison between two RPO methods. \subref{fig5a} Non-maneuver case. \subref{fig5b} Maneuver case.}
	\label{fig5}
\end{figure*}

First, we investigate the solution space of the NLP optimization problem in Eq.~\eqref{eq9}. The confidence level $\alpha_x$ is set as 0.9. The proposed RPO method is employed to solve for the optimal solution $\delta \boldsymbol{x}_0^ * $ (the threshold is set as $\eta  = {10^{ - 6}}$). Figure~\ref{fig5} compares the convergence performances of two RPO methods with the norm of $\delta {\boldsymbol{\tilde z}_1}$ (\emph{i.e.}, $\left\| {\delta {{\boldsymbol{\tilde z}}_1}} \right\|$) as an index to represent the optimization objective. As shown in Fig.~\ref{fig5}, the convergence processes of the proposed RPO and the RPO in Ref.~\cite{Pavanello2024IEEE} are plotted using red dashed and blue solid lines, respectively. The blue numbers (1-5) indicate the highest polynomial orders used in different iterations of the competitive method. Notably, the proposed RPO converges within 3 and 4 iterations in the non-maneuver and maneuver cases, respectively. This is because the proposed method utilizes all polynomial orders (except for the first iteration, which only uses the constant and linear terms), resulting in fewer iterations. Additionally, it’s worth noting that there is a visible difference between the convergence results of the two methods (as shown in Fig.~\ref{fig5a}). This is due to the convergence threshold being set as $10^{-6}$, while the results in Fig.~\ref{fig5a} converge to approximately $10^{-12}$, which is within an acceptable error range.

\begin{figure*}[!h]
	\centering
	\subfigure[]
	{
		\label{fig6a}
		\centering
		\includegraphics[width=0.3\textwidth]{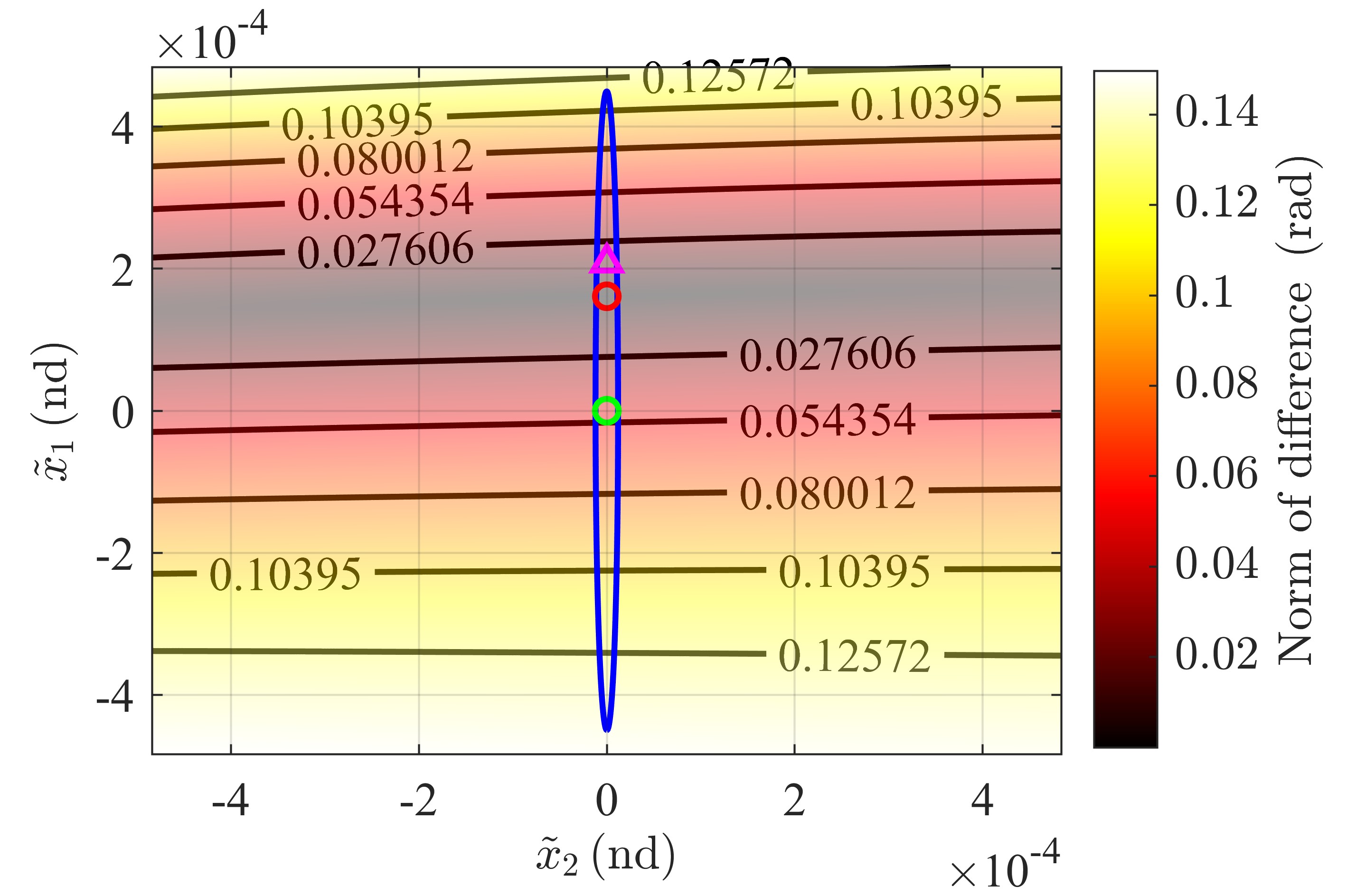}
	}
	\subfigure[]
	{
		\label{fig6b}
		\centering
		\includegraphics[width=0.3\textwidth]{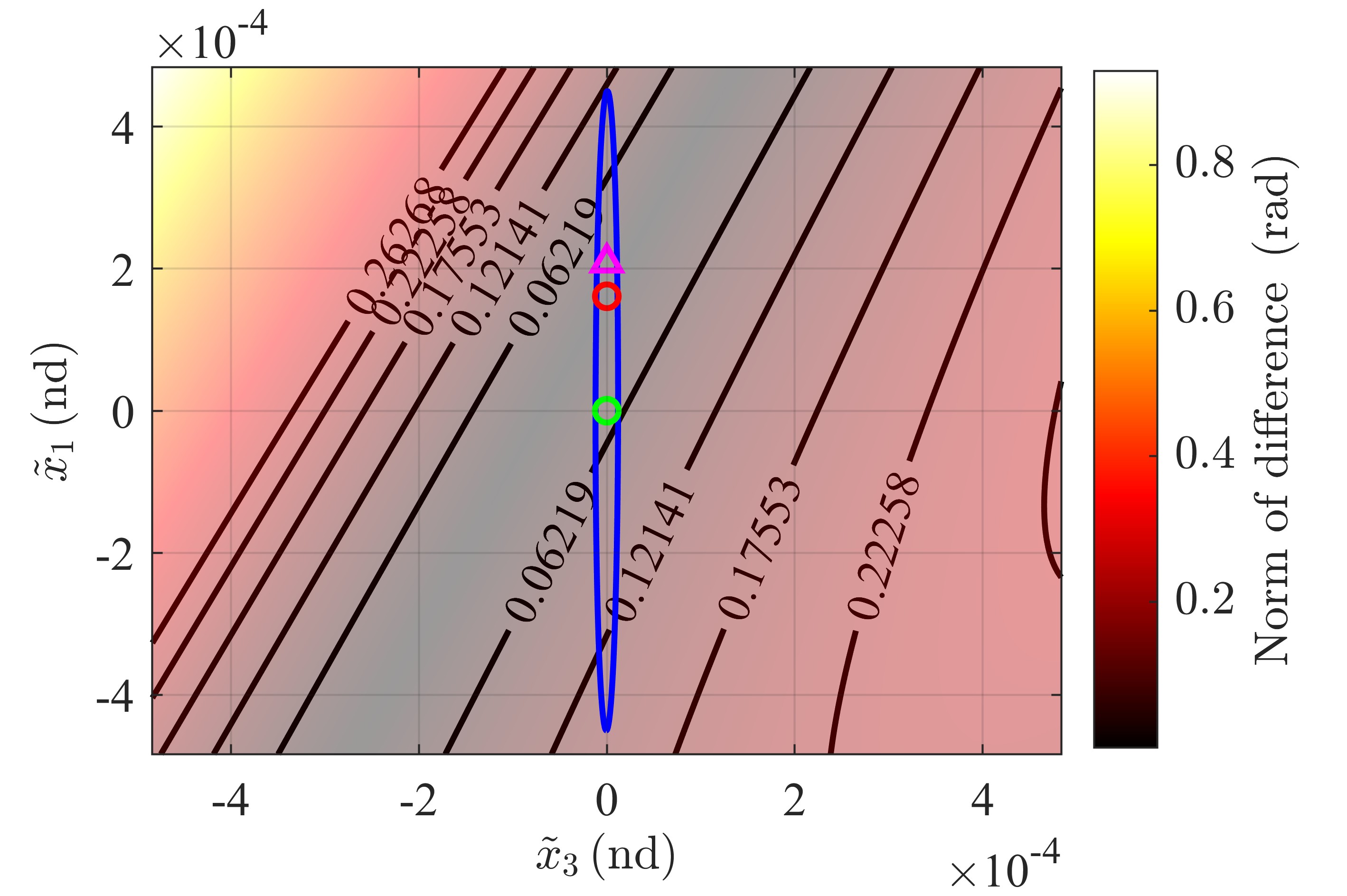}
	}
    \subfigure[]
	{
		\label{fig6c}
		\centering
		\includegraphics[width=0.3\textwidth]{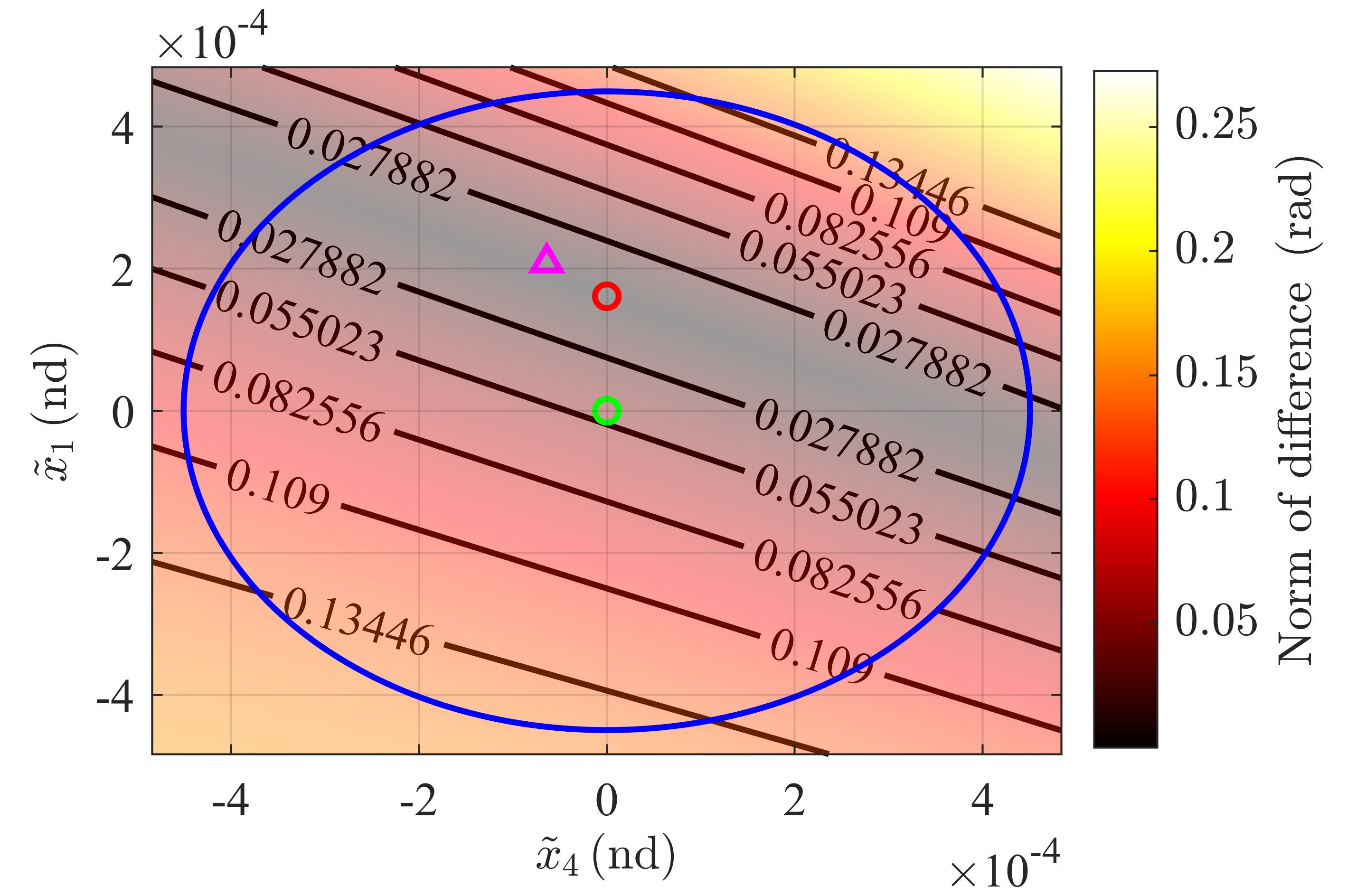}
	}
    \subfigure[]
	{
		\label{fig6d}
		\centering
		\includegraphics[width=0.3\textwidth]{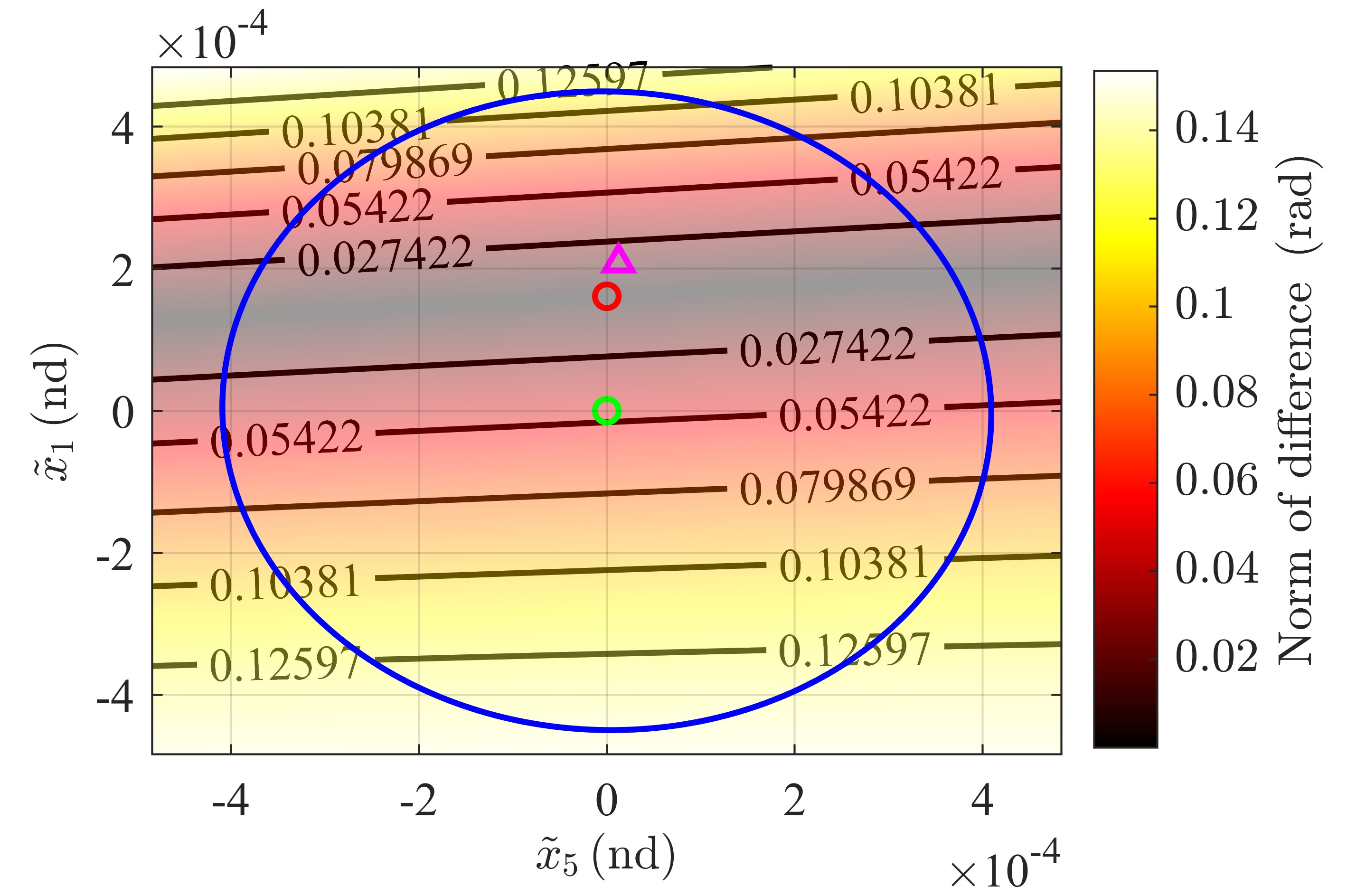}
	}
    \subfigure[]
	{
		\label{fig6e}
		\centering
		\includegraphics[width=0.3\textwidth]{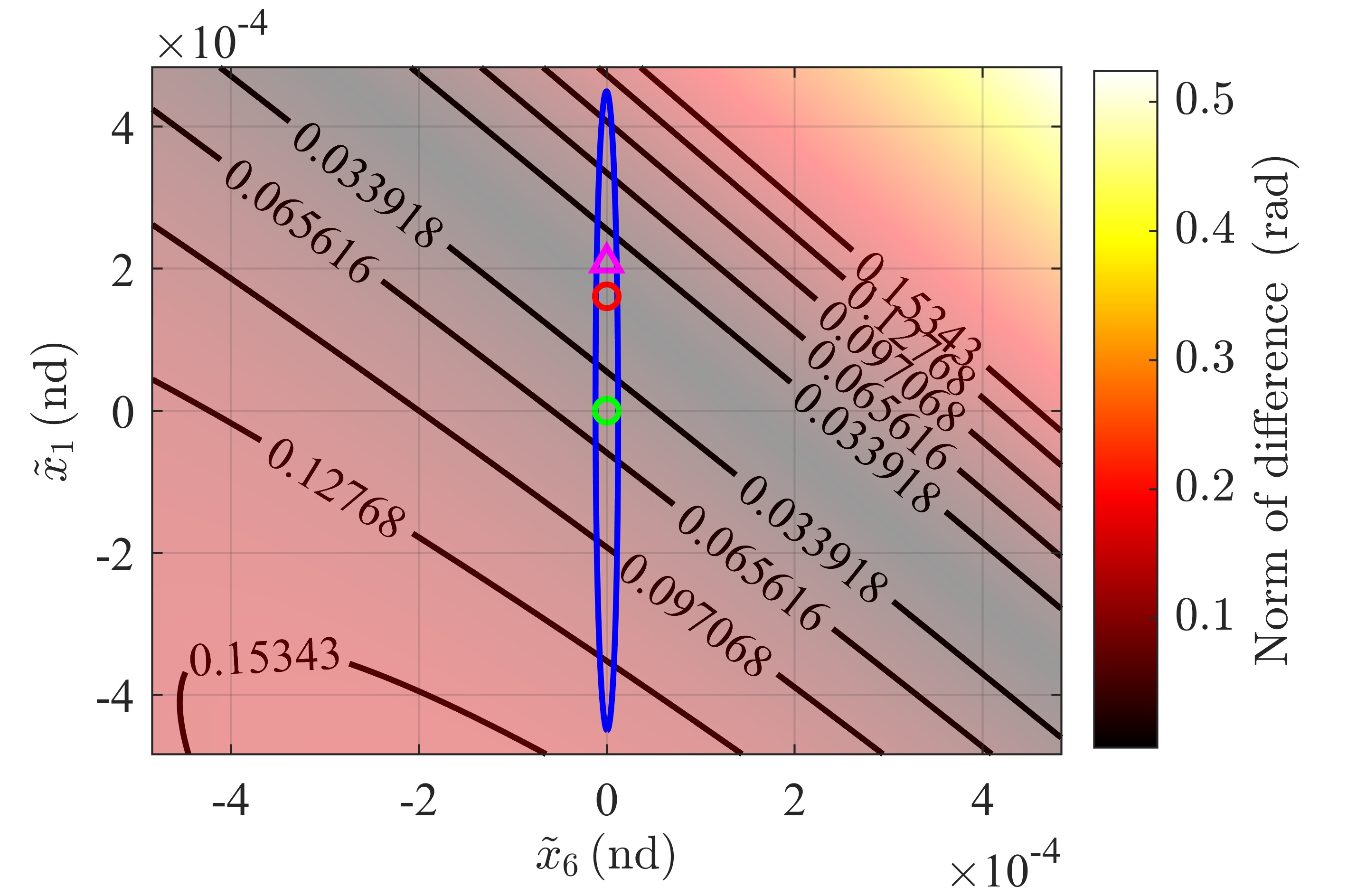}
	}
    \subfigure[]
	{
		\label{fig6f}
		\centering
		\includegraphics[width=0.3\textwidth]{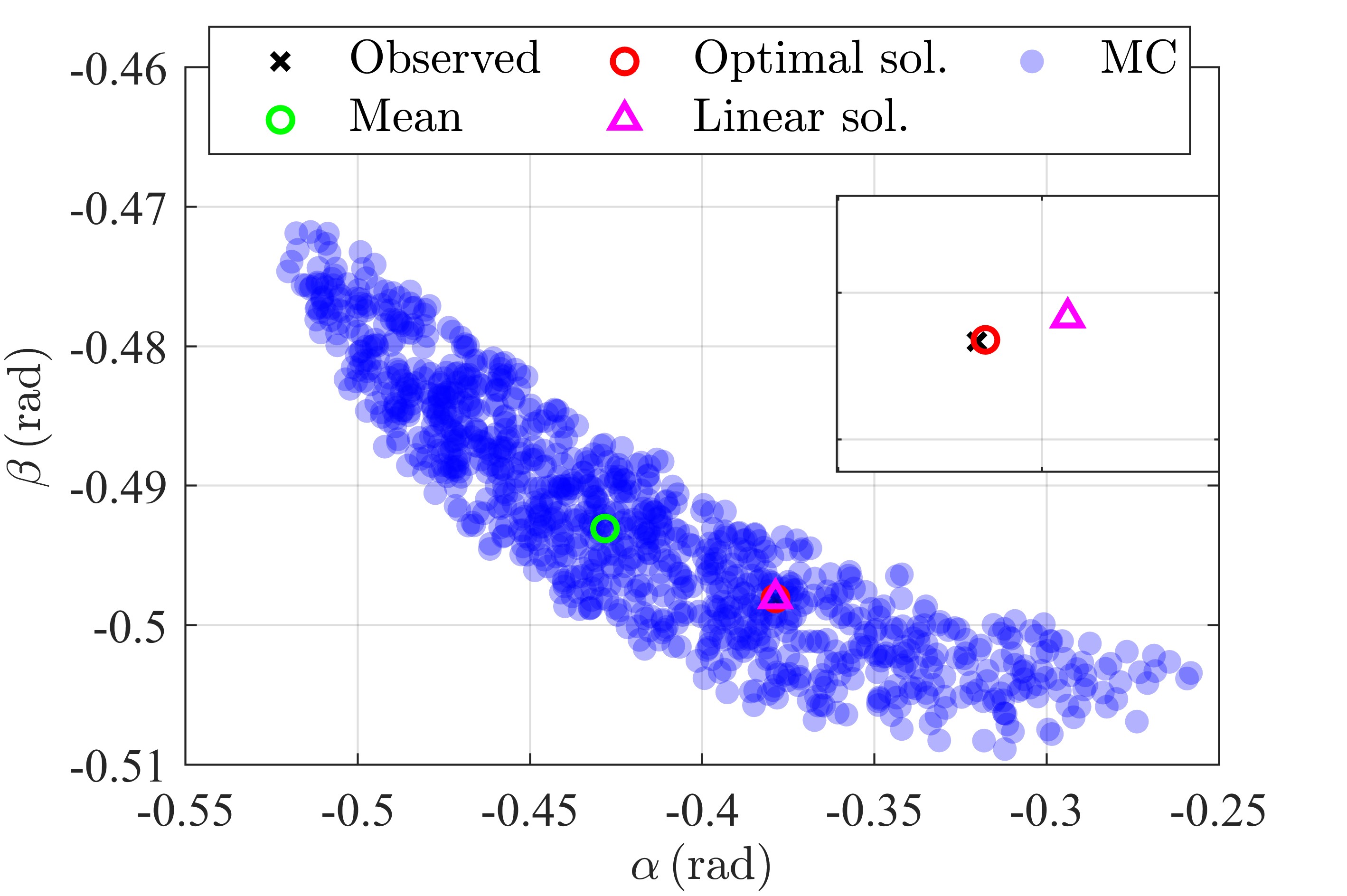}
	}
	\caption{Solution space of the non-maneuver case. \subref{fig6a} ${\tilde x_1} - {\tilde x_2}$. \subref{fig6b} ${\tilde x_1} - {\tilde x_3}$. \subref{fig6c} ${\tilde x_1} - {\tilde x_4}$. \subref{fig6d} ${\tilde x_1} - {\tilde x_5}$. \subref{fig6e} ${\tilde x_1} - {\tilde x_6}$. \subref{fig6d} Observation space. }
	\label{fig6}
\end{figure*}

Then, to facilitate the visualization of the objective function values within the six-dimensional space (the design variable is $\delta {\boldsymbol{x}_0} \in {\mathbb{R}^6}$), we performed a coordinate transformation, converting the orbital state coordinate $\boldsymbol{x} = [x,y,z,\dot x,\dot y,\dot z] \in {\mathbb{R}^6}$ (in the non-dimensional Earth-Moon rotating frame under the CRTBP dynamics) into $\boldsymbol{\tilde x} = [{\tilde x_1},{\tilde x_2},{\tilde x_3},{\tilde x_4},{\tilde x_5},{\tilde x_6}] \in {\mathbb{R}^6}$. In the new coordinate, the origin is located at the estimated mean ${\boldsymbol{\hat x}_0}$ (at the initial time), ${\tilde x_1}$-axis is aligned with the direction from ${\boldsymbol{\hat x}_0}$ to ${\boldsymbol{\hat x}_0} + \delta \boldsymbol{x}_0^ * $, while the other five axes are obtained through a Schmidt orthogonalization process. According to this coordinate definition, the optimal solution (\emph{i.e.}, $\delta \boldsymbol{x}_0^ * $) lies on the ${\tilde x_1}$-axis. Figures~\ref{fig6a}-\ref{fig6e} show the projection of the six-dimensional solution space in the non-maneuver case. The blue ellipse denotes the constraint corresponding to the CCSR (\emph{i.e.}, $\frac{1}{2}\delta \boldsymbol{x}_0^T\boldsymbol{P}_0^{ - 1}\delta {\boldsymbol{x}_0} \le {q_{{\chi ^2}}}({\alpha _x} = 0.9,6)$). The green dot represents the initial estimated mean in the new coordinate, which serves as the starting point for RPO. The red circle marks the optimal solution, while the pink triangle indicates the linear solution obtained by considering only the constant and linear terms of the polynomial (\emph{i.e.}, the result from the first iteration of the RPO method). Figure~\ref{fig6f} illustrates the angle measurements of the linear and optimal solutions. In Fig.~\ref{fig6f}, the observed measurement (the one with the noises) is marked by the black cross. Recall that the goal of the RPO is to determine an initial state within the CCSR such that its corresponding measurement is closest to the observed measurement (\emph{i.e.}, minimize the value of $\left\| {\delta {{\boldsymbol{\tilde z}}_1}} \right\|$). Additionally, 1,000 sample points are randomly generated within the CCSR to approximate the CCMR (\emph{i.e.}, ${\cal Z}({\alpha _x} = 0.9,{t_1})$, shown by blue points in Fig.~\ref{fig6f}). It can be seen from Fig.~\ref{fig6f} that the optimal solution is closer to the observed measurement (\emph{i.e.}, the black cross) than the linear solution and meets the constraint (\emph{i.e.}, it is located within the CCMR ${\cal Z}({\alpha _x} = 0.9,{t_1})$).

\begin{figure*}[!h]
	\centering
	\subfigure[]
	{
		\label{fig7a}
		\centering
		\includegraphics[width=0.3\textwidth]{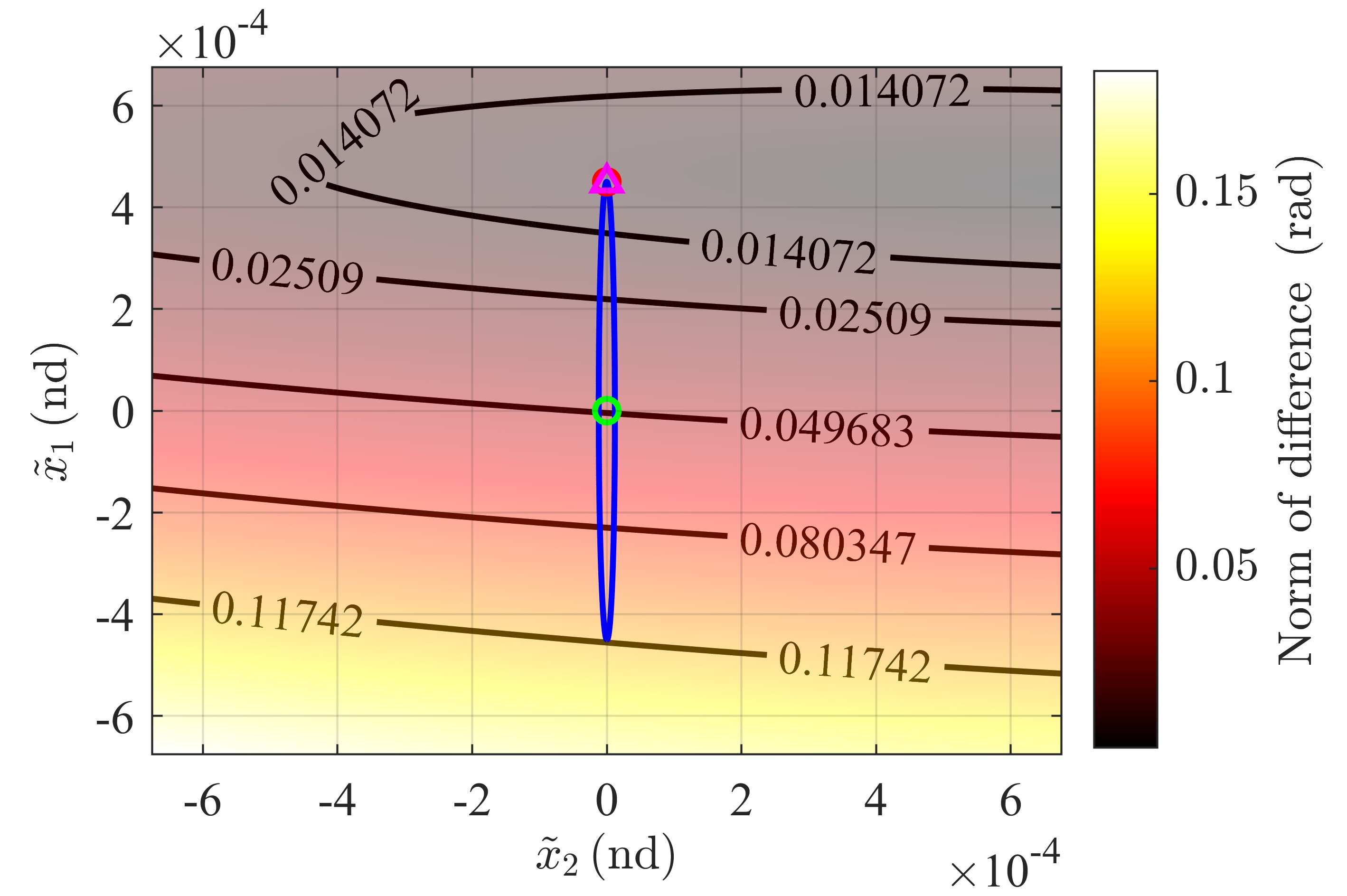}
	}
	\subfigure[]
	{
		\label{fig7b}
		\centering
		\includegraphics[width=0.3\textwidth]{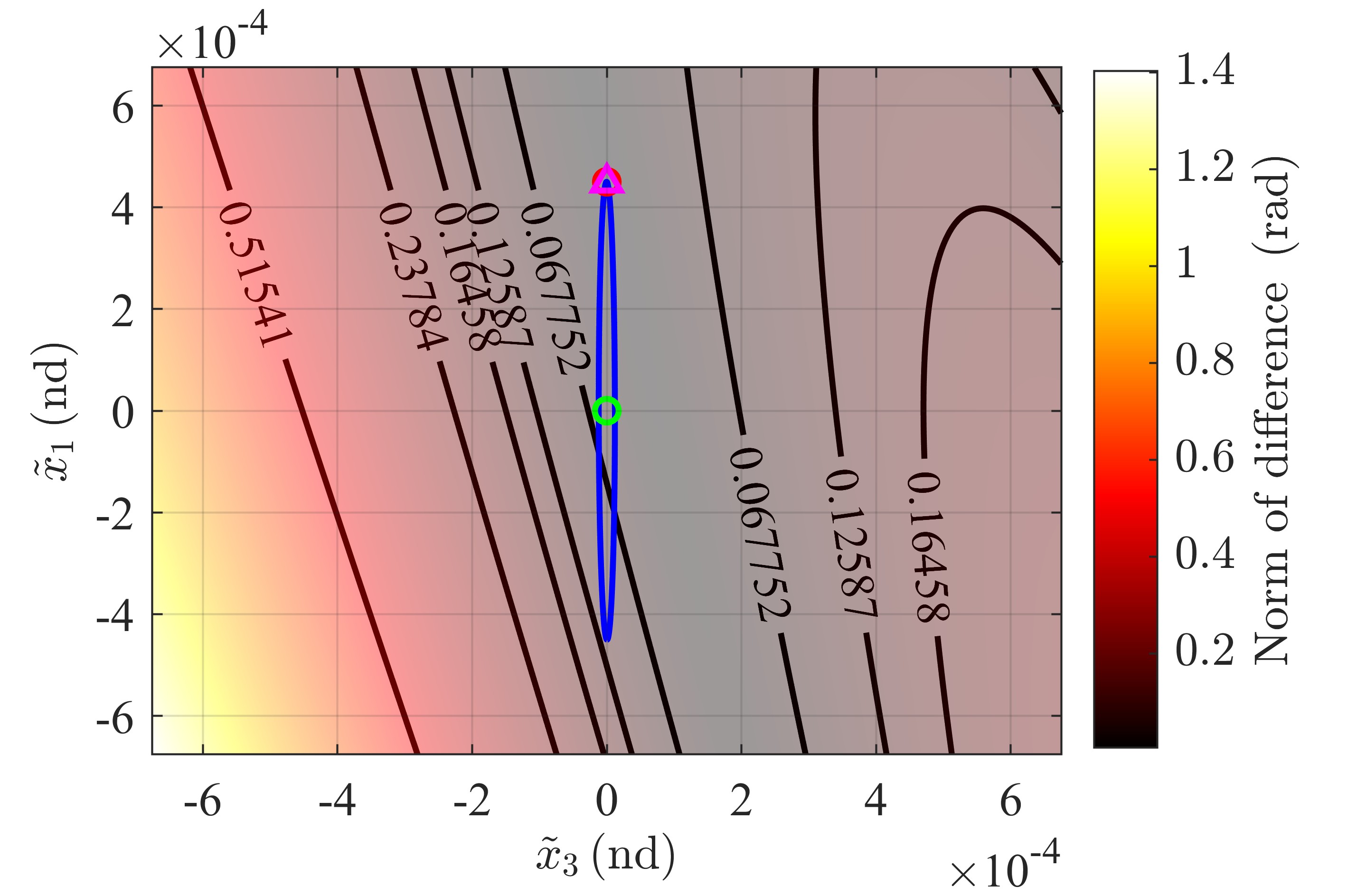}
	}
    \subfigure[]
	{
		\label{fig7c}
		\centering
		\includegraphics[width=0.3\textwidth]{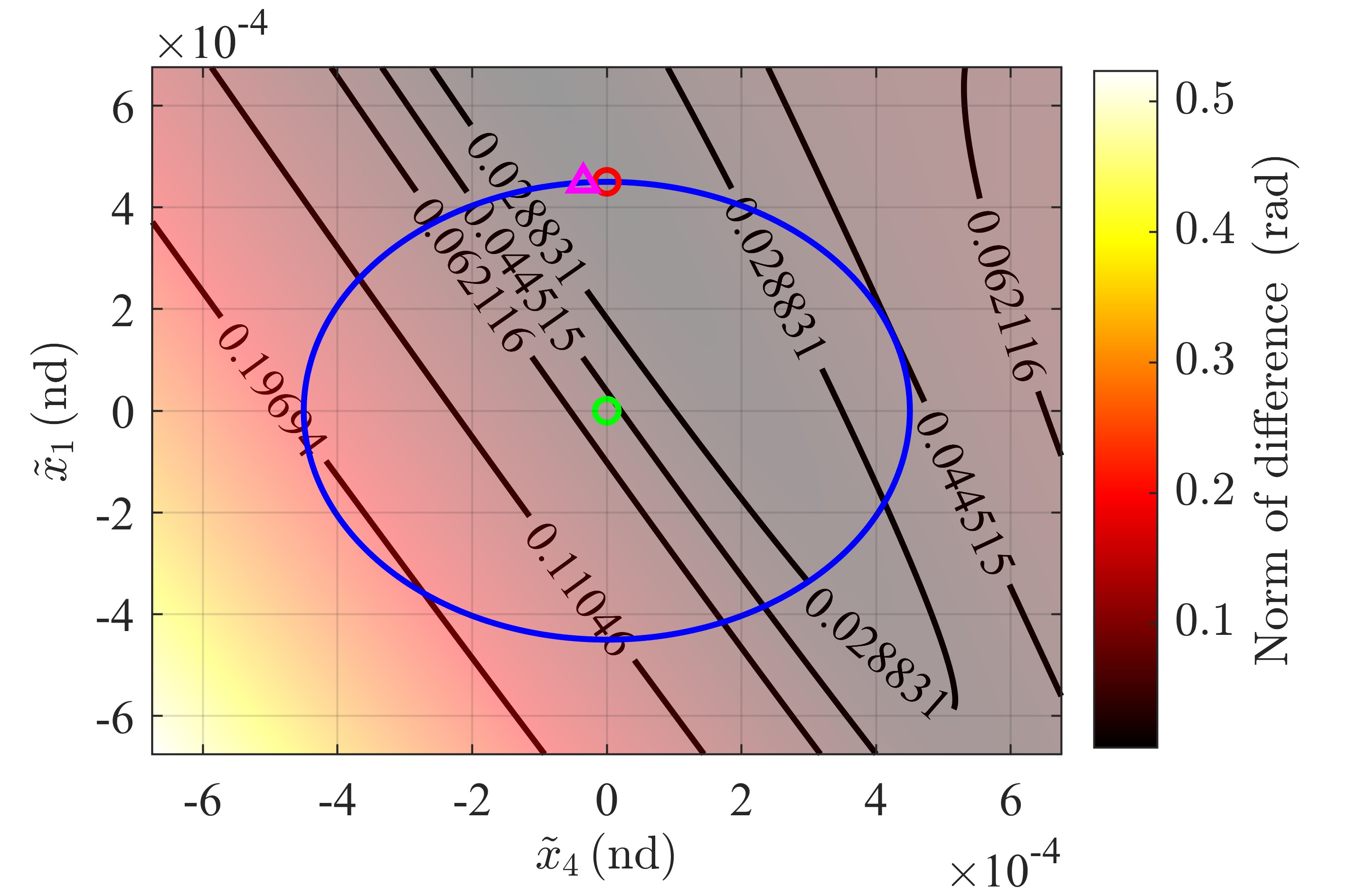}
	}
    \subfigure[]
	{
		\label{fig7d}
		\centering
		\includegraphics[width=0.3\textwidth]{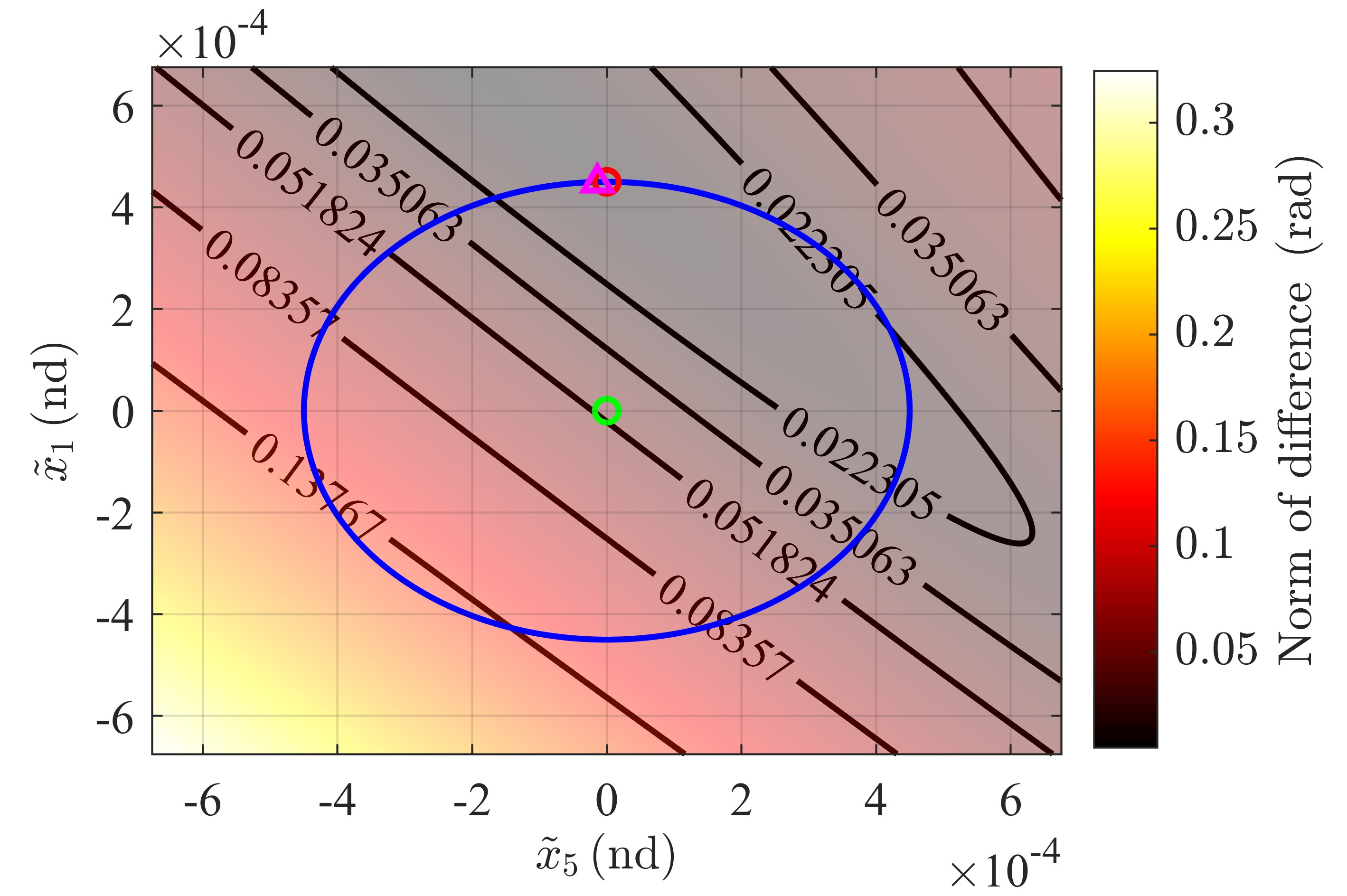}
	}
    \subfigure[]
	{
		\label{fig7e}
		\centering
		\includegraphics[width=0.3\textwidth]{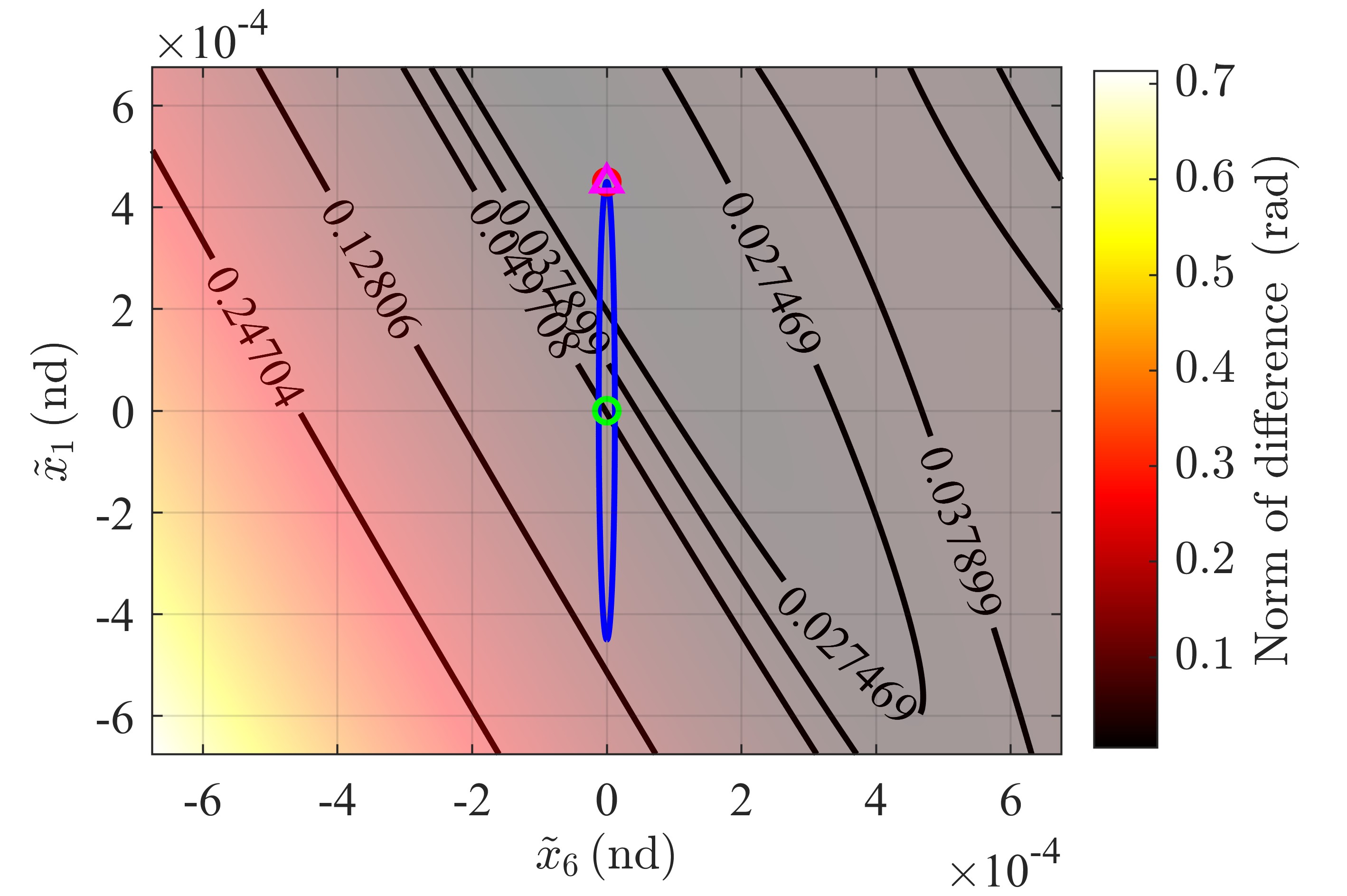}
	}
    \subfigure[]
	{
		\label{fig7f}
		\centering
		\includegraphics[width=0.3\textwidth]{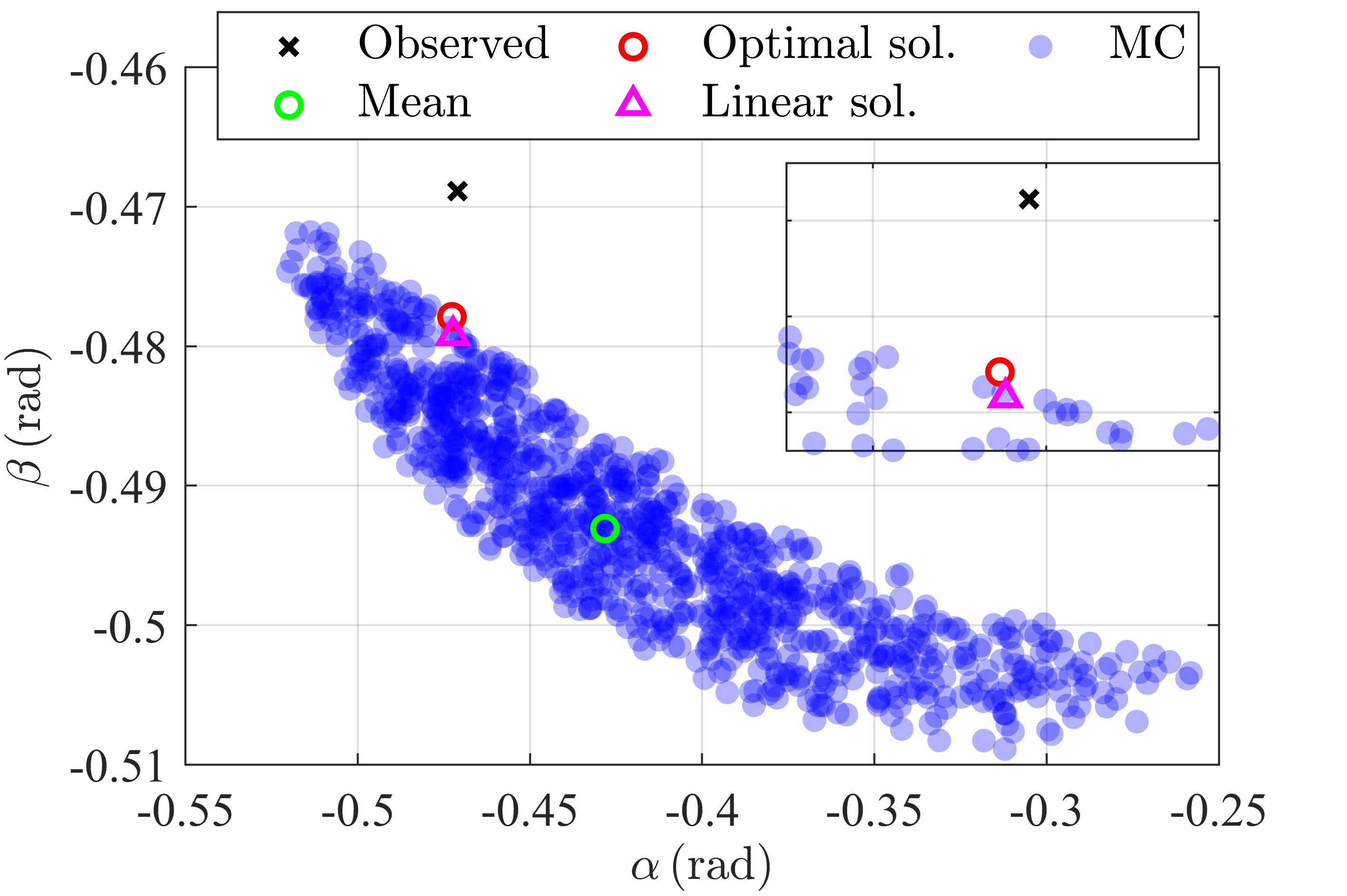}
	}
	\caption{Solution space of the maneuver case. \subref{fig7a} ${\tilde x_1} - {\tilde x_2}$. \subref{fig7b} ${\tilde x_1} - {\tilde x_3}$. \subref{fig7c} ${\tilde x_1} - {\tilde x_4}$. \subref{fig7d} ${\tilde x_1} - {\tilde x_5}$. \subref{fig7e} ${\tilde x_1} - {\tilde x_6}$. \subref{fig7d} Observation space. }
	\label{fig7}
\end{figure*}

Figure~\ref{fig7} presents the solution space results of the maneuver case (parameters given in Table~\ref{tab4}). One can see from Fig.~\ref{fig7} that the optimal solution obtained by the RPO method satisfies the constraint (the red circles are within the blue ellipse, as shown in Figs.~\ref{fig7a}-\ref{fig7f}) and has better optimality than the linear solution (its associated angle measurement is closer to the observed one compared to that of the linear solution, as shown in Fig~\ref{fig7f}). 

\begin{figure*}[!h]
	\centering
	\subfigure[]
	{
		\label{fig8a}
		\centering
		\includegraphics[width=0.4\textwidth]{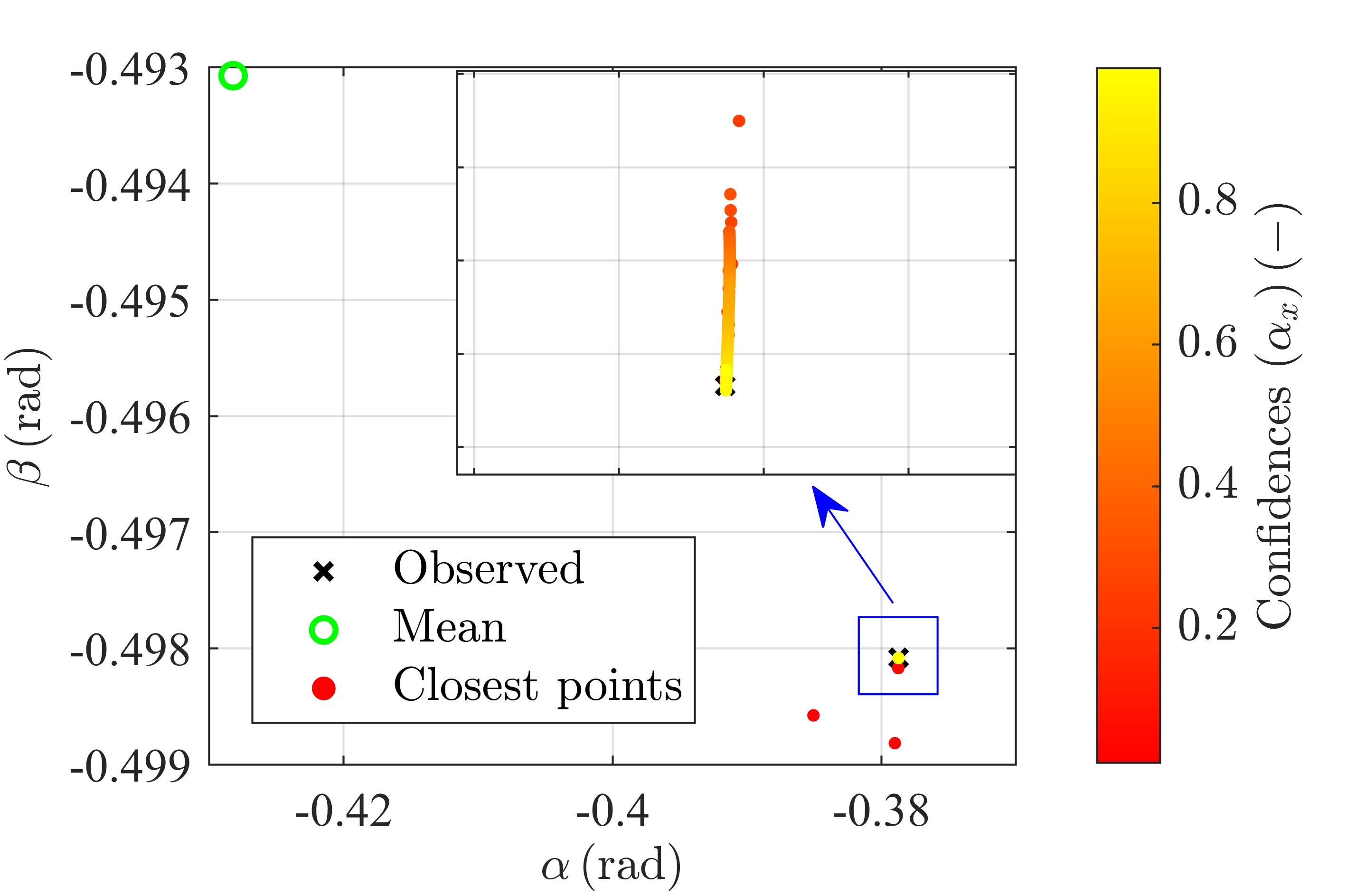}
	}
	\subfigure[]
	{
		\label{fig8b}
		\centering
		\includegraphics[width=0.4\textwidth]{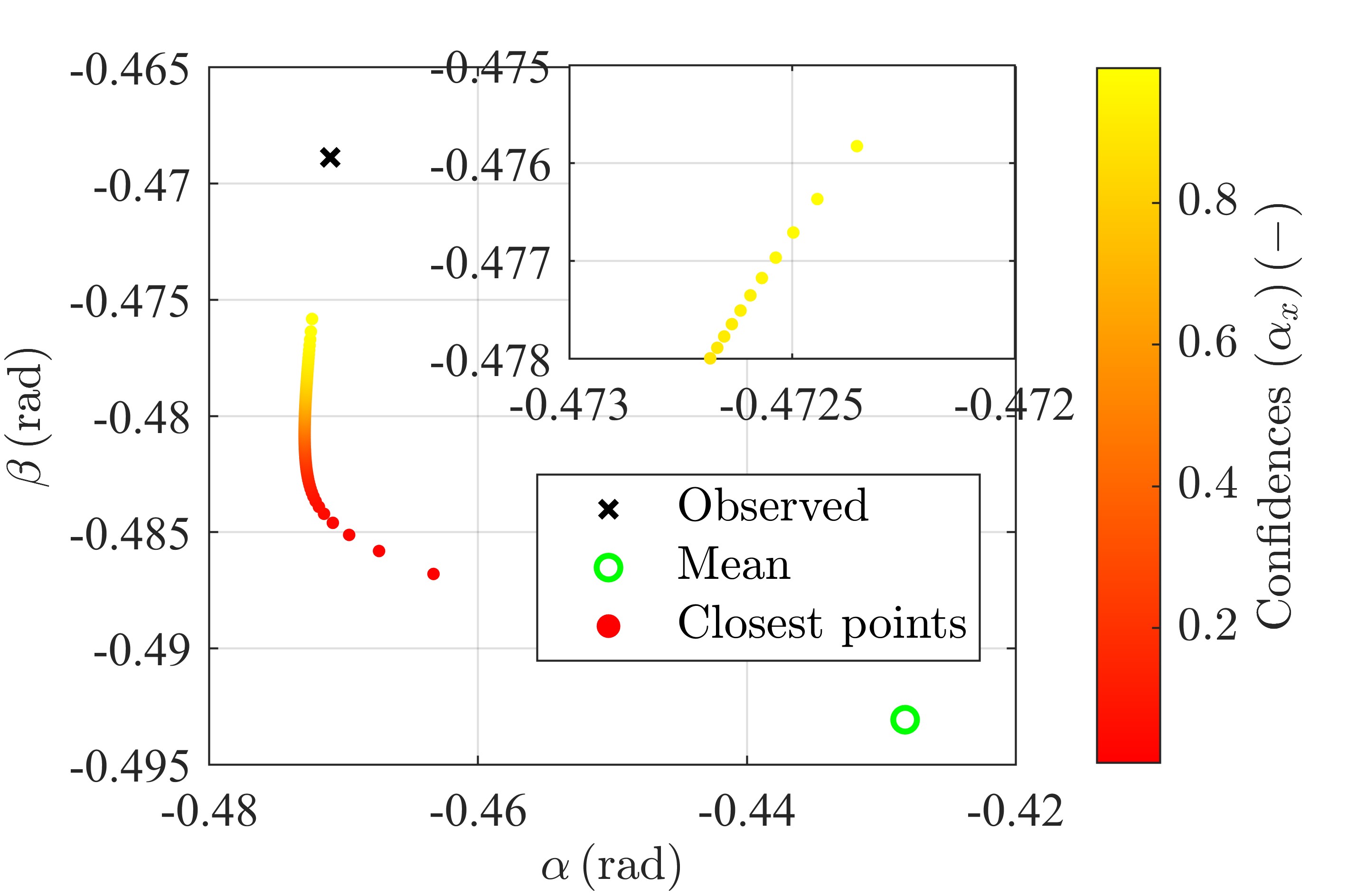}
	}
	\caption{Closest points under different confidence $\alpha_x$. \subref{fig8a} Non-maneuver case. \subref{fig8b} Maneuver case.}
	\label{fig8}
\end{figure*}

\begin{figure*}[!h]
	\centering
	\subfigure[]
	{
		\label{fig9a}
		\centering
		\includegraphics[width=0.4\textwidth]{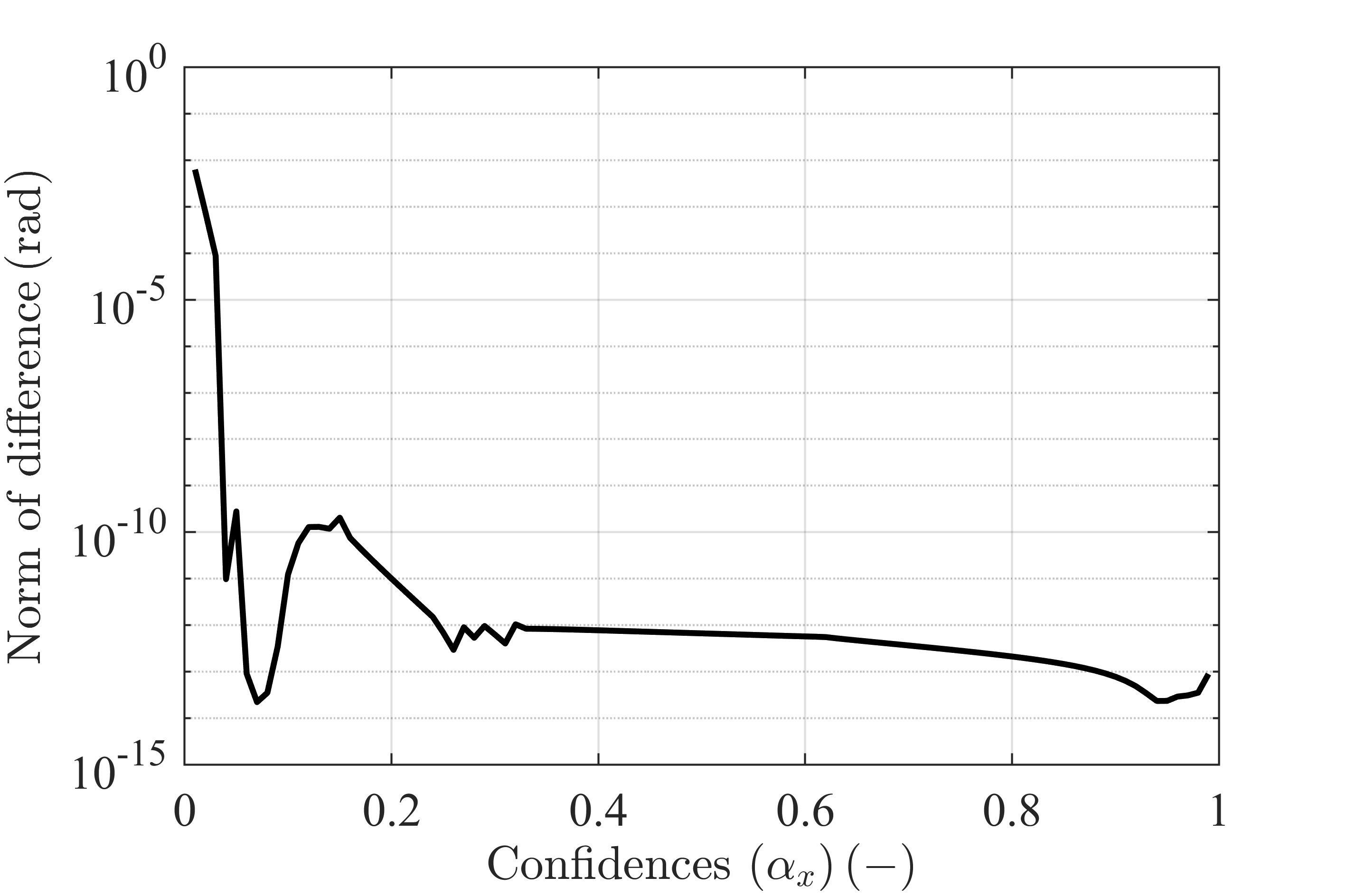}
	}
	\subfigure[]
	{
		\label{fig9b}
		\centering
		\includegraphics[width=0.4\textwidth]{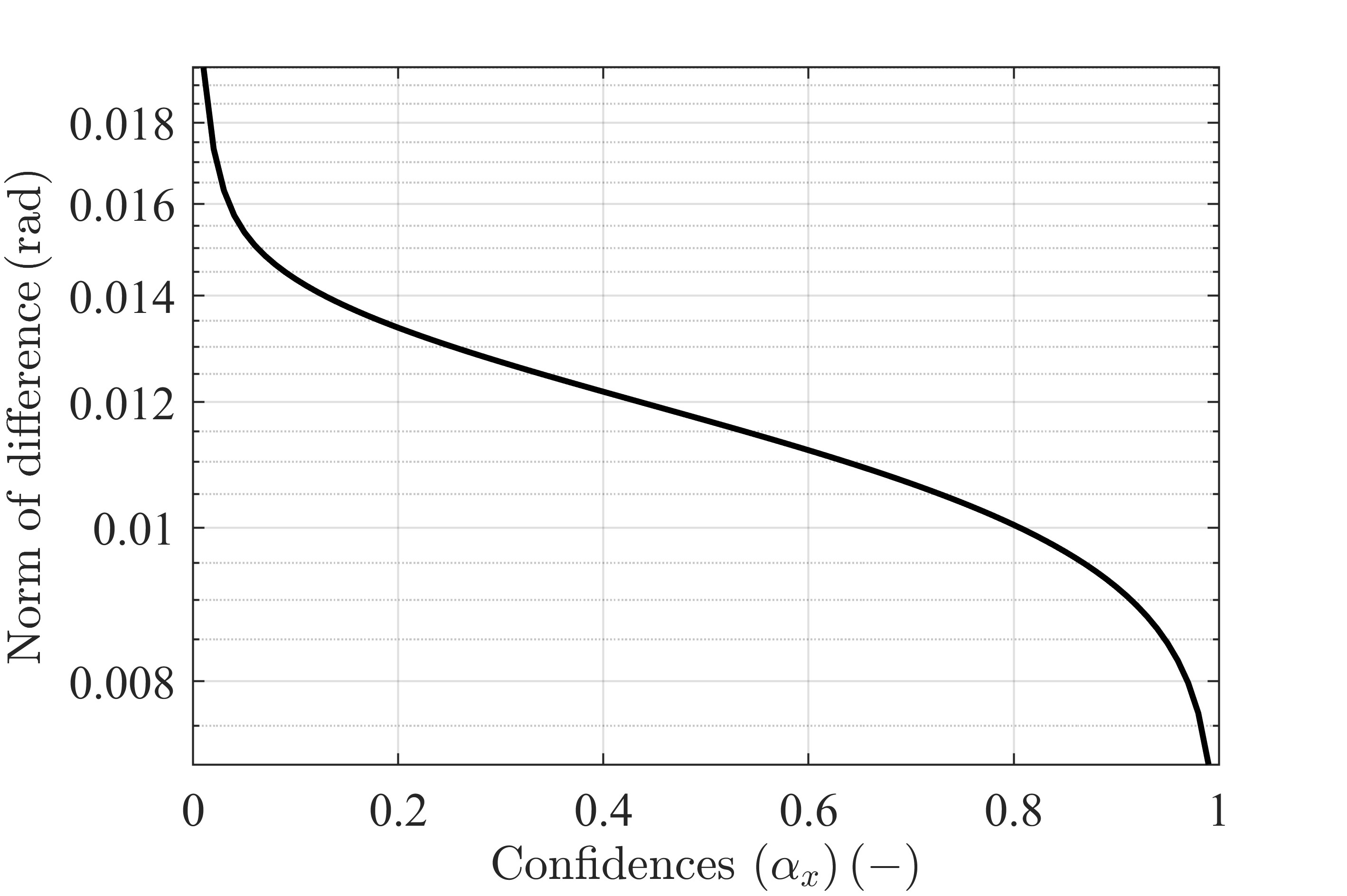}
	}
	\caption{Norm of difference between the closest and observed measurements. \subref{fig9a} Non-maneuver case. \subref{fig9b} Maneuver case.}
	\label{fig9}
\end{figure*}

Figure~\ref{fig8} plots the optimized closest points (using the proposed RPO) under different initial state confidence $\alpha_x$, and Fig.~\ref{fig9} shows the corresponding norm of differences between these optimized closest points and the observed measurements (represented by the black crosses in Fig.~\ref{fig8}). Generally, as the initial state confidence $\alpha_x$ increases, the CCSR expands, leading to a larger CCMR. This should result in the optimized closest point being closer to the observed measurement, as seen in Fig.~\ref{fig8b} and Fig.~\ref{fig9b}. However, the results in Fig.~\ref{fig9a} indicate that in the non-maneuver case, the norm of difference does not decrease with increasing initial confidence but instead fluctuates around $10^{-10}$. Our explanation is as follows. In the non-maneuver case, the observed measurement is likely already within the CCMR (as shown in Fig.~\ref{fig6f}). Ideally, the optimized closest point should coincide with the observed measurement, meaning the norm of difference should be zero. However, due to numerical errors and the convergence threshold set at $10^{-6}$, MOSEK produces a solution with a slight deviation from the observed measurements, approximately in the range of $10^{-10}$-$10^{-15}$.

\begin{figure*}[!h]
	\centering
	\subfigure[]
	{
		\label{fig10a}
		\centering
		\includegraphics[width=0.4\textwidth]{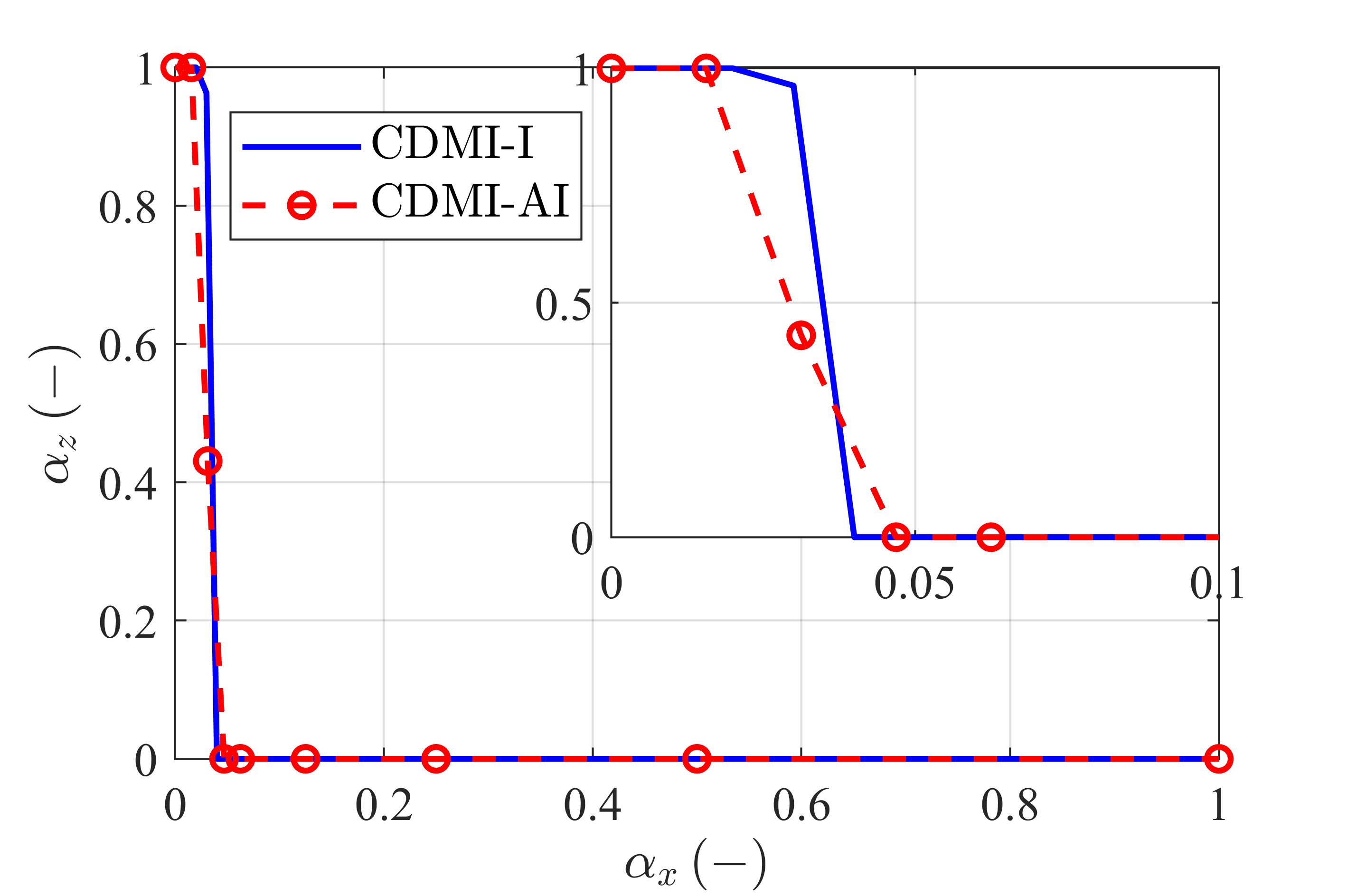}
	}
	\subfigure[]
	{
		\label{fig10b}
		\centering
		\includegraphics[width=0.4\textwidth]{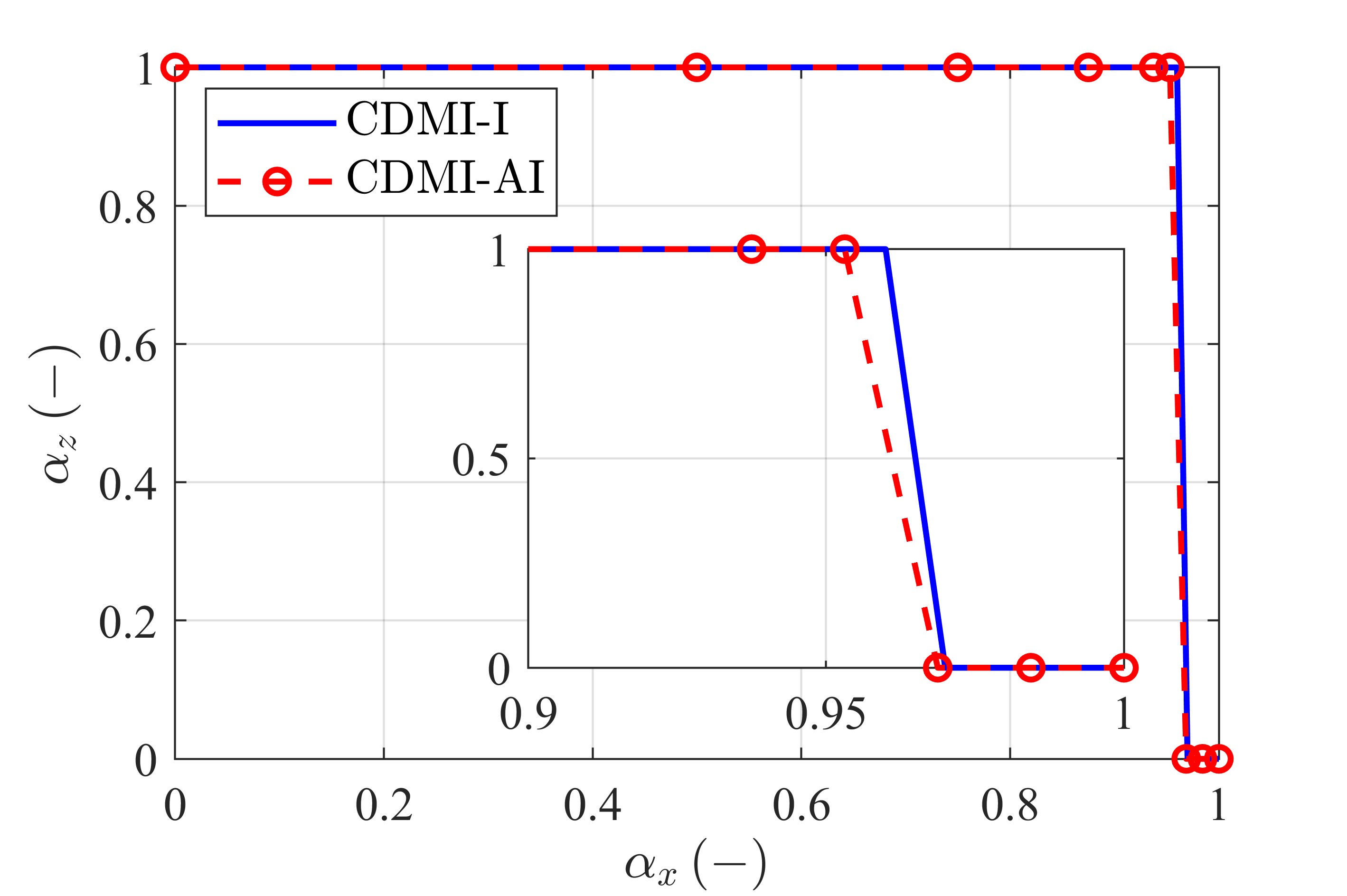}
	}
	\caption{$\alpha_x$-$\alpha_z$ curves of the one-run simulations. \subref{fig10a} Non-maneuver case. \subref{fig10b} Maneuver case.}
	\label{fig10}
\end{figure*}

After presenting the results of the proposed RPO, we now implement maneuver detection for the cases in Table~\ref{tab4}, beginning with the integrated-indicator approaches (\emph{i.e.}, the CDMI-AI and CDMI-I methods, as discussed in Table~\ref{tab3}). The thresholds of the adaptive sampling strategy are set as ${\varepsilon _1} = 0.01$ and ${\varepsilon _2} = 0.02$. Figure~\ref{fig10} presents the confidence curves (\emph{i.e.}, ${\alpha _z} - {\alpha _x}$) obtained by two methods. For the CDMI-I method (represented by blue lines), the initial state confidences $\alpha_x$ are equally spaced, with an interval of 0.01 (\emph{i.e.}, a total of 101 confidences are sampled). In both the non-maneuver and maneuver cases, the CDMI-AI requires only nine samples (shown by the circles on the red dashed lines in Fig.~\ref{fig10}). That means that only six more samples are adaptively added by the CDMI-AI (recall that $\alpha_x=0$, $\alpha_x=0.5$, and $\alpha_x=1$ are three initial samples). One can see from Fig.~\ref{fig10} that the red dashed lines and blue solid lines are in good agreement. Moreover, Table~\ref{tab5} compares the performances of the two strategies. With the adaptive sampling strategy, the maneuver probability errors (with respect to the non-adaptive strategy) remain below 0.005, even with a 91\% (from 101 to 9) reduction in sampling points, successfully detecting maneuvers in both cases.

\begin{table}[!h]
	\centering
	\caption{A comparison between the adaptive and non-adaptive strategies in the one-run simulation}
	\label{tab5}
	\begin{tabular}{|l|l|l|l|}
        \hline
        Case                               & Index                & CDMI-AI  & CDMI-I   \\ \hline
        \multirow{2}{*}{Non-maneuver case} & Number of samples    & 9      & 101    \\ \cline{2-4} 
                                           & Maneuver probability & 0.0301 & 0.0346 \\ \hline
        \multirow{2}{*}{Maneuver case}     & Number of samples    & 9      & 101    \\ \cline{2-4} 
                                           & Maneuver probability & 0.9609 & 0.9649 \\ \hline
    \end{tabular}
\end{table}

\begin{figure}[!h]
	\centering
	\subfigure[]
	{
		\label{fig11a}
		\centering
		\includegraphics[width=0.4\textwidth]{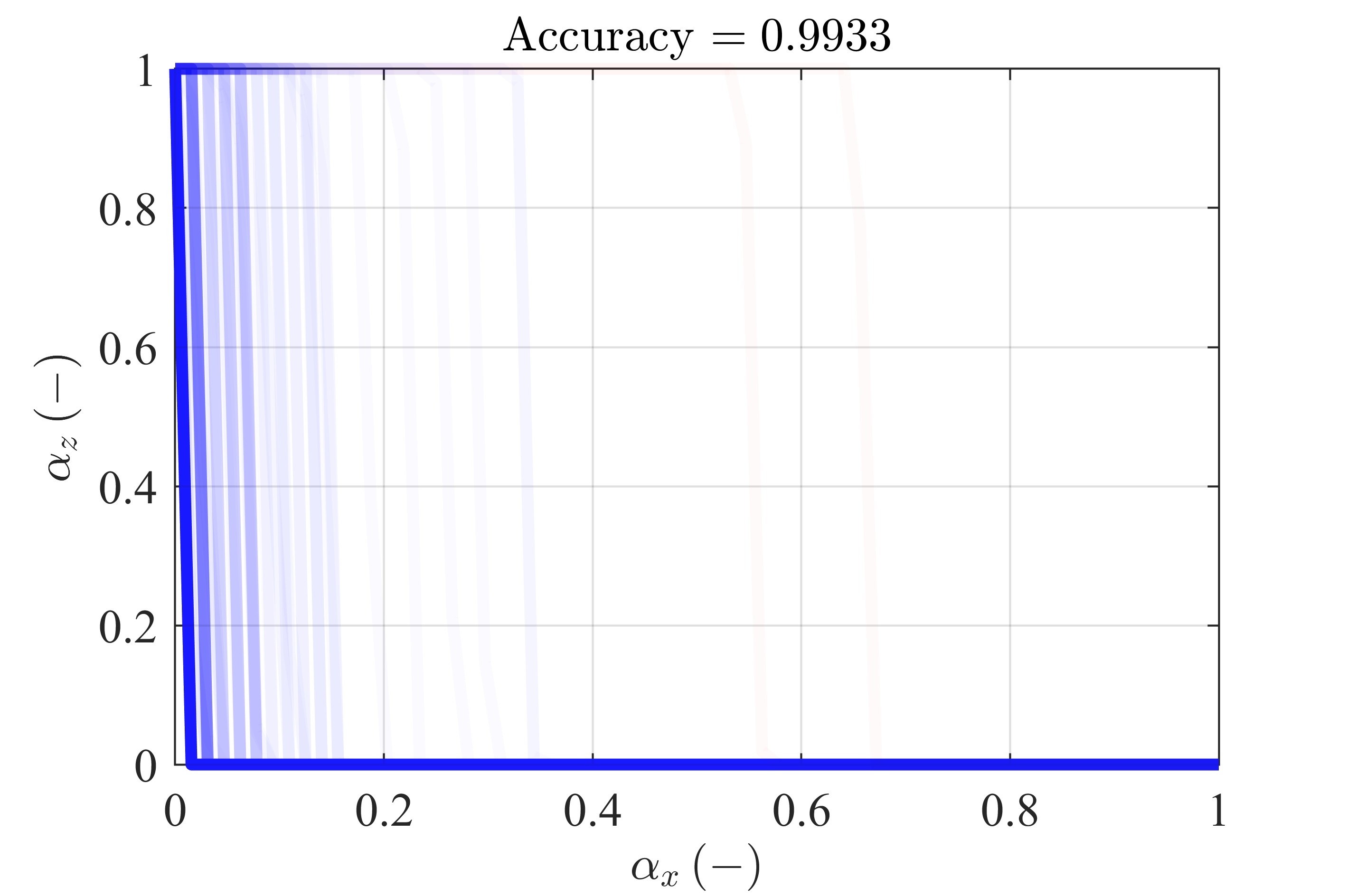}
	}
	\subfigure[]
	{
		\label{fig11b}
		\centering
		\includegraphics[width=0.4\textwidth]{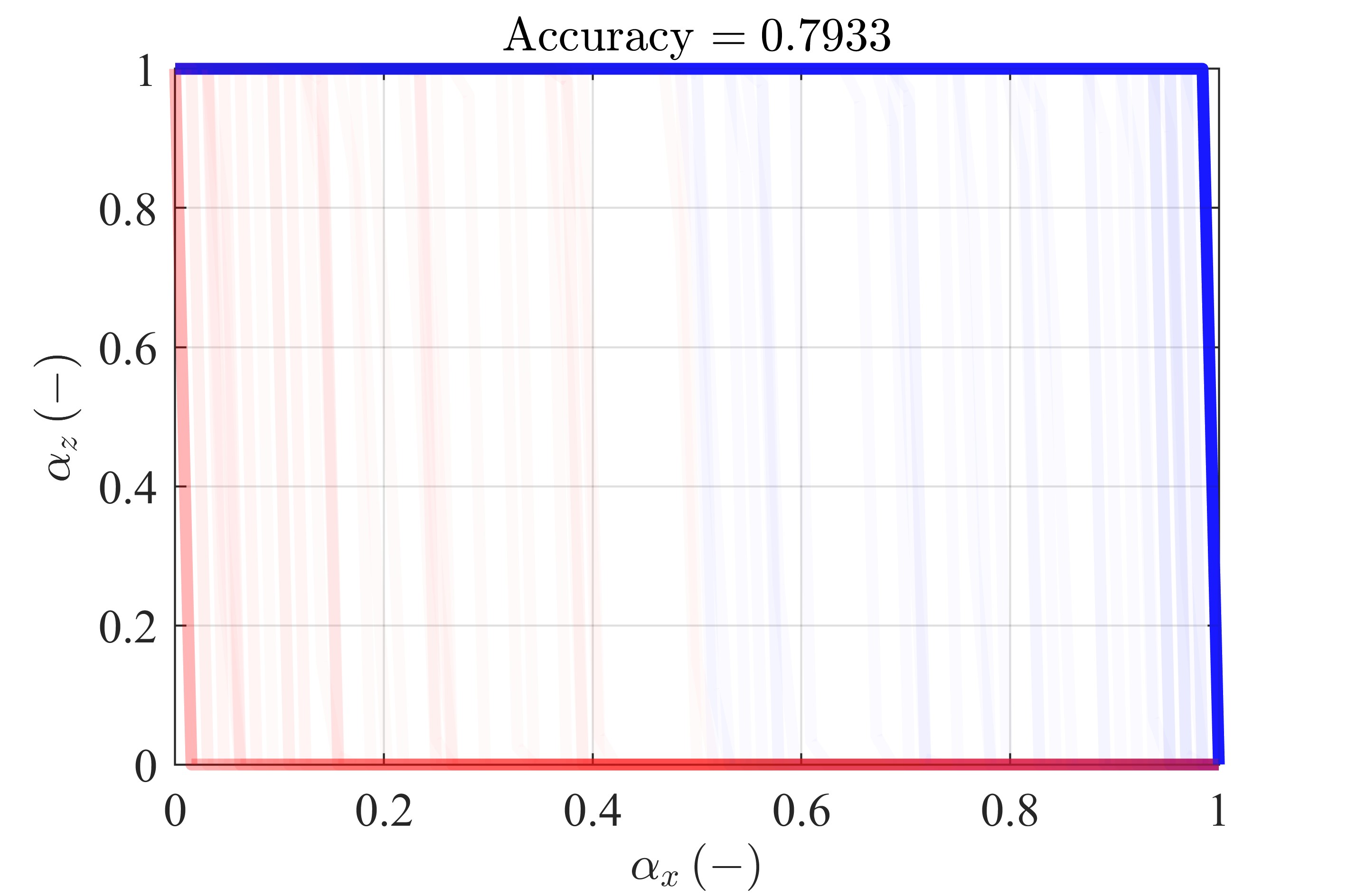}
	}
	\caption{$\alpha_x$-$\alpha_z$ curves obtained by the CDMI-AI with single-epoch angle. \subref{fig11a} Non-maneuver case. \subref{fig11b} Maneuver case.}
	\label{fig11}
\end{figure}

\begin{figure}[!h]
	\centering
	\subfigure[]
	{
		\label{fig12a}
		\centering
		\includegraphics[width=0.4\textwidth]{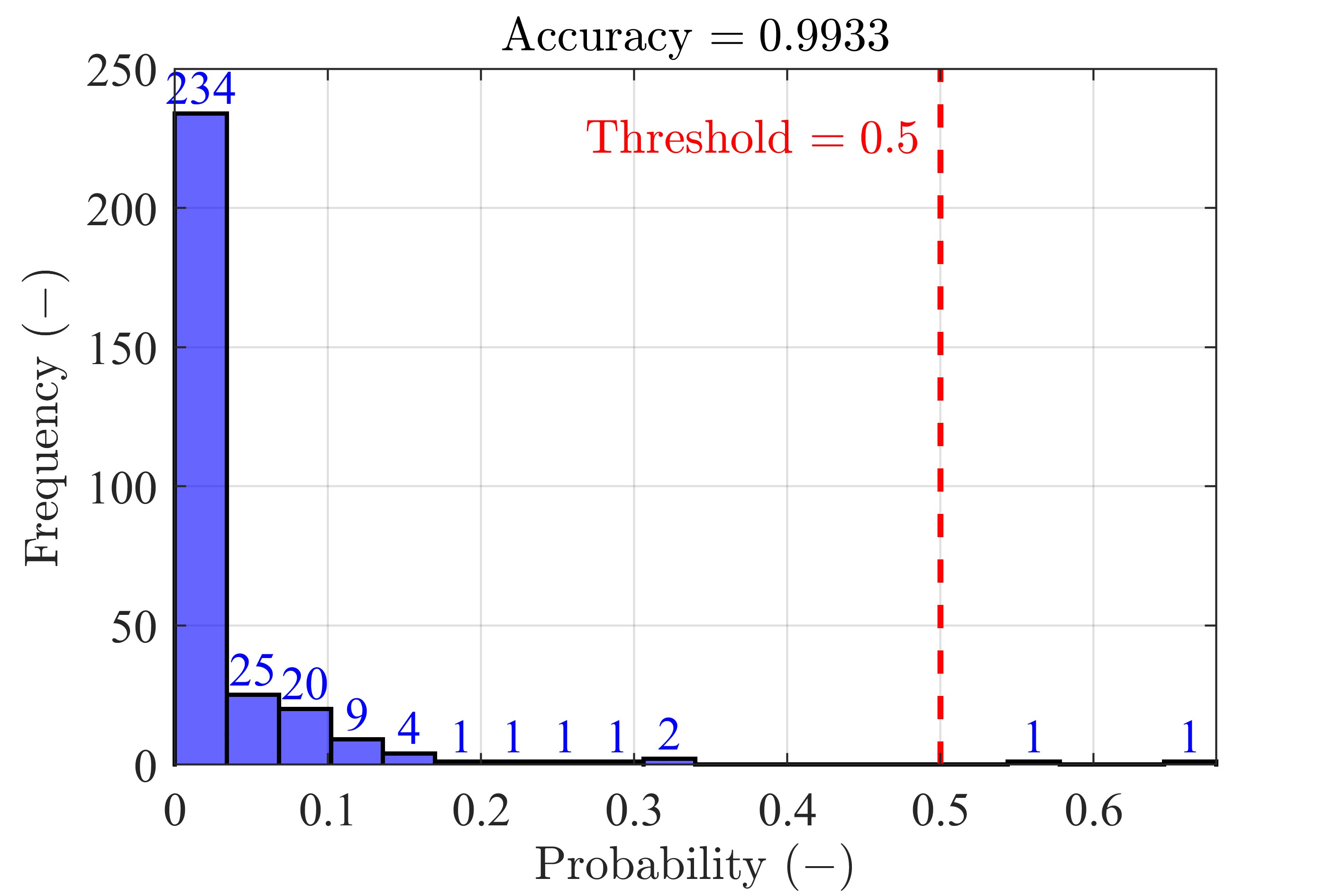}
	}
	\subfigure[]
	{
		\label{fig12b}
		\centering
		\includegraphics[width=0.4\textwidth]{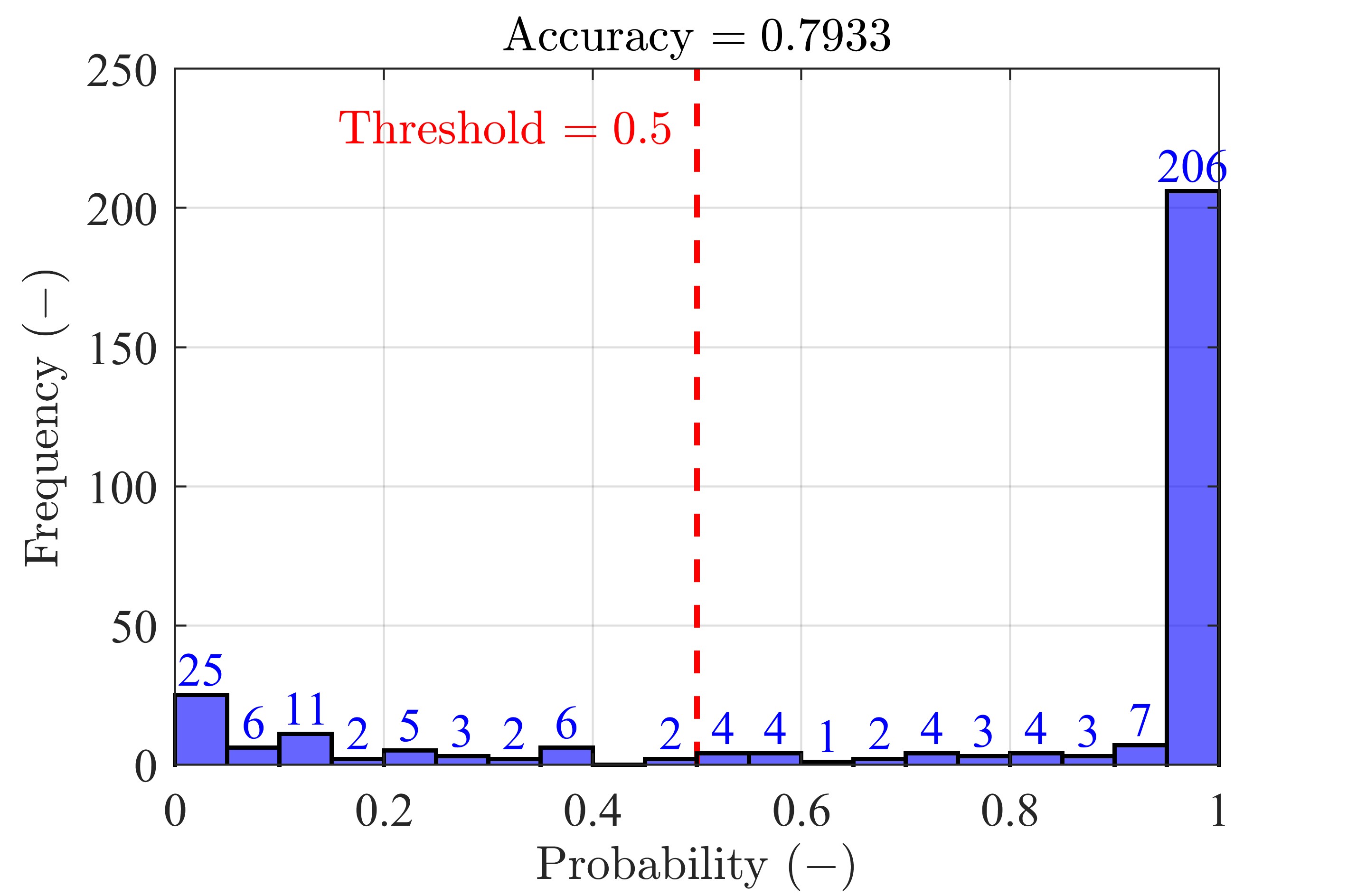}
	}
	\caption{Maneuver probabilities obtained by the CDMI-AI with single-epoch angle. \subref{fig12a} Non-maneuver case. \subref{fig12b} Maneuver case.}
	\label{fig12}
\end{figure}

Then, MC simulations are implemented using the methods listed in Table~\ref{tab3}. For each maneuver detection method, 300 MC runs are tested, where, in each MC run, the initial estimated errors and measurement noises are randomly generated based on the given covariance (\emph{i.e.}, Eq.~\eqref{eq48} for the initial uncertainties and $5''$ for the measurement noises). For maneuver cases, the magnitudes of the impulsive maneuvers are fixed as 1 m/s, while their directions are randomly generated. The confidence curves obtained by the proposed CDMI-AI method in the 300 MC runs are depicted in Fig.~\ref{fig11}, where blue and red curves correspond to true (\emph{i.e.}, no maneuver is detected in the non-maneuver cases or maneuvers are detected in the maneuver cases) and false (\emph{i.e.}, maneuvers are detected in the non-maneuver cases or no maneuver is detected in the maneuver cases) detections. Since the confidence curves from different MC simulations may overlap, transparency is applied to the curve colors in Fig.~\ref{fig11}, with darker shades indicating higher repetition. Additionally, Fig.~\ref{fig12} presents the maneuver detection probability distribution in 300 MC runs. The proposed CDMI-AI method achieves a detection accuracy of 99.33\% for the non-maneuver cases, with only two misdetections (as shown by red lines in Fig.~\ref{fig11a}). In the maneuver cases, the detection accuracy is 79.33\%, resulting in an overall accuracy of 89.33\%.

\begin{figure}[!h]
	\centering
	\subfigure[]
	{
		\label{fig13a}
		\centering
		\includegraphics[width=0.4\textwidth]{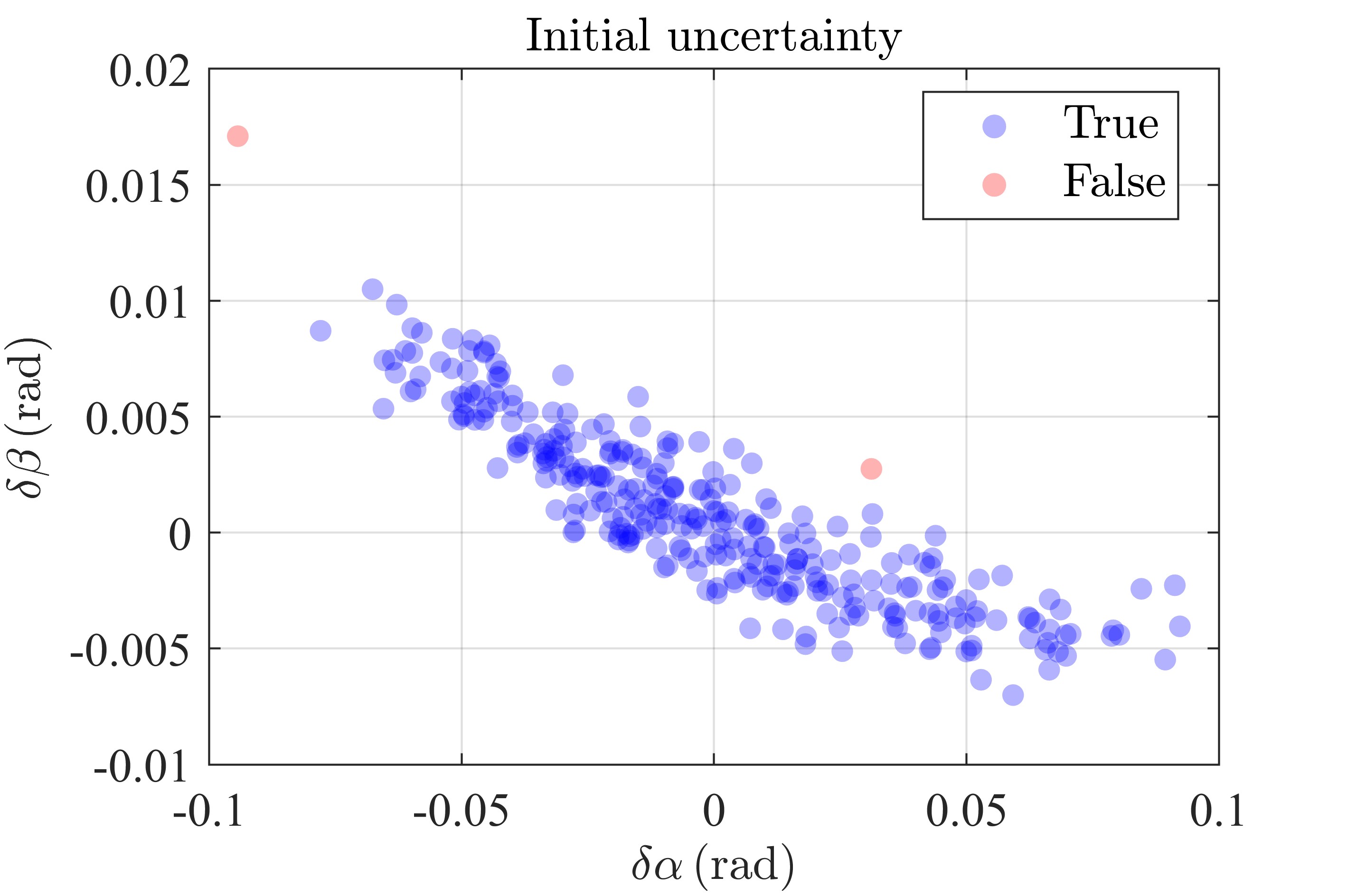}
	}
	\subfigure[]
	{
		\label{fig13b}
		\centering
		\includegraphics[width=0.4\textwidth]{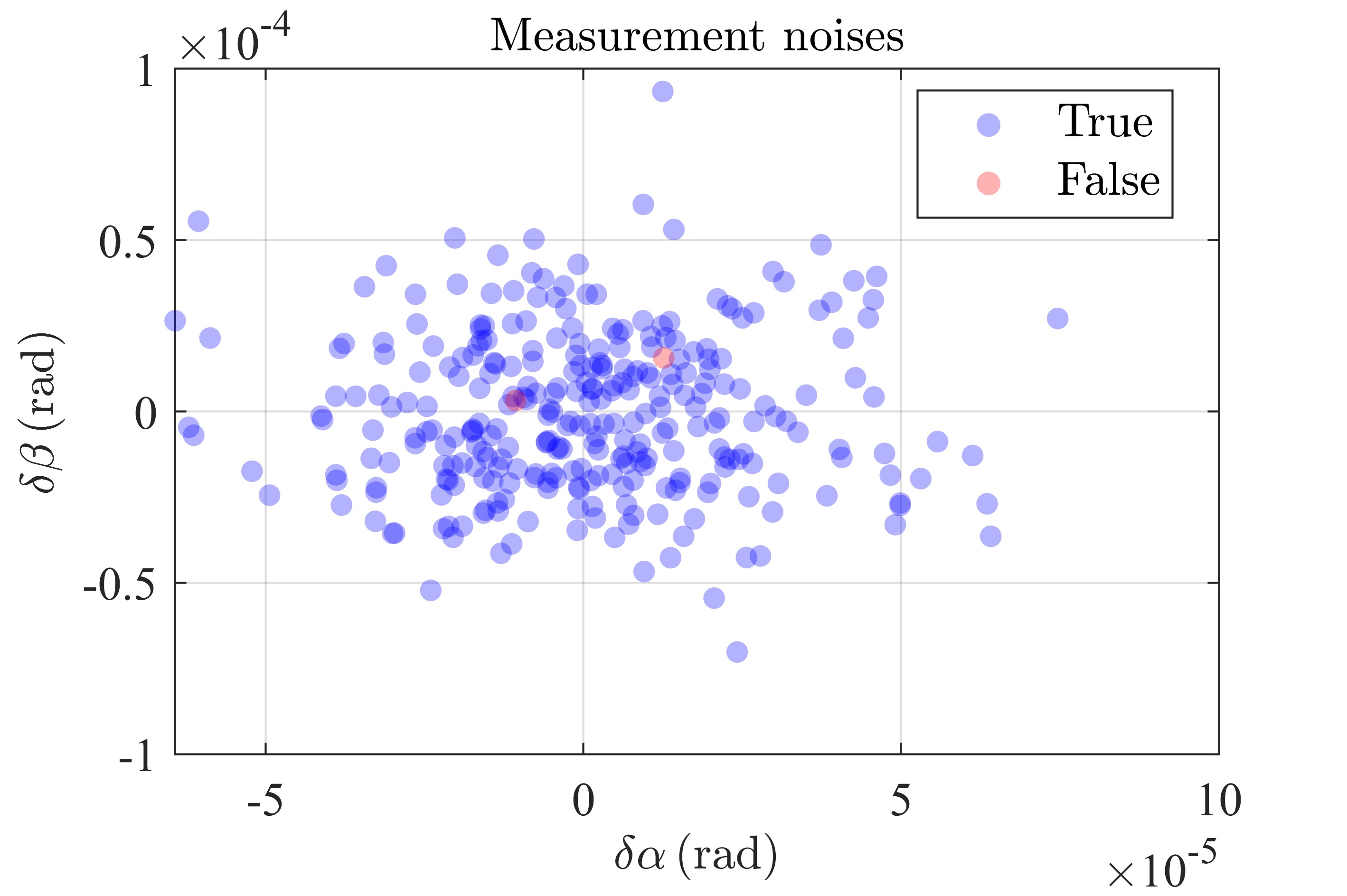}
	}
	\caption{Effects of initial estimated errors and measurement noises on detection accuracy in the non-maneuver cases. \subref{fig13a} Initial estimated errors. \subref{fig13b} Measurement noises.}
	\label{fig13}
\end{figure}

\begin{figure}[!h]
	\centering
	\subfigure[]
	{
		\label{fig14a}
		\centering
		\includegraphics[width=0.4\textwidth]{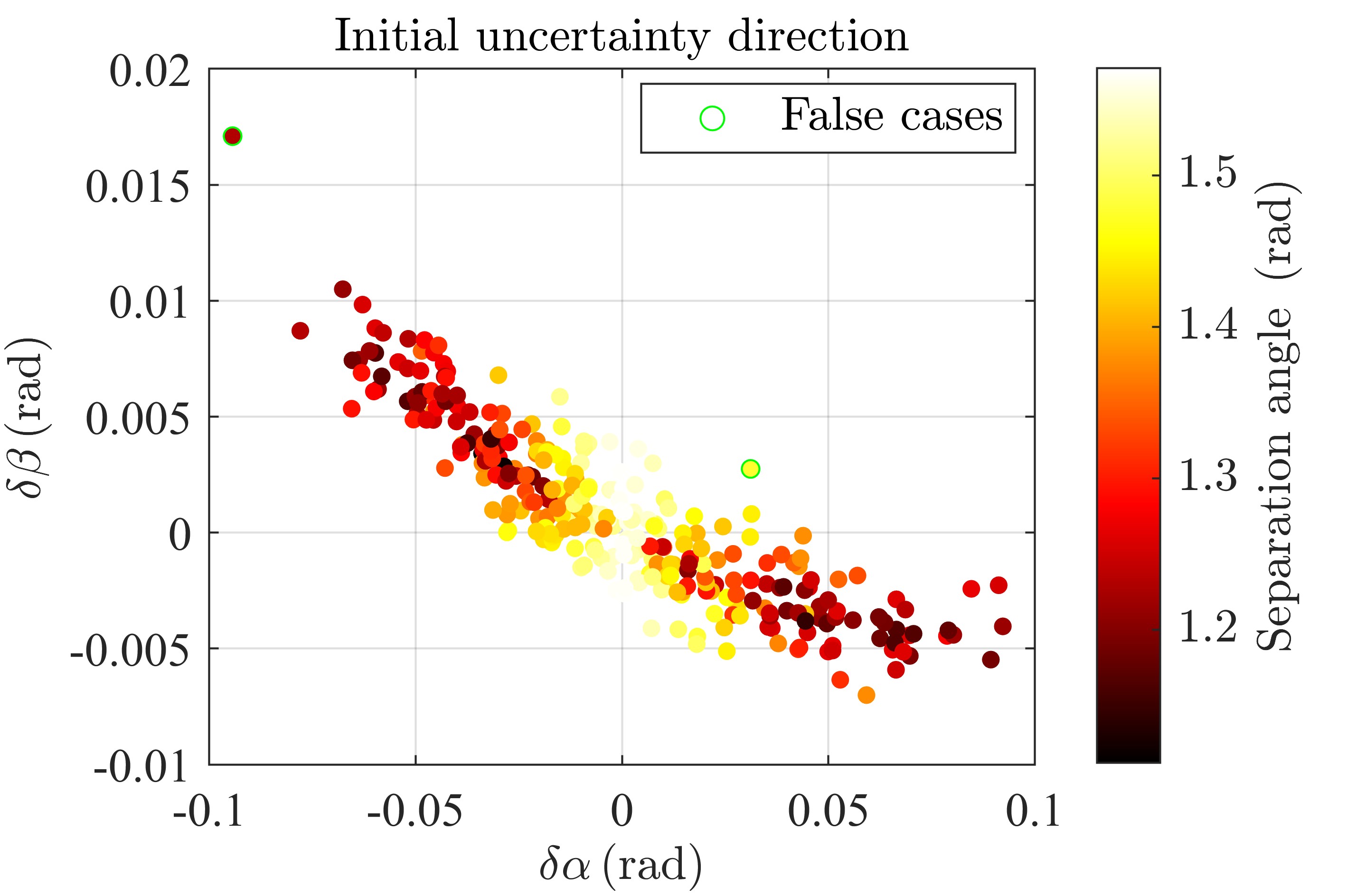}
	}
	\subfigure[]
	{
		\label{fig14b}
		\centering
		\includegraphics[width=0.4\textwidth]{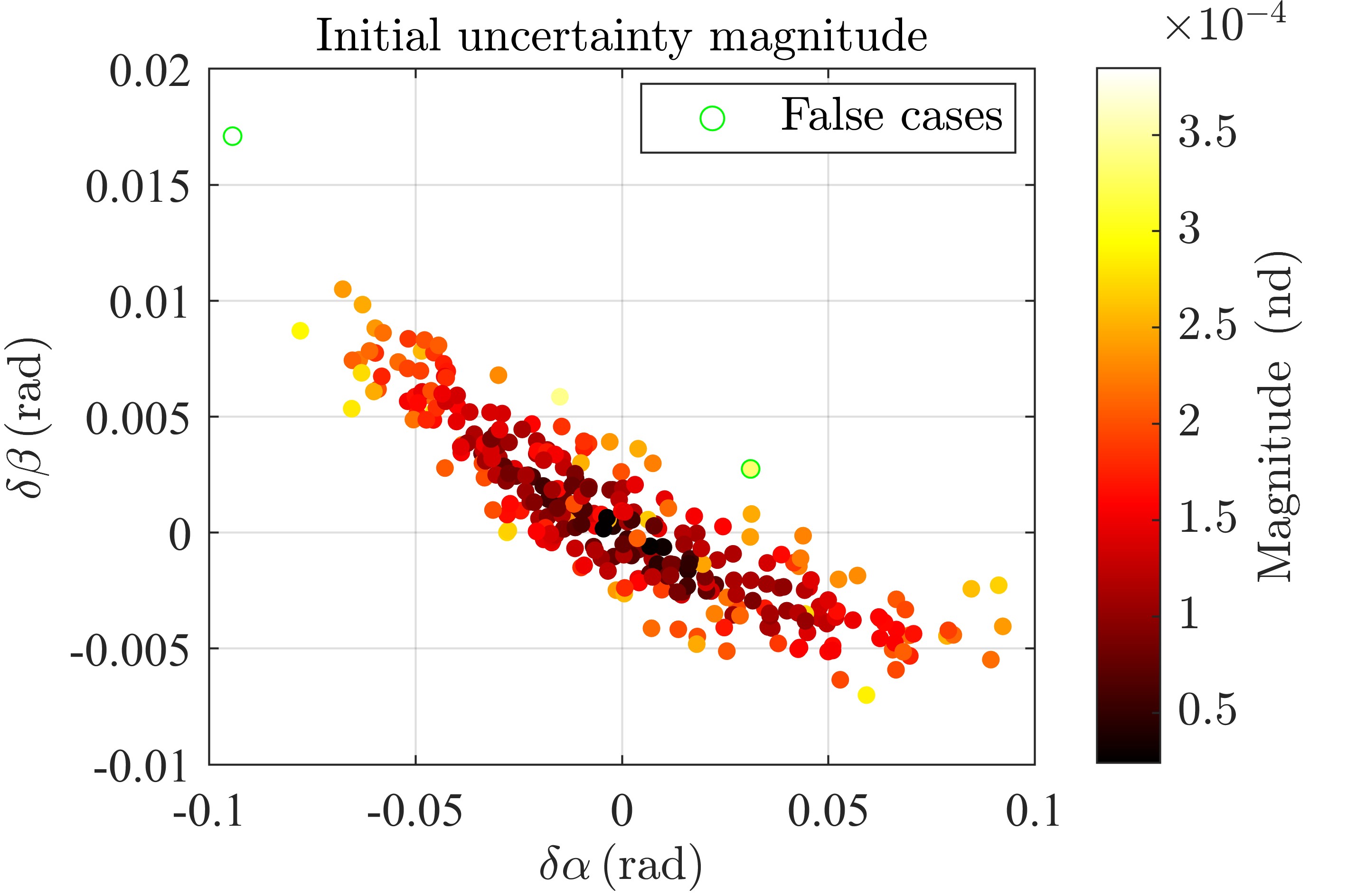}
	}
	\caption{Effects of the directions and magnitudes of initial estimated errors on detection accuracy in the non-maneuver cases. \subref{fig14a} Direction. \subref{fig14b} Magnitude.}
	\label{fig14}
\end{figure}

The factors affecting the maneuver detection accuracy in non-maneuver cases are analyzed. During the implementation of MC simulations, the randomly generated initial estimated errors and measurement noises are recorded. Figure~\ref{fig13a} illustrates the effects of the initial estimated errors on the angle measurements. In Fig.~\ref{fig13a}, blue points represent the cases that are correctly detected (true), while red points indicate misdetections (false) where non-maneuver cases were mistakenly identified as maneuver cases. It can be seen from Fig.~\ref{fig13a} that the two cases are wrongly detected due to the significant effect of their initial estimated errors on the angle measurements. In these two false cases, the measurements deviate considerably from the nominal ones, despite the deviation being mainly caused by initial estimated errors (barely caused by the measurement noises shown in Fig.~\ref{fig13b}), ultimately leading to false detection by the proposed CDMI-AI algorithm. Figure~\ref{fig13b} examines the impact of measurement noises on maneuver detection performance (in non-maneuver cases). The results in Fig.~\ref{fig13b} show that, since the effect of measurement noises is about three orders of magnitude smaller than that of initial estimated errors, it doesn’t impact the maneuver detection accuracy. Furthermore, Fig.~\ref{fig14} demonstrates how the magnitude and direction of initial estimated errors influence the detection performance in non-maneuver cases. Similar to the results in Fig.~\ref{fig14}, the directions of initial estimated errors are represented by the separation angles with respect to the most sensitive direction (obtained by the CGT approach). As shown in Fig.~\ref{fig14}, among the two misclassified runs, one is due to a large estimated error (see Fig.~\ref{fig14b}), while the other is caused by both a large magnitude and a sensitive direction.

\begin{figure}[!h]
	\centering
	\subfigure[]
	{
		\label{fig15a}
		\centering
		\includegraphics[width=0.4\textwidth]{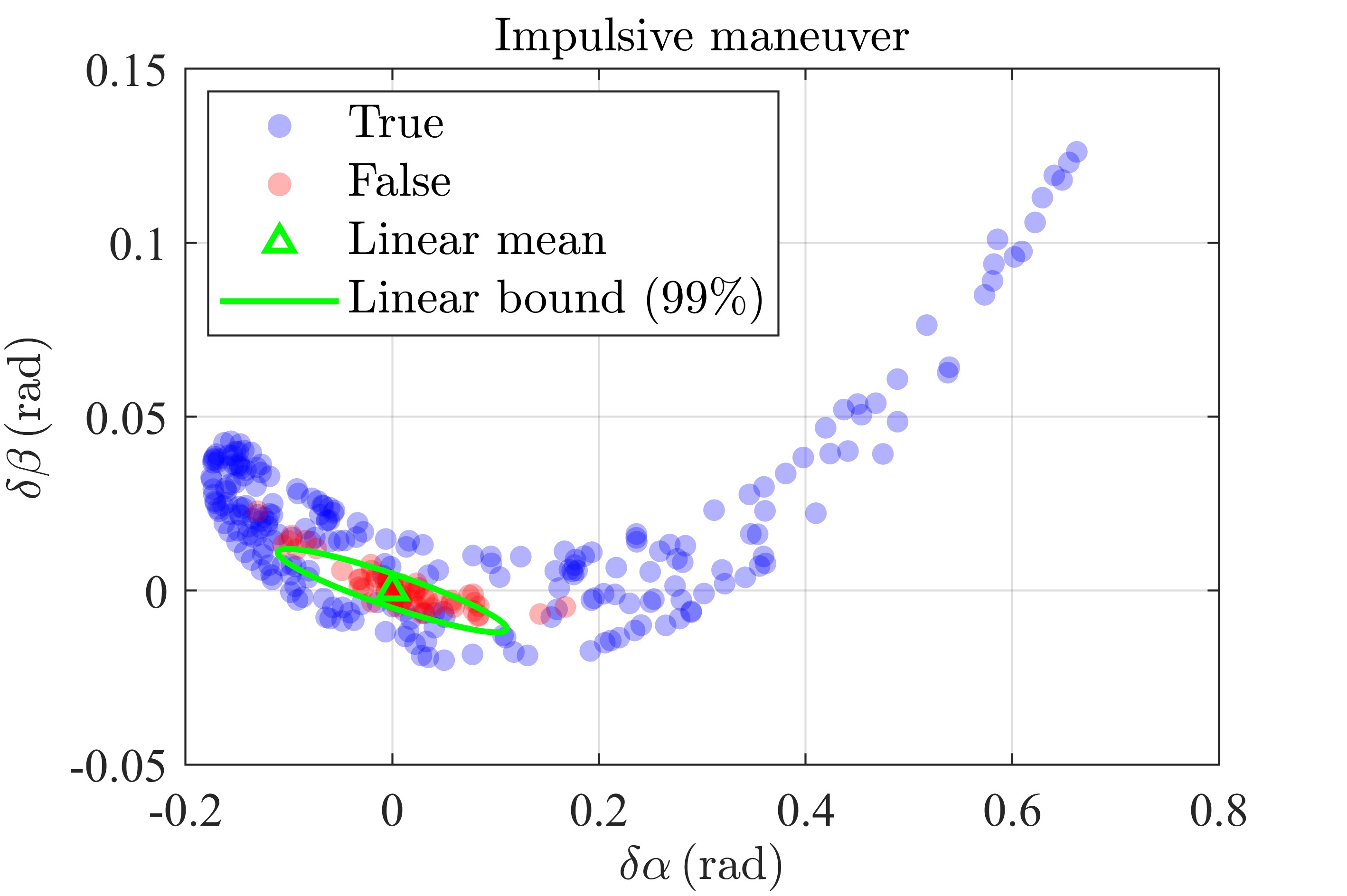}
	}
	\subfigure[]
	{
		\label{fig15b}
		\centering
		\includegraphics[width=0.4\textwidth]{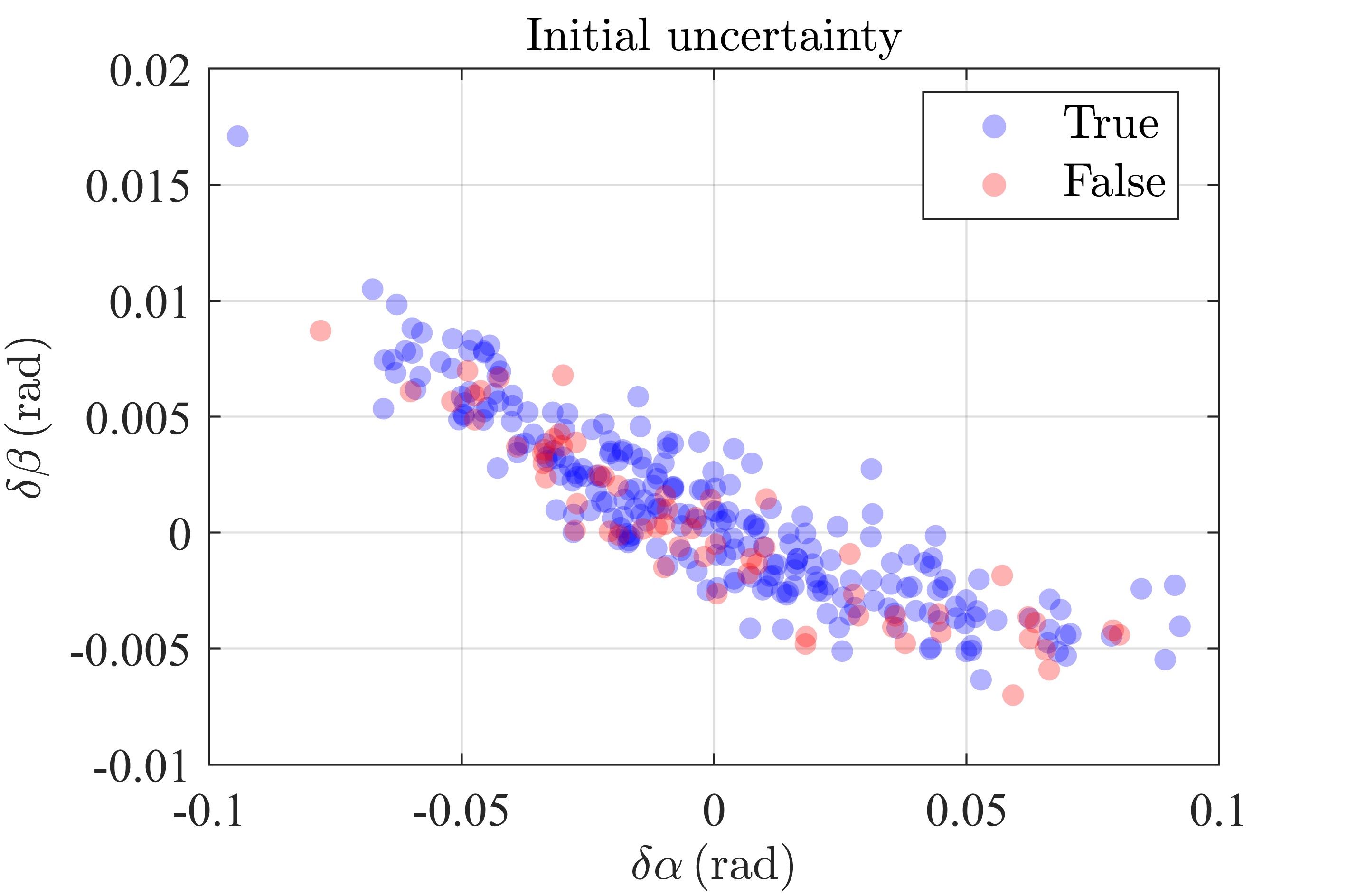}
	}
	\caption{Effects of maneuvers and initial estimated errors on detection accuracy in the maneuver cases. \subref{fig15a} Maneuvers. \subref{fig15b} Initial estimated errors.}
	\label{fig15}
\end{figure}

Then, we analyze the effects of impulsive maneuvers and initial estimated errors on CDMI-AI’s maneuver detection performances in maneuver cases. Since the previous simulations (\emph{i.e.}, Fig.~\ref{fig14}) have shown that measurement noises have a much smaller effect on the angle measurements than the initial estimated errors, the influences of measurement noises are not presented here. Figure~\ref{fig15a} shows the angle measurements from 300 MC runs, considering only the effect of impulsive maneuvers, where red and blue points indicate misdetections and correct detections, respectively. The green ellipse from Fig.~\ref{fig4} (\emph{i.e.}, linear bound) is also added, representing the region where angle measurement deviations can be attributed to initial estimated errors. If an angle measurement affected by the impulsive maneuver falls within this green boundary, it suggests that the resulting measurement deviation is likely caused by initial estimated errors, making it more prone to being detected as a non-maneuver case (leading to a false detection). As expected, in Fig.~\ref{fig15a}, most of the misdetected maneuver cases occur when the maneuver-induced measurement deviation is insignificant (\emph{i.e.}, most red points fall within the green elliptical boundary). Additionally, Fig.~\ref{fig15b} shows the effects of initial estimated errors, which indicates that, for the simulated cases, initial estimated errors do not significantly impact the maneuver detection performance. Figure~\ref{fig6} further analyzes how the impulsive maneuver direction (shown by the separation angle with respect to the sensitive direction obtained by the CGT approach) affects detection accuracy. The results indicate that if an impulsive maneuver exists but is not detected by the algorithm, it is likely oriented along an insensitive direction, resulting in a weak impact on the angle measurements. In such cases, the maneuver-induced measurement deviation can be easily obscured by the one induced by initial estimated errors, leading to misdetections.

\begin{figure}[!h]
	\centering
	\subfigure[]
	{
		\label{fig16a}
		\centering
		\includegraphics[width=0.4\textwidth]{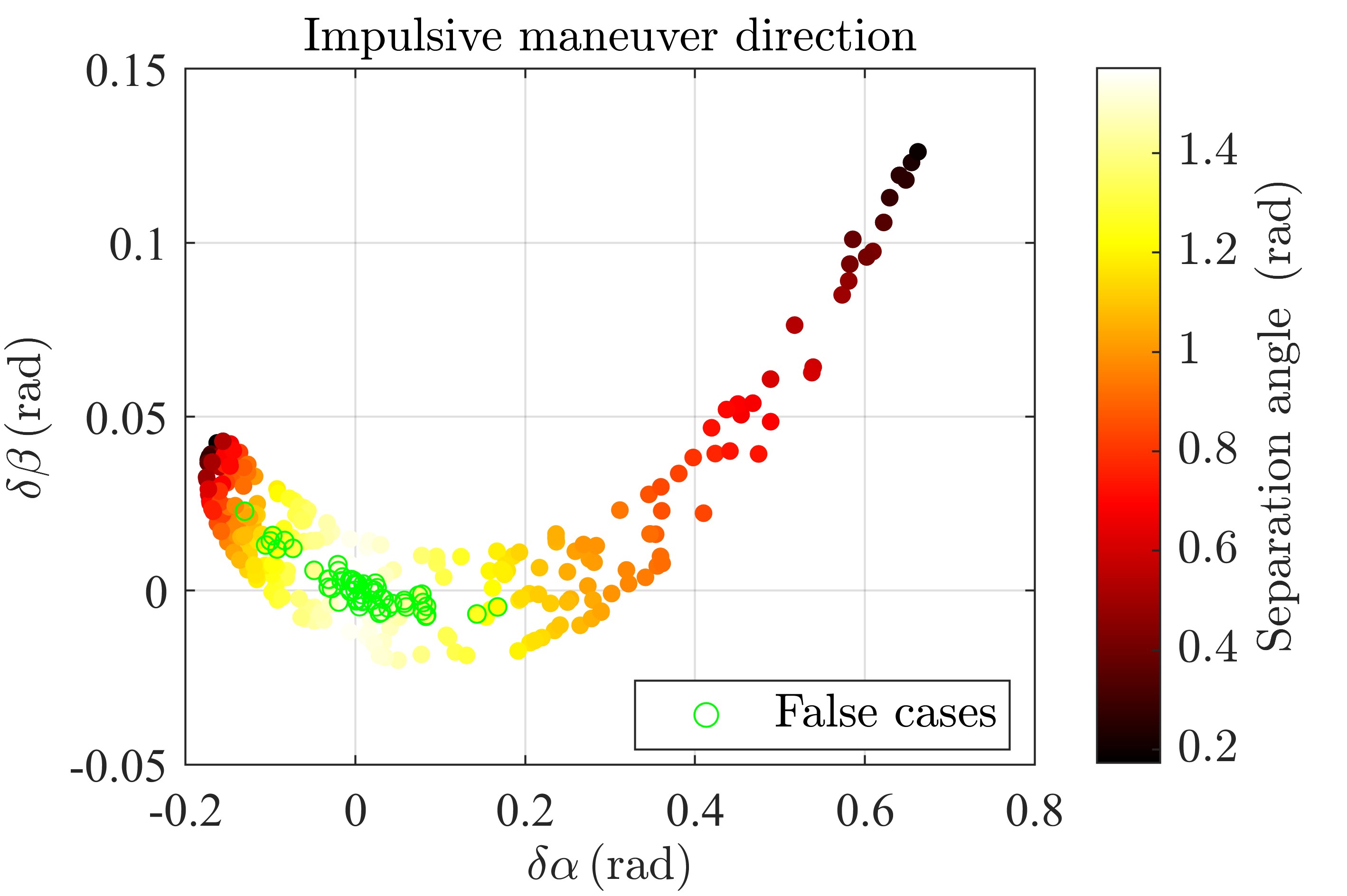}
	}
	\subfigure[]
	{
		\label{fig16b}
		\centering
		\includegraphics[width=0.4\textwidth]{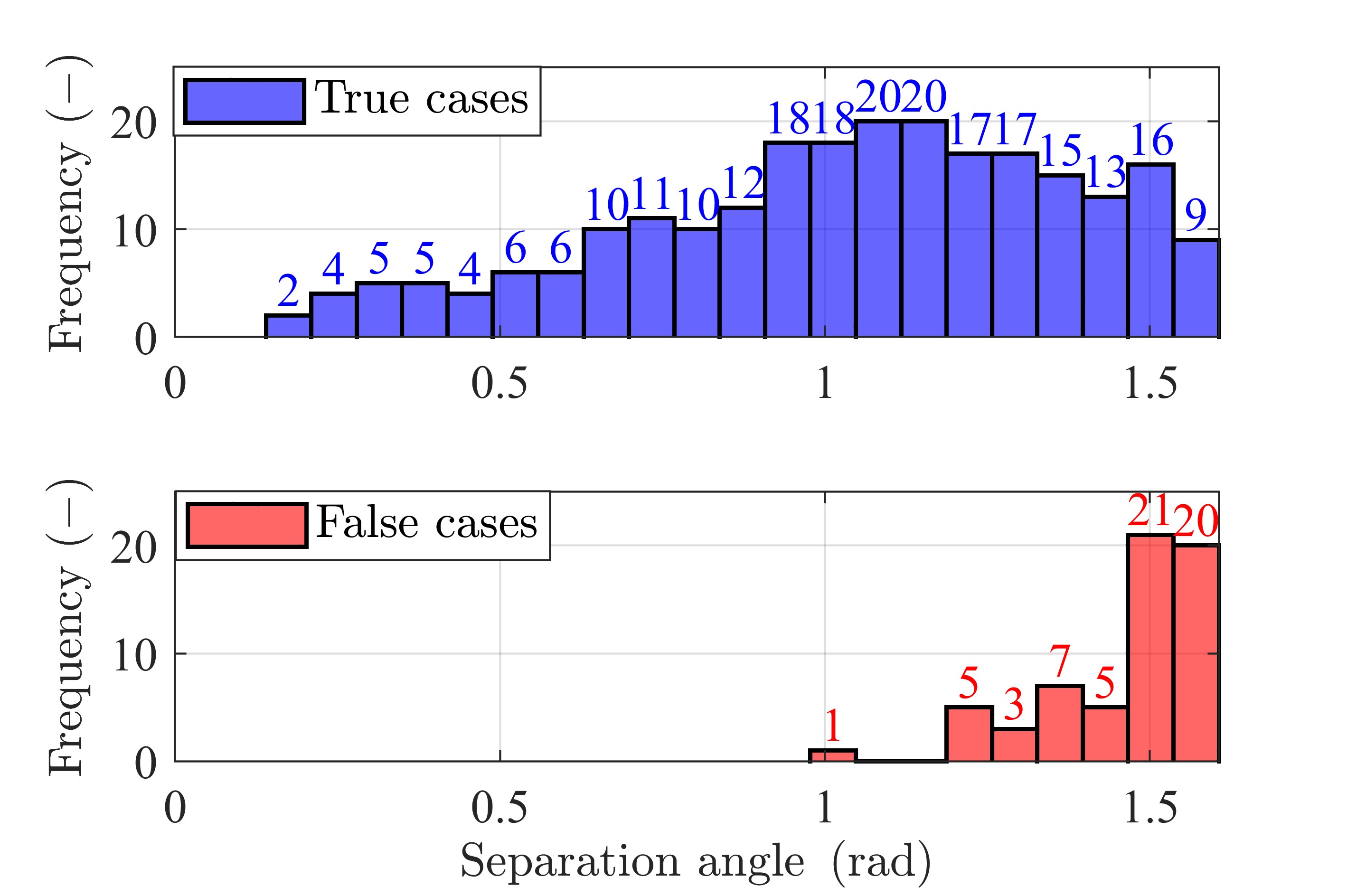}
	}
	\caption{Effects of the directions and magnitudes of initial estimated errors on detection accuracy in the non-maneuver cases. \subref{fig16a} Effects of maneuver directions. \subref{fig16b} Separation angle distributions.}
	\label{fig16}
\end{figure}

\begin{figure}[!h]
	\centering
	\subfigure[]
	{
		\label{fig17a}
		\centering
		\includegraphics[width=0.4\textwidth]{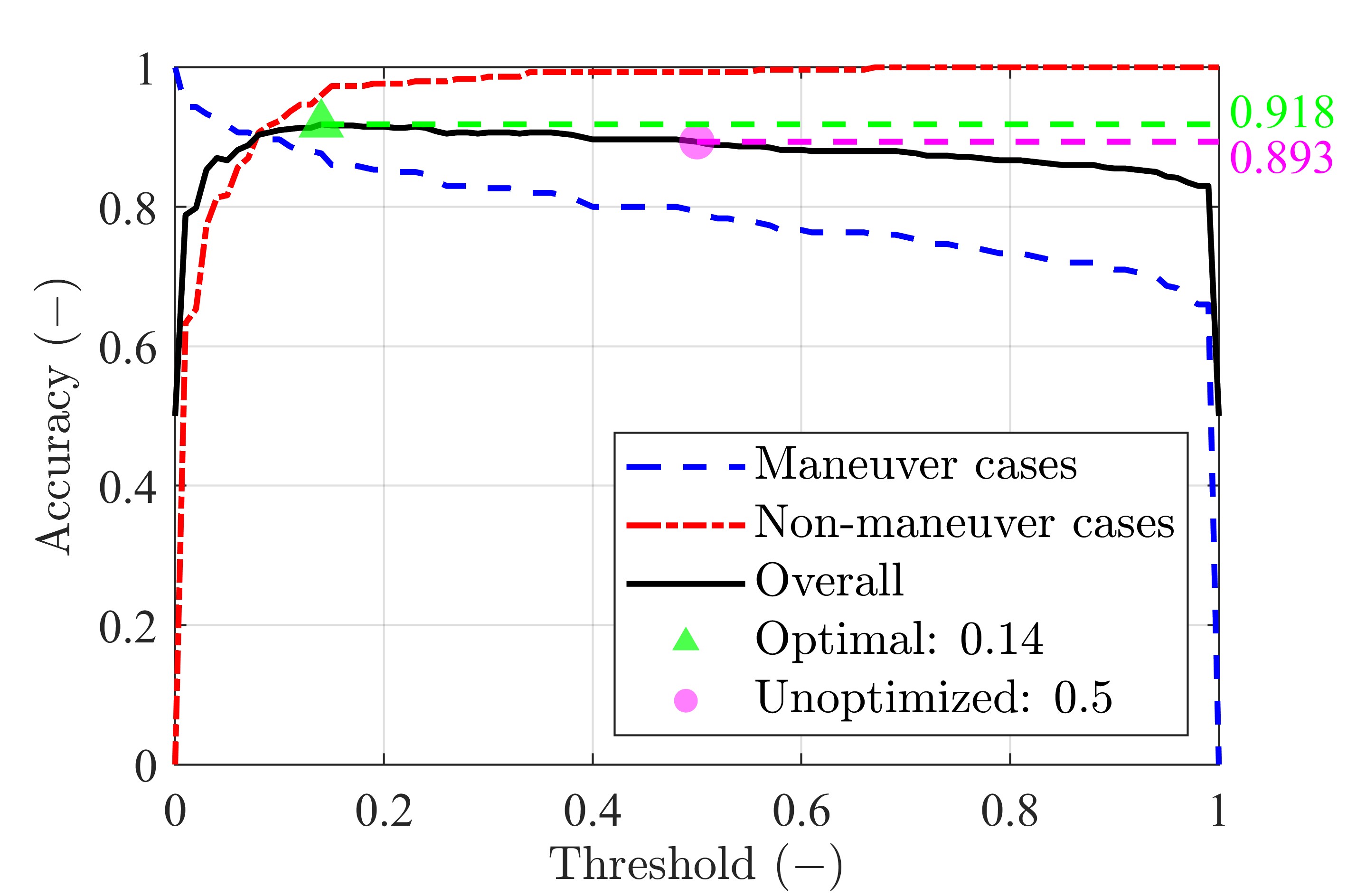}
	}
	\subfigure[]
	{
		\label{fig17b}
		\centering
		\includegraphics[width=0.4\textwidth]{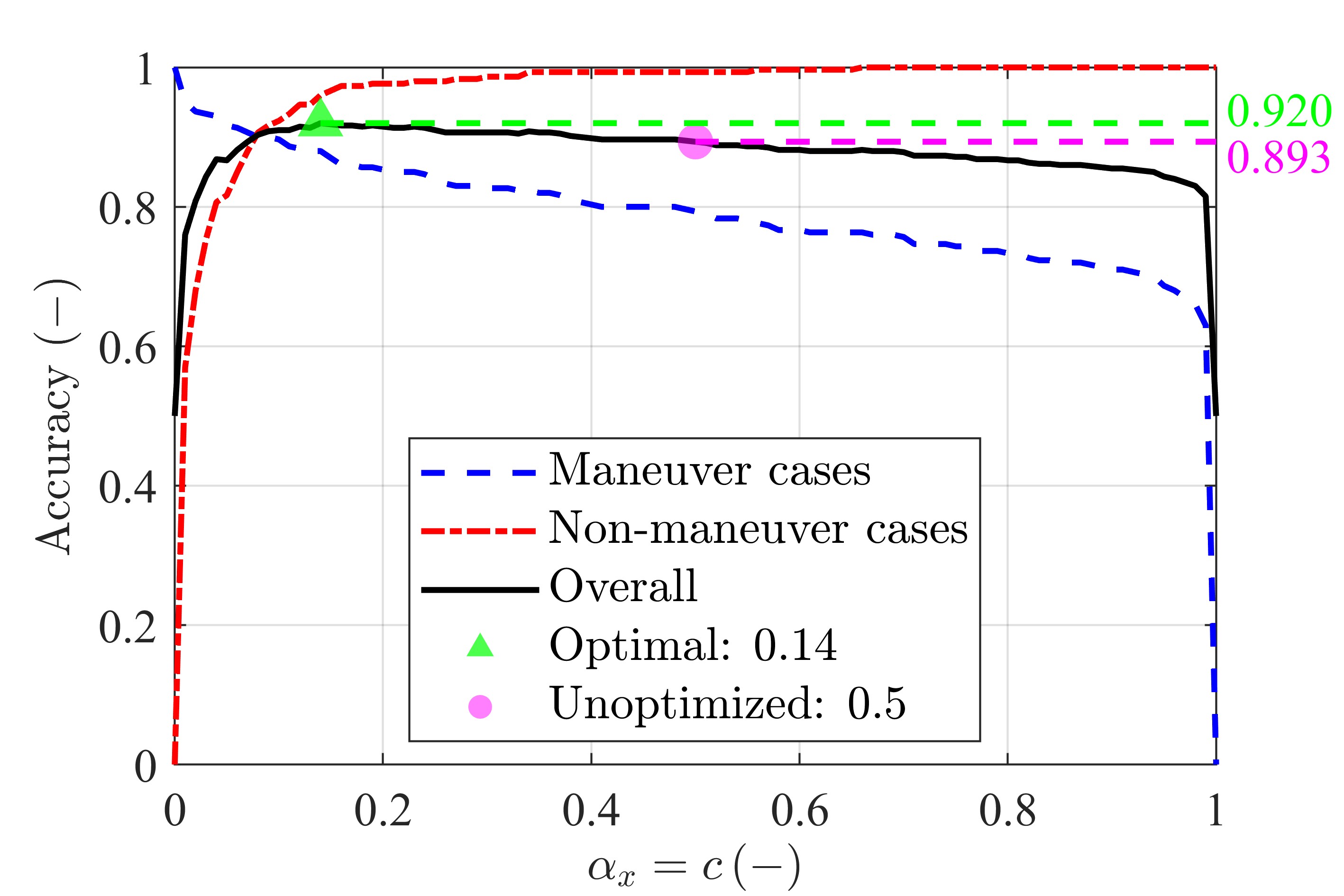}
	}
	\caption{Effects of user-defined parameters on detection accuracy (single measurement). \subref{fig17a} CDMI-AI. \subref{fig17b} CDMI-S.}
	\label{fig17}
\end{figure}

Next, the effects of the threshold on the CDMI-AI’s maneuver detection performances are investigated. Figure~\ref{fig17a} illustrates the maneuver detection accuracy of the CDMI-AI method under different threshold settings. In the above simulations (\emph{i.e.}, Figs.~\ref{fig10}-\ref{fig16}), an empirically chosen threshold of 0.5 is used (as discussed in Sec.~\ref{sec:Maneuver Detection Approaches}), yielding an overall detection accuracy of 0.893 (see Figs.~\ref{fig11}-\ref{fig12}). After evaluating various threshold values, it is found that setting the threshold to 0.14 results in the highest detection accuracy of 0.918. However, it’s worth noting that this “optimal” threshold is determined through retrospective analysis and, therefore, cannot be directly applied to a real-world maneuver detection task. Similarly, the proposed CDMI-S method relies on a pre-defined confidence level (\emph{i.e.}, $\alpha_x=c$), which serves as its “threshold” (see the discussion in Sec.~\ref{sec:Indicator with Measurement at Single Epoch}). As shown in Fig.~\ref{fig11b}, when using an unoptimized confidence level $c$ of 0.5, the overall accuracy is 0.893. The highest accuracy of 0.920 is achieved when the confidence level $c$ is set to 0.14. Recall that, although in this specific simulation, the CDMI-S achieves a detection accuracy comparable to or even slightly better than the CDMI-AI, its capability is limited to binary maneuver detection. The CDMI-AI method can not only detect maneuvers but also provide an indication of the likelihood of a maneuver (as shown in Fig.~\ref{fig12}).

Finally, Table~\ref{tab6} and Table~\ref{tab7} present the maneuver detection accuracy and computational costs of the proposed and competitive methods listed in Table~\ref{tab3}. In Table~\ref{tab6}, “NA” means “not applicable”, as the competitive OCDM and MBA methods don’t require an optimization process. For the proposed methods, Table~\ref{tab7} provides two types of CPU times: one is the total CPU time (including deriving the polynomial in the DA framework and implementing the optimization to find the closest point), and the other is the CPU time of the optimization (within brackets). The sensitive directions obtained by the CGT approach are taken as splitting directions for the MBAs. As the seven-component univariate library is employed for each splitting direction, the MBA$_1$ and MBA$_2$ use 7 ($7=7^1$) and 49 ($49=7^2$) components to approximate the initial state distributions, respectively. Three pre-defined confidence levels (\emph{i.e.}, $\alpha_x=c$) are chosen and investigated for the CDMI-S method: 0.5 (a reasonable value adopted from our preliminary analysis), 0.7 (a threshold value adopted from Ref.~\cite{Montilla2025}), and 0.95 (a threshold value adopted from Ref.~\cite{Pirovano2024}). 

\begin{table}[!h]
	\centering
	\caption{Maneuver detection accuracy with single-epoch measurement}
	\label{tab6}
	\begin{tabular}{|l|l|lll|}
    \hline
    \multirow{2}{*}{Method} & \multirow{2}{*}{\makecell[l]{Optimization\\ approach}} & \multicolumn{3}{l|}{Accuracy (-)} \\ \cline{3-5} 
     & & \multicolumn{1}{l|}{Non-maneuver} & \multicolumn{1}{l|}{Maneuver} & Overall \\ \hline
    \multirow{2}{*}{CDMI-S$_{0.5}$} & RPO & \multicolumn{1}{l|}{0.9933} & \multicolumn{1}{l|}{0.7933} & 0.8933 \\ \cline{2-5} 
     & SLSQP & \multicolumn{1}{l|}{0.8467} & \multicolumn{1}{l|}{0.8200} & 0.8333 \\ \hline
    \multirow{2}{*}{CDMI-S$_{0.7}$} & RPO & \multicolumn{1}{l|}{1} & \multicolumn{1}{l|}{0.7567} & 0.8783  \\ \cline{2-5} 
     & SLSQP & \multicolumn{1}{l|}{0.8567} & \multicolumn{1}{l|}{0.7867} & 0.8217 \\ \hline
    \multirow{2}{*}{CDMI-S$_{0.95}$} & RPO & \multicolumn{1}{l|}{1} & \multicolumn{1}{l|}{0.6867} & 0.8433  \\ \cline{2-5} 
     & SLSQP & \multicolumn{1}{l|}{0.8667} & \multicolumn{1}{l|}{0.7167} & 0.7917 \\ \hline
    \multirow{2}{*}{CDMI-AI$_{0.5}$} & RPO & \multicolumn{1}{l|}{0.9933} & \multicolumn{1}{l|}{0.7933}   & 0.8933  \\ \cline{2-5} 
     & SLSQP  & \multicolumn{1}{l|}{0.8533} & \multicolumn{1}{l|}{0.8200} & 0.8367 \\ \hline
    \multirow{2}{*}{CDMI-I$_{0.5}$} & RPO & \multicolumn{1}{l|}{0.9933} & \multicolumn{1}{l|}{0.7933} & 0.8933  \\ \cline{2-5} 
     & SLSQP & \multicolumn{1}{l|}{0.8466} & \multicolumn{1}{l|}{0.8233} & 0.8350 \\ \hline
    OCDM & NA & \multicolumn{1}{l|}{0.6367} & \multicolumn{1}{l|}{0.8900} & 0.7633 \\ \hline
    MBA$_1$ & NA & \multicolumn{1}{l|}{0.3533} & \multicolumn{1}{l|}{0.9633} & 0.6583 \\ \hline
    MBA$_2$ & NA & \multicolumn{1}{l|}{0.6600} & \multicolumn{1}{l|}{0.9300} & 0.7950 \\ \hline
    \end{tabular}
\end{table}

\begin{table*}[!h]
	\centering
	\caption{Computational costs with single-epoch measurement}
	\label{tab7}
	\begin{tabular}{|l|l|lll|}
    \hline
    \multirow{2}{*}{Method}        & \multirow{2}{*}{\makecell[l]{Optimization \\approach}} & \multicolumn{3}{l|}{Averaged CPU time (s)}                                                          \\ \cline{3-5} 
                                   &                                        & \multicolumn{1}{l|}{Non-maneuver}      & \multicolumn{1}{l|}{Maneuver}          & Overall           \\ \hline
    \multirow{2}{*}{CDMI-S}          & RPO                                    & \multicolumn{1}{l|}{3.0084 (0.0131)}   & \multicolumn{1}{l|}{3.0185 (0.0145)}   & 3.0135 (0.0138)   \\ \cline{2-5} 
                                   & SLSQP                                  & \multicolumn{1}{l|}{3.3719 (0.3708)}   & \multicolumn{1}{l|}{3.4382 (0.4251)}   & 3.4050 (0.3979)   \\ \hline
    \multirow{2}{*}{CDMI-AI$_{0.5}$} & RPO                                    & \multicolumn{1}{l|}{3.2644 (0.1520)}   & \multicolumn{1}{l|}{3.1578 (0.1273)}   & 3.2111 (0.1397)   \\ \cline{2-5} 
                                   & SLSQP                                  & \multicolumn{1}{l|}{18.1826 (15.0666)} & \multicolumn{1}{l|}{10.3366 (7.3004)}  & 14.2596 (11.1835) \\ \hline
    \multirow{2}{*}{CDMI-I$_{0.5}$}  & RPO                                    & \multicolumn{1}{l|}{4.3294 (1.3341)}   & \multicolumn{1}{l|}{4.47985 (1.4757)}  & 4.40465 (1.4049)  \\ \cline{2-5} 
                                   & SLSQP                                  & \multicolumn{1}{l|}{40.8295 (37.8285)} & \multicolumn{1}{l|}{46.3744 (43.3613)} & 43.6020 (40.5949) \\ \hline
    OCDM                           & NA                                     & \multicolumn{1}{l|}{6.7730}            & \multicolumn{1}{l|}{6.7525}            & 6.7628            \\ \hline
    MBA$_1$                        & NA                                     & \multicolumn{1}{l|}{3.2108}            & \multicolumn{1}{l|}{3.1252}            & 3.1680            \\ \hline
    MBA$_2$                        & NA                                     & \multicolumn{1}{l|}{3.7887}            & \multicolumn{1}{l|}{3.7005}            & 3.7446            \\ \hline
    \end{tabular}
\end{table*}

One can see from Table~\ref{tab6} that the proposed CDMI-AI (with RPO) ranks the best in overall accuracy. It is more than 20\% (overall accuracy of 0.8933) better than the OCDM (overall accuracy of 0.7633) and 10\% better than the MBAs (MBA$_1$: 0.6583, MBA$_2$: 0.7950). The pre-defined confidence level $c$ plays a crucial role in the maneuver detection accuracy of the CDMI-S method, highlighting one of its key limitations. When applying the proposed methods, the RPO consistently outperforms the SLSQP in both accuracy and efficiency. Leveraging the convex optimization’s breakneck computational speed, the average CPU time of the RPO is more than an order of magnitude lower than that of the SLSQP. Moreover, it is observed that the SLSQP often converges to a local solution rather than the desired one (\emph{i.e.}, the optimized closest point is a bit farther from the observed measurement than expected), leading to an underestimation of the measurement confidence level $\alpha_z$. As a result, compared to proposed methods utilizing RPO, the ones employing SLSQP are more likely to misdetect the non-maneuver cases, yielding higher detection accuracy for maneuver cases but lower accuracy for the non-maneuver ones.

In addition, one can see from Table~\ref{tab6} and Table~\ref{tab7} that the proposed adaptive sampling strategy (\emph{i.e.}, CDMI-AI$_{0.5}$) can significantly reduce the computational cost in the optimization process, compared with the non-adaptive one (\emph{i.e.}, CDMI-AI$_{0.5}$), while achieving the same level of accuracy in maneuver detection. The CDMI-S method is the most efficient among the proposed methods. However, the generation of the high-order Taylor polynomials requires most of the computational burden. The average CPU time for deriving the polynomials in the DA scheme ranges from 2.9 s to 3.1 s \footnote{Although the random seed is fixed at the beginning of the MC simulations to ensure that the generated random sequences are the same, allowing for a fair comparison where random parameters do not affect the results, slight differences in CPU time are still observed. Since all methods share the same random parameters in each MC run, the Taylor polynomials generated should, in theory, require the same computational time. However, minor discrepancies arise in practice due to factors such as CPU thermal throttling and memory access latency.}. Therefore, the CDMI-S method does not have a significant advantage in computational efficiency over the CDMI-AI method. The CDMI-S and CDMI-AI$_{0.5}$ methods are much more efficient than the OCDM and MBA$_2$. The MBA$_1$ has a computational burden similar to that of the CDMI-AI$_{0.5}$, but with much lower detection accuracy.

\subsection{Maneuver Detection Results Using Multiple-Epoch Angle} \label{sec:Maneuver Detection Results Using Multiple-Epoch Angle}
\begin{figure}[!h]
	\centering
	\subfigure[]
	{
		\label{fig18a}
		\centering
		\includegraphics[width=0.4\textwidth]{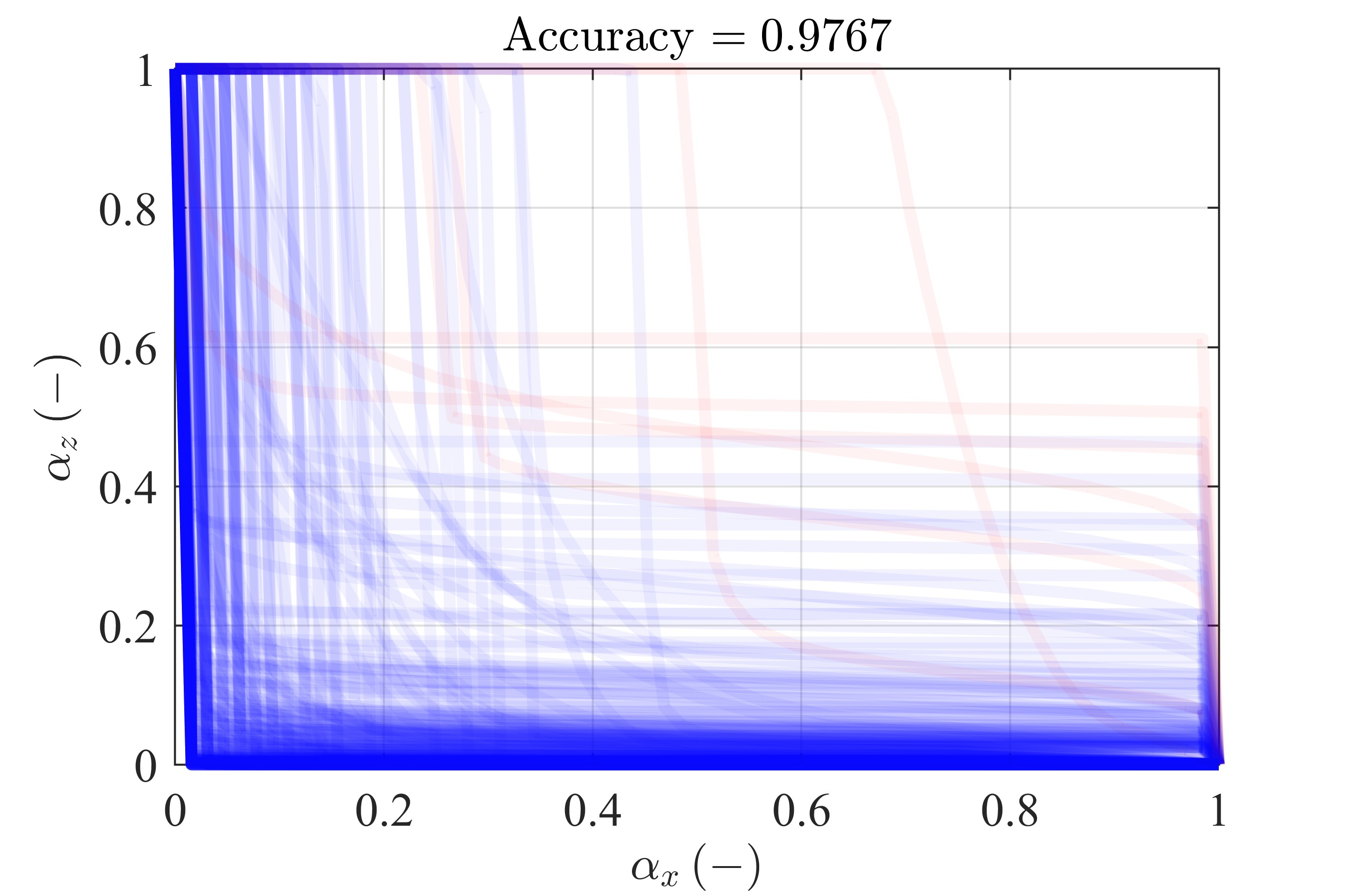}
	}
	\subfigure[]
	{
		\label{fig18b}
		\centering
		\includegraphics[width=0.4\textwidth]{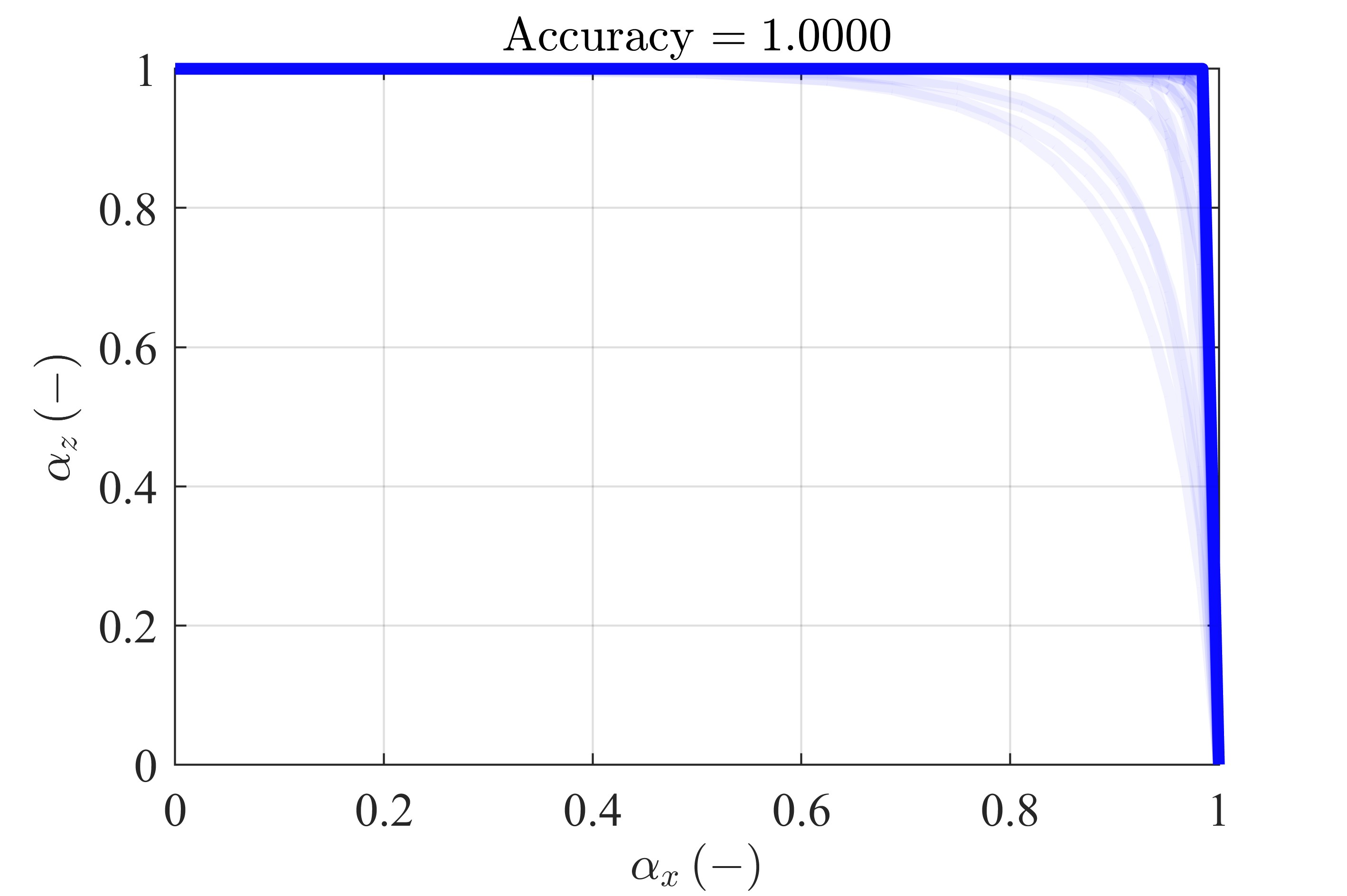}
	}
	\caption{$\alpha_x$-$\alpha_z$ curves obtained by the CDMI-AI with multiple-epoch angle. \subref{fig18a} Non-maneuver case. \subref{fig18b} Maneuver case.}
	\label{fig18}
\end{figure}

\begin{figure}[!h]
	\centering
	\subfigure[]
	{
		\label{fig19a}
		\centering
		\includegraphics[width=0.4\textwidth]{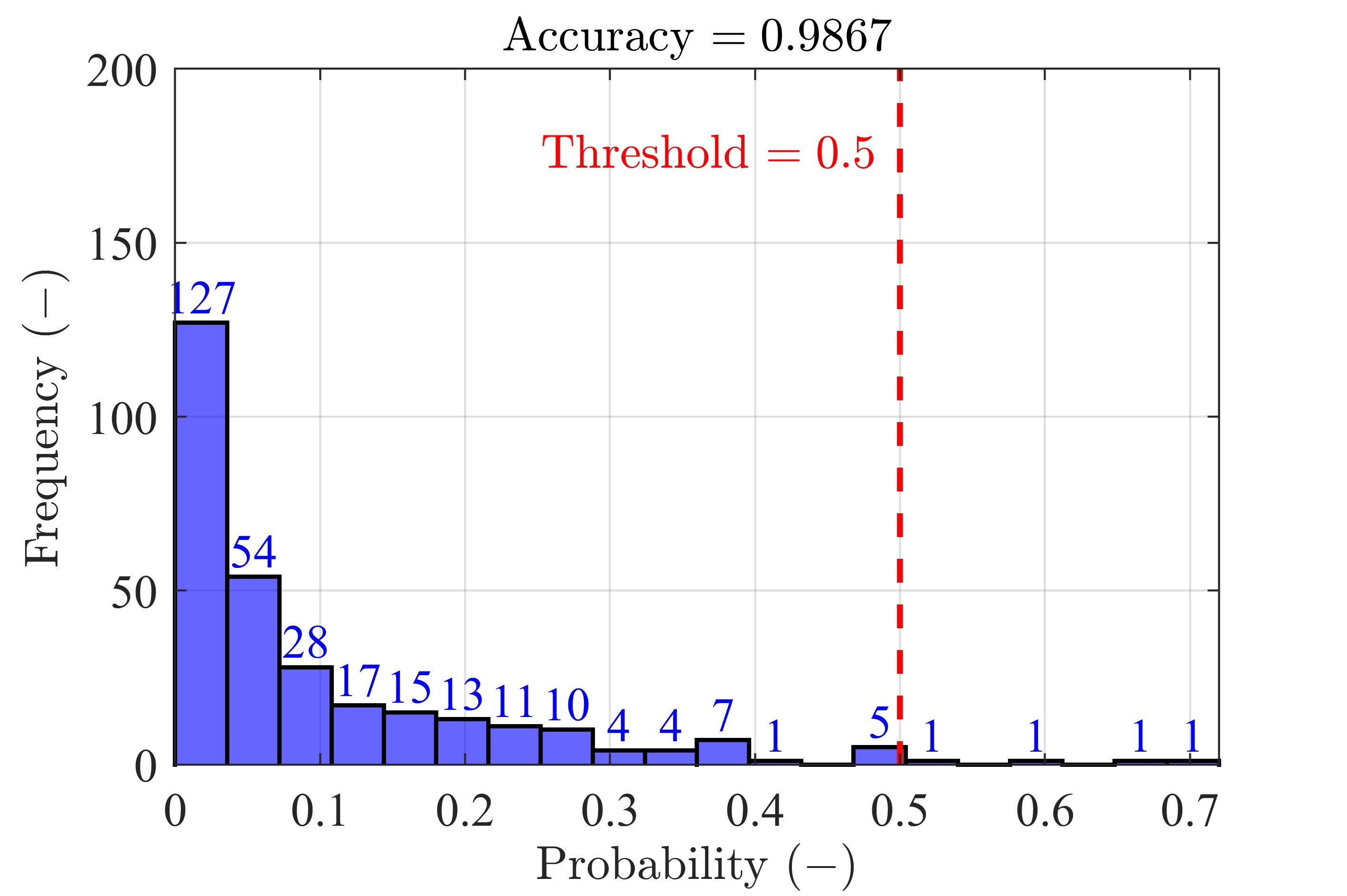}
	}
	\subfigure[]
	{
		\label{fig19b}
		\centering
		\includegraphics[width=0.4\textwidth]{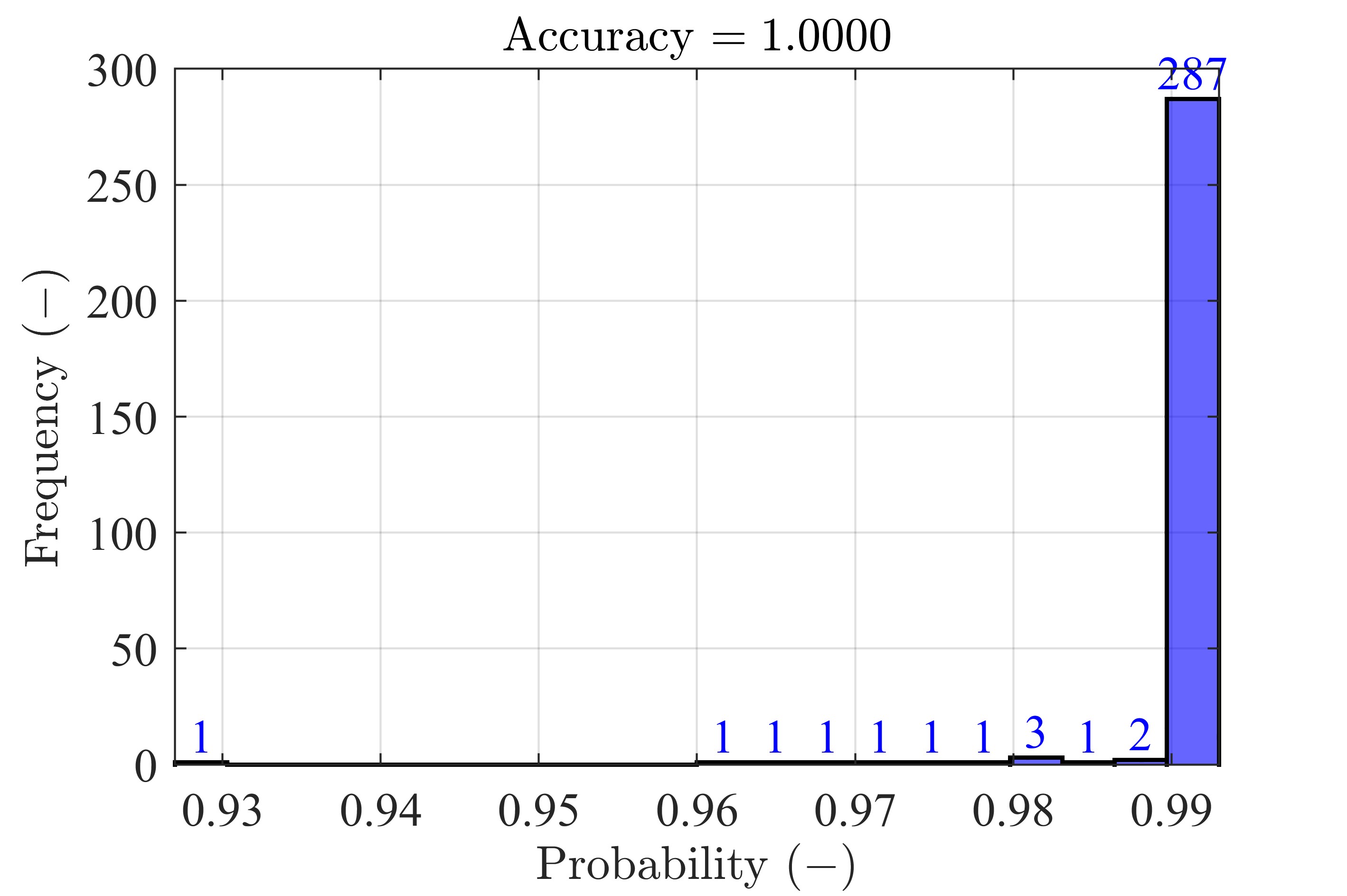}
	}
	\caption{Maneuver probabilities obtained by the CDMI-AI with multiple-epoch angle. \subref{fig18a} Non-maneuver case. \subref{fig18b} Maneuver case.}
	\label{fig19}
\end{figure}

In this subsection, three angle measurements are employed for maneuver detection. With the first observation epoch set at $t_1=3T$ (the same as that in the single-epoch angle case), the other two observation epochs are set as $t_2=t_1+0.01T$ and $t_3=t_2+0.01T$ ($T$ is the period of the target). The user-defined parameters, such as the initial estimated covariance, measurement noise STD, and convergence threshold of the RPO $\eta$, remain the same as those given in Sec.~\ref{sec:Maneuver Detection Results Using Single-Epoch Angle}. Three hundred MC runs are implemented using the CDMI-AI method, with the obtained confidence curves and maneuver probabilities presented in Figs.~\ref{fig18}-\ref{fig19}, respectively. By adding two additional angle measurements, the detection accuracy is significantly improved, with the accuracies for the non-maneuver and maneuver cases being 98.67\% and 100\%, respectively, achieving an overall accuracy of 99.3\% (recall that the overall accuracy under one single measurement is 89.3\%). Only four non-maneuver cases are misdetected. 

\begin{figure}[!h]
	\centering
	\subfigure[]
	{
		\label{fig20a}
		\centering
		\includegraphics[width=0.4\textwidth]{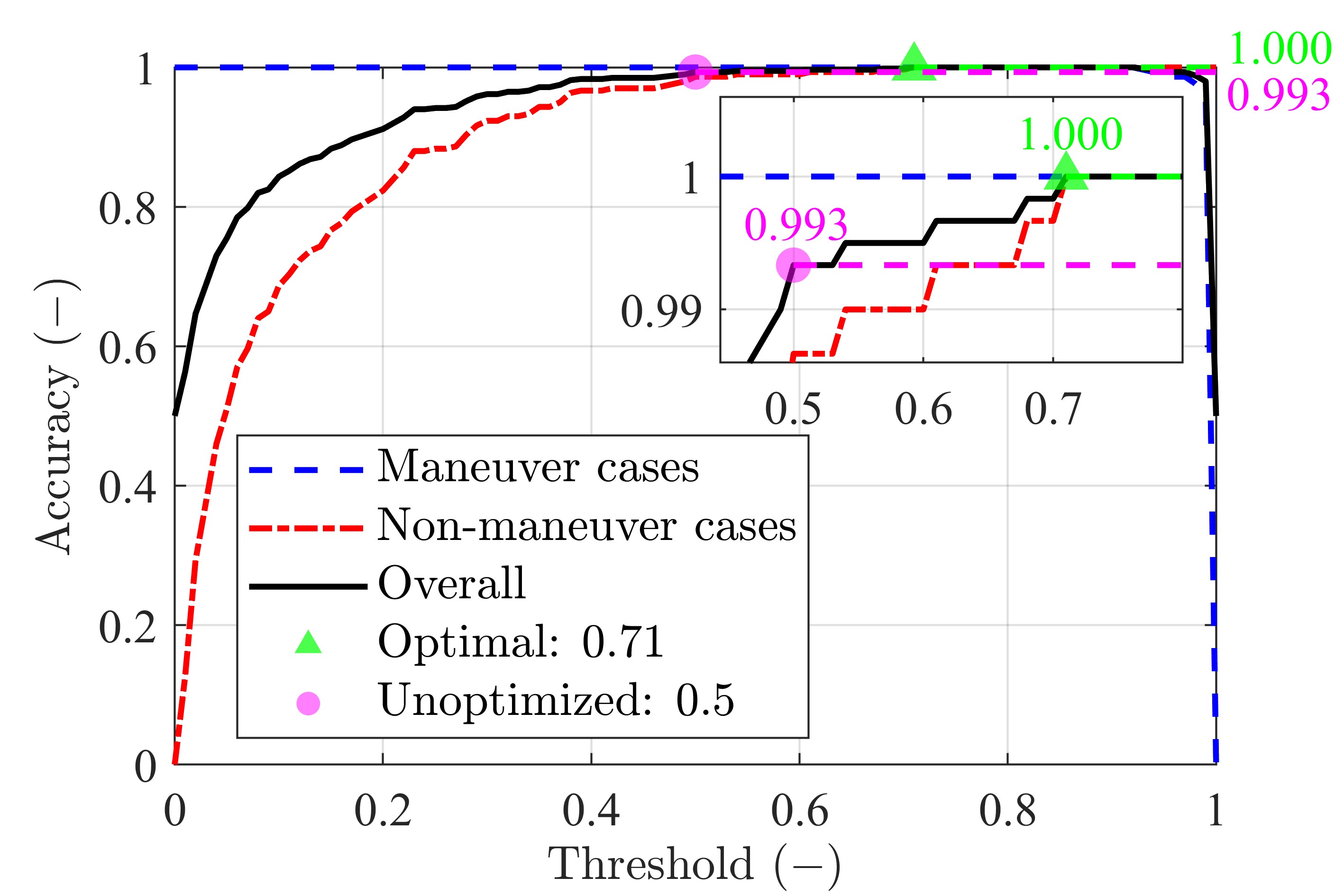}
	}
	\subfigure[]
	{
		\label{fig20b}
		\centering
		\includegraphics[width=0.4\textwidth]{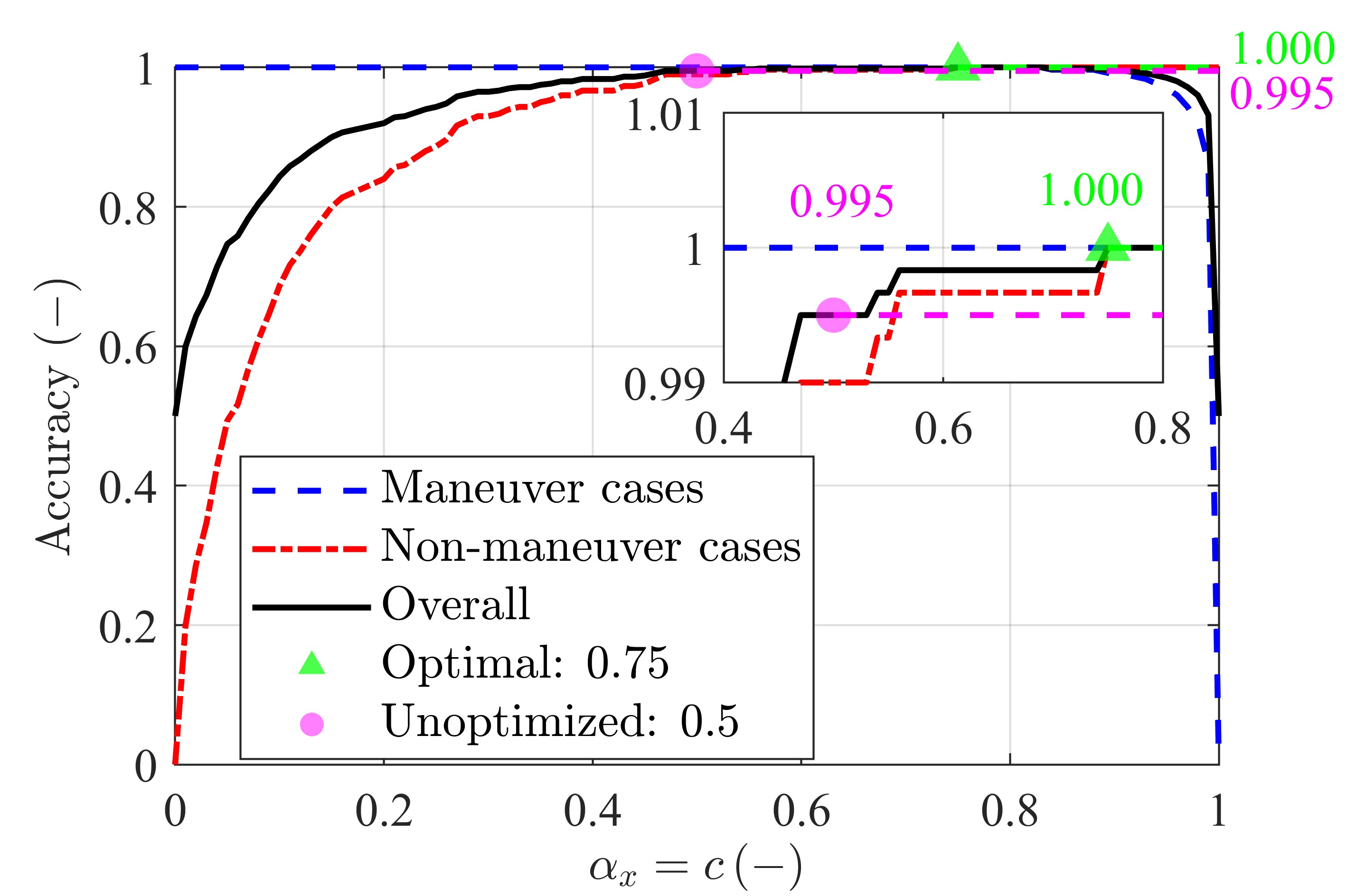}
	}
	\caption{Effects of user-defined parameters on detection accuracy (multiple measurements). \subref{fig20a} CDMI-AI. \subref{fig20b} CDMI-S.}
	\label{fig20}
\end{figure}

Figure~\ref{fig20} presents the detection accuracy of the CDMI-AI and CDMI-S methods under different threshold settings. Compared with Fig.~\ref{fig17}, it can be observed that the optimal thresholds are different. Taking the CDMI-AI method as an example, the optimal threshold under the single-angle measurement is 0.14 (see Fig.~\ref{fig17a}), while that under the three-angle measurements is 0.71 (see Fig.~\ref{fig20a}). However, using a threshold of 0.5 consistently yields satisfactory results across different cases (though not necessarily optimal), demonstrating the reasonableness of selecting 0.5 as the threshold.

Table~\ref{tab8} and Table~\ref{tab9} compare the maneuver detection accuracy and computational cost of the proposed methods and competitive MBAs. Note that the OCDM method is not able to address multiple measurements; thus, it is not simulated here. The second MBA in Ref.~\cite{Montilla2025}, the mixture-based cost variation metric, is simulated in this subsection, as it is designed for cases with multiple-epoch measurements. As shown in Table~\ref{tab8}, the CDMI-S$_{0.5}$ and CDMI-I$_{0.5}$ (using RPO) have the highest detection accuracy (0.995), which is marginally higher than that of the CDMI-AI$_{0.5}$ method (0.9933) and is much better than that of the competitive MBAs. Recall that the CDMI-AI$_{0.5}$ method has the advantage of providing maneuver probability information (over the CDMI-S$_{0.5}$ method) and is more efficient than the CDMI-I$_{0.5}$ method (saving more than 45\% of the CPU time). In addition, as expected, the SLSQP is worse than the RPO in terms of both detection accuracy and efficiency.

\begin{table}[!h]
	\centering
	\caption{Maneuver detection accuracy with multiple-epoch measurements}
	\label{tab8}
	\begin{tabular}{|l|l|lll|}
        \hline
        \multirow{2}{*}{Method} & \multirow{2}{*}{\makecell[l]{Optimization\\ approach}} & \multicolumn{3}{l|}{Accuracy (-)} \\ \cline{3-5} 
         & & \multicolumn{1}{l|}{Non-maneuver} & \multicolumn{1}{l|}{Maneuver} & Overall \\ \hline
        \multirow{2}{*}{CDMI-S$_{0.5}$} & RPO & \multicolumn{1}{l|}{0.9900} & \multicolumn{1}{l|}{1}  & 0.9950 \\ \cline{2-5} 
         & SLSQP & \multicolumn{1}{l|}{0.7033} & \multicolumn{1}{l|}{1} & 0.8516 \\ \hline
        \multirow{2}{*}{CDMI-S$_{0.7}$} & RPO & \multicolumn{1}{l|}{0.9967} & \multicolumn{1}{l|}{1}  & 0.9983 \\ \cline{2-5} 
         & SLSQP & \multicolumn{1}{l|}{0.7467} & \multicolumn{1}{l|}{1} & 0.8733 \\ \hline
        \multirow{2}{*}{CDMI-S$_{0.95}$} & RPO & \multicolumn{1}{l|}{1} & \multicolumn{1}{l|}{0.9700} & 0.9850 \\ \cline{2-5} 
         & SLSQP & \multicolumn{1}{l|}{0.7867} & \multicolumn{1}{l|}{1} & 0.8933 \\ \hline
        \multirow{2}{*}{CDMI-AI$_{0.5}$} & RPO & \multicolumn{1}{l|}{0.9867} & \multicolumn{1}{l|}{1} & 0.9933 \\ \cline{2-5} 
         & SLSQP & \multicolumn{1}{l|}{0.6333} & \multicolumn{1}{l|}{1} & 0.8167 \\ \hline
        \multirow{2}{*}{CDMI-I$_{0.5}$} & RPO & \multicolumn{1}{l|}{0.9900} & \multicolumn{1}{l|}{1}  & 0.9950 \\ \cline{2-5} 
         & SLSQP & \multicolumn{1}{l|}{0.6366} & \multicolumn{1}{l|}{1} & 0.8183 \\ \hline
        MBA$_1$ & NA & \multicolumn{1}{l|}{0.9666} & \multicolumn{1}{l|}{0.6900} & 0.8283 \\ \hline
        MBA$_2$ & NA & \multicolumn{1}{l|}{0.9833} & \multicolumn{1}{l|}{0.5967} & 0.7900 \\ \hline
    \end{tabular}
\end{table}

\begin{table*}[!h]
	\centering
	\caption{Computational costs with multiple-epoch measurements}
	\label{tab9}
	\begin{tabular}{|l|l|lll|}
        \hline
        \multirow{2}{*}{Method}        & \multirow{2}{*}{\makecell[l]{Optimization \\approach}} & \multicolumn{3}{l|}{Averaged CPU time (s)}                                                          \\ \cline{3-5} 
                                       &                                        & \multicolumn{1}{l|}{Non-maneuver}      & \multicolumn{1}{l|}{Maneuver}          & Overall           \\ \hline
        \multirow{2}{*}{CDMI-S}          & RPO                                    & \multicolumn{1}{l|}{3.0150 (0.0301)}   & \multicolumn{1}{l|}{3.0172 (0.0297)}   & 3.0161 (0.0299)   \\ \cline{2-5} 
                                       & SLSQP                                  & \multicolumn{1}{l|}{3.6946 (0.7138)}   & \multicolumn{1}{l|}{3.8750 (0.8814)}   & 3.7848 (0.7976)   \\ \hline
        \multirow{2}{*}{CDMI-AI$_{0.5}$} & RPO                                    & \multicolumn{1}{l|}{3.2774 (0.3797)}   & \multicolumn{1}{l|}{3.2781 (0.2118)}   & 3.2777 (0.2958)   \\ \cline{2-5} 
                                       & SLSQP                                  & \multicolumn{1}{l|}{23.9033 (20.9982)} & \multicolumn{1}{l|}{8.4577 (5.3882)}   & 16.1805 (13.1932) \\ \hline
        \multirow{2}{*}{CDMI-I$_{0.5}$}  & RPO                                    & \multicolumn{1}{l|}{5.9981 (3.0132)}   & \multicolumn{1}{l|}{5.9673 (2.9799)}   & 5.9827 (2.9966)   \\ \cline{2-5} 
                                       & SLSQP                                  & \multicolumn{1}{l|}{74.3664 (71.3856)} & \multicolumn{1}{l|}{91.1375 (88.1440)} & 82.7520 (79.7648) \\ \hline
        MBA$_1$                        & NA                                     & \multicolumn{1}{l|}{12.6747}           & \multicolumn{1}{l|}{13.2748}           & 12.9748           \\ \hline
        MBA$_2$                        & NA                                     & \multicolumn{1}{l|}{20.5584}           & \multicolumn{1}{l|}{22.4347}           & 21.4966           \\ \hline
    \end{tabular}
\end{table*}

\subsection{Robustness Analysis} \label{sec:Robustness Analysis}
\begin{figure}[!h]
	\centering
	\subfigure[]
	{
		\label{fig21a}
		\centering
		\includegraphics[width=0.4\textwidth]{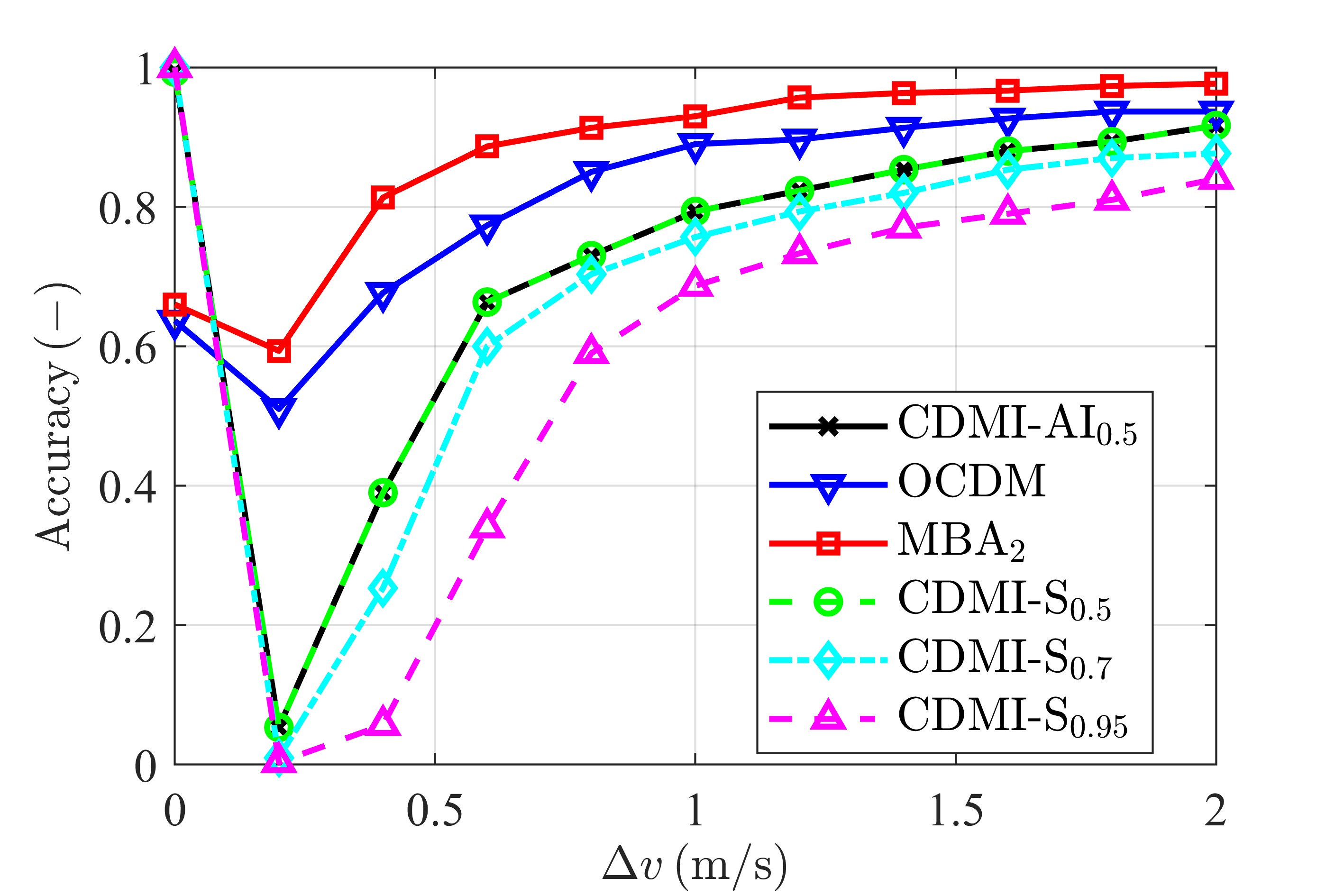}
	}
	\subfigure[]
	{
		\label{fig21b}
		\centering
		\includegraphics[width=0.4\textwidth]{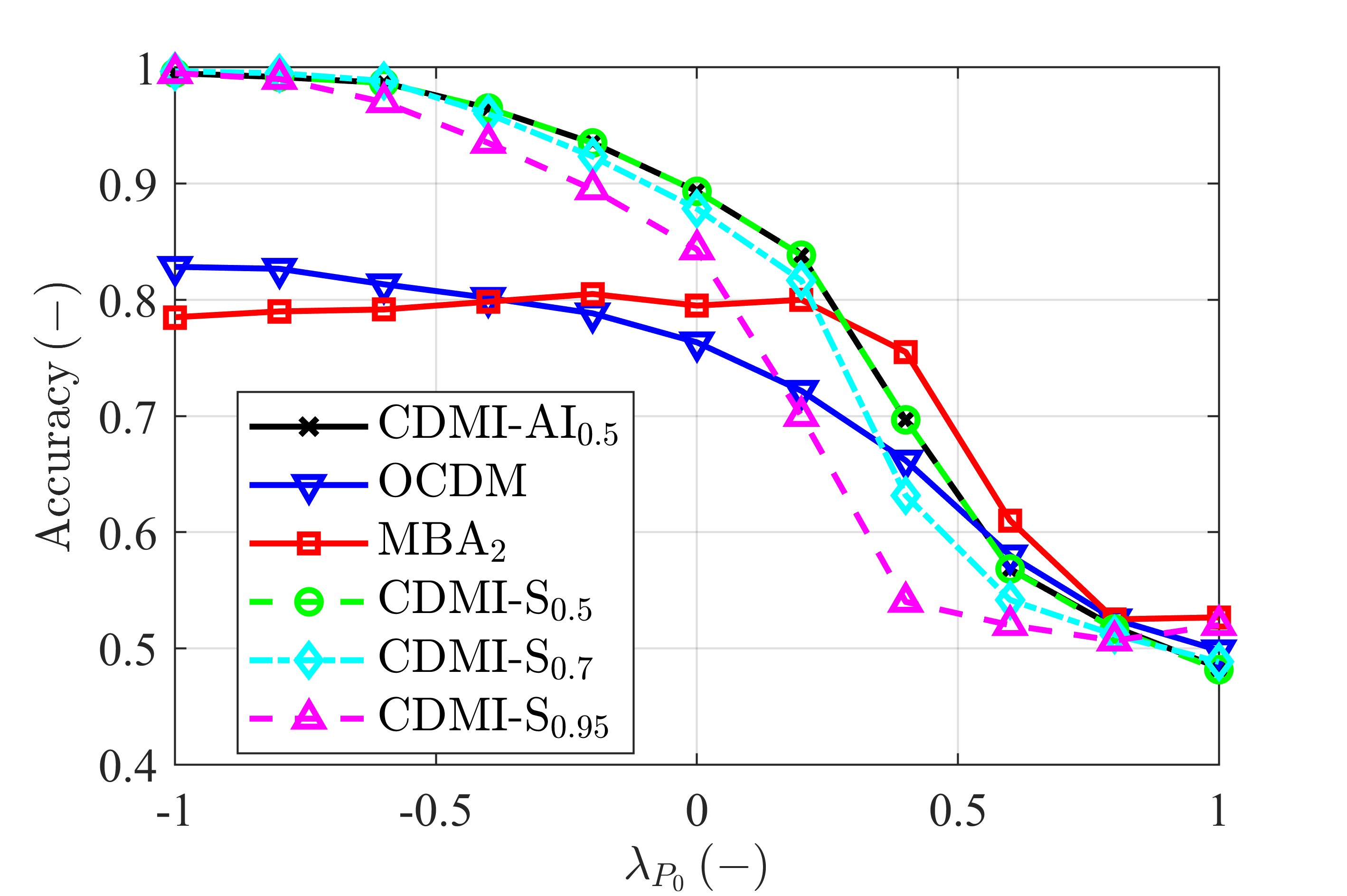}
	}
    \subfigure[]
	{
		\label{fig21c}
		\centering
		\includegraphics[width=0.4\textwidth]{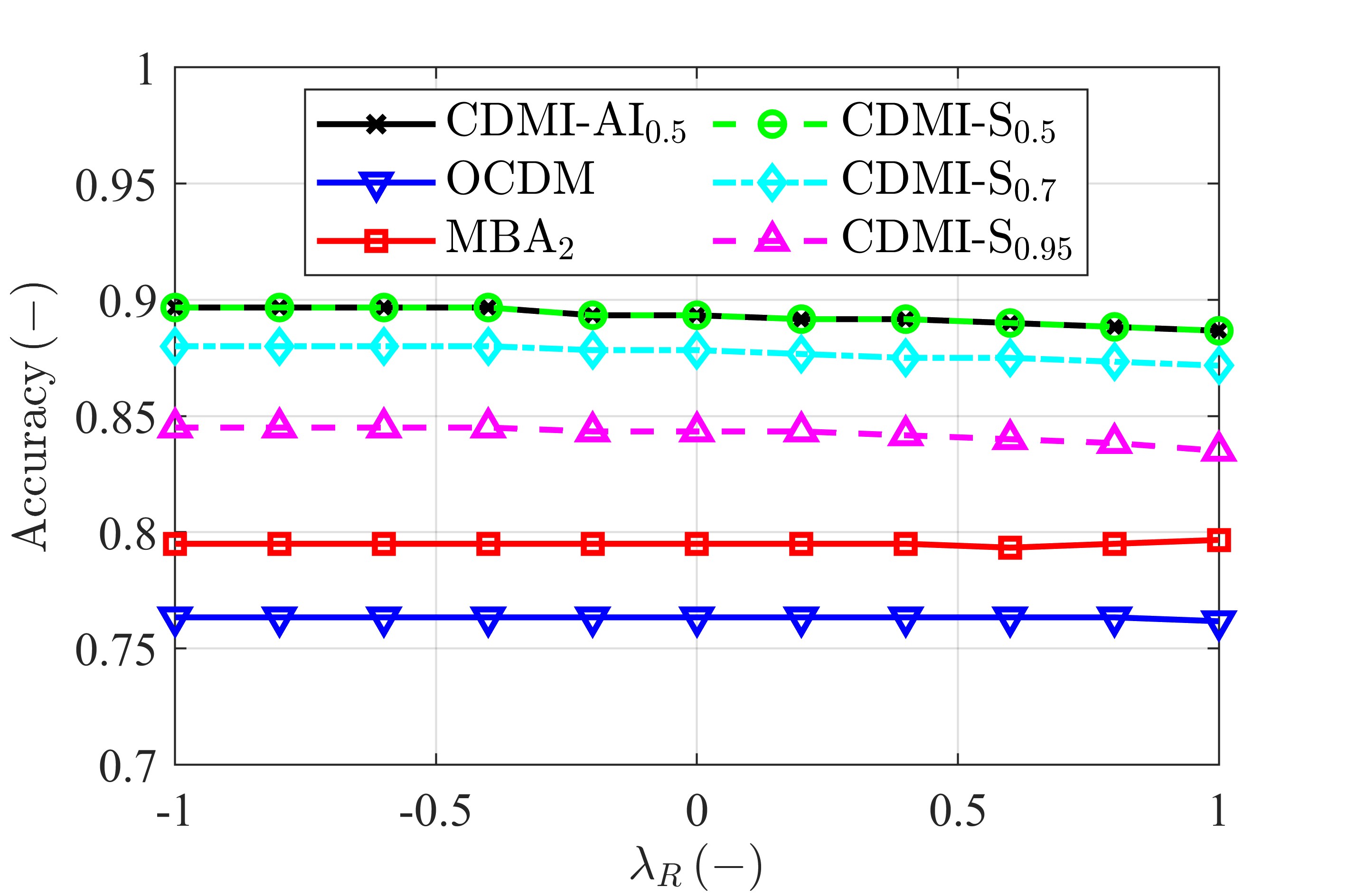}
	}
	\caption{Robustness analysis results for the single-epoch measurement cases. \subref{fig21a} $\Delta v$. \subref{fig21b} $\boldsymbol{P}_0$. \subref{fig21c} $\boldsymbol{R}$.}
	\label{fig21}
\end{figure}

\begin{figure}[!h]
	\centering
	\subfigure[]
	{
		\label{fig22a}
		\centering
		\includegraphics[width=0.4\textwidth]{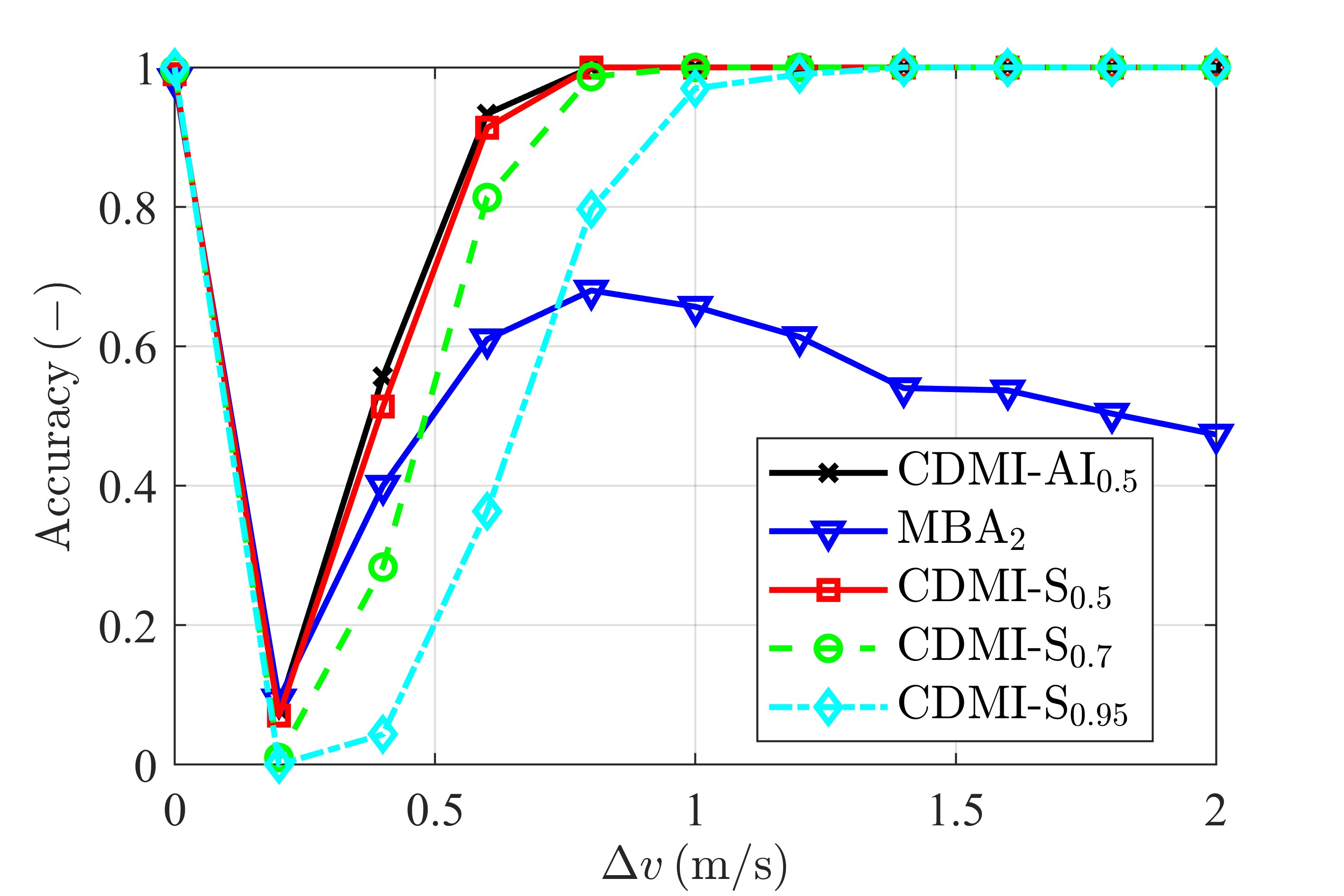}
	}
	\subfigure[]
	{
		\label{fig22b}
		\centering
		\includegraphics[width=0.4\textwidth]{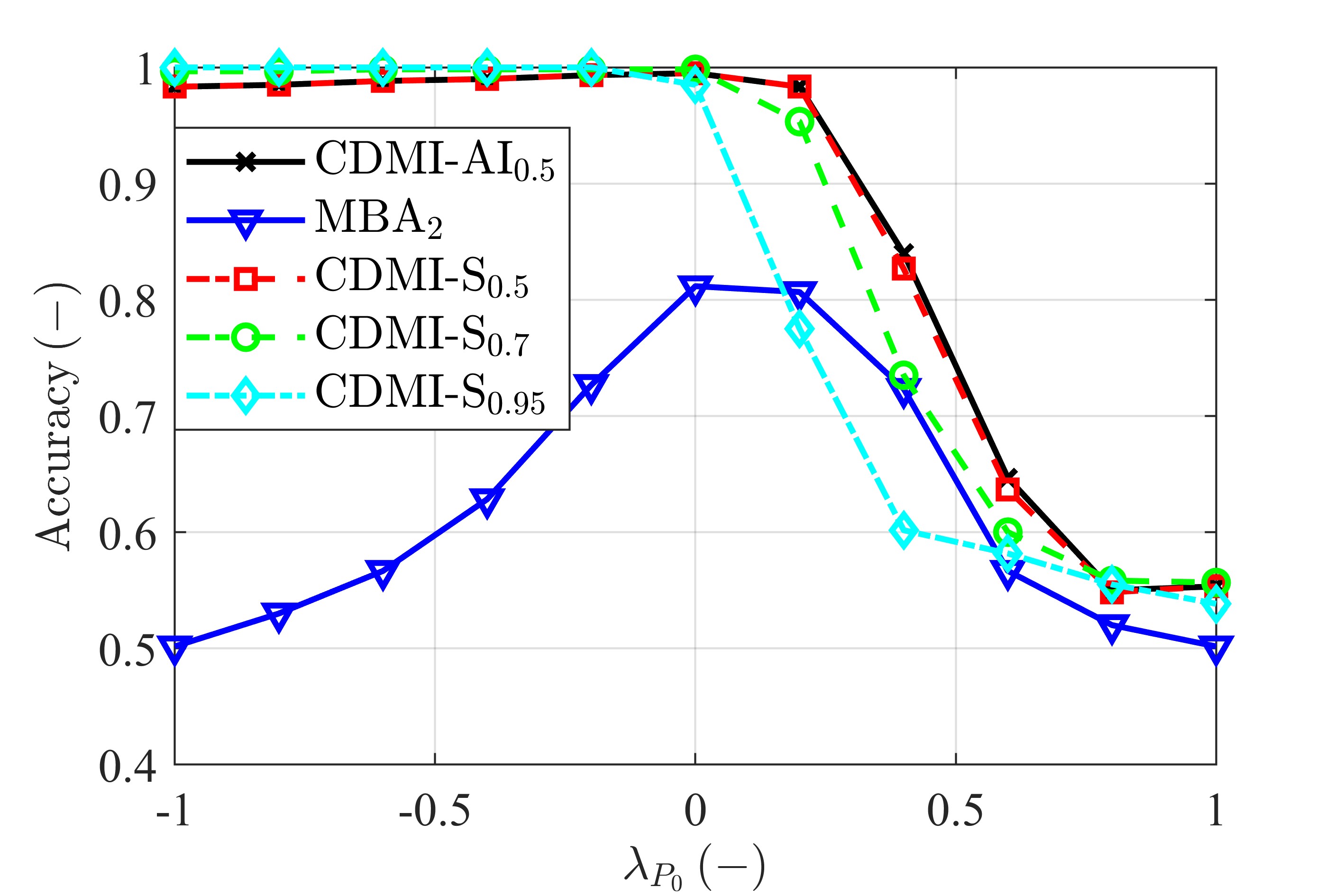}
	}
    \subfigure[]
	{
		\label{fig22c}
		\centering
		\includegraphics[width=0.4\textwidth]{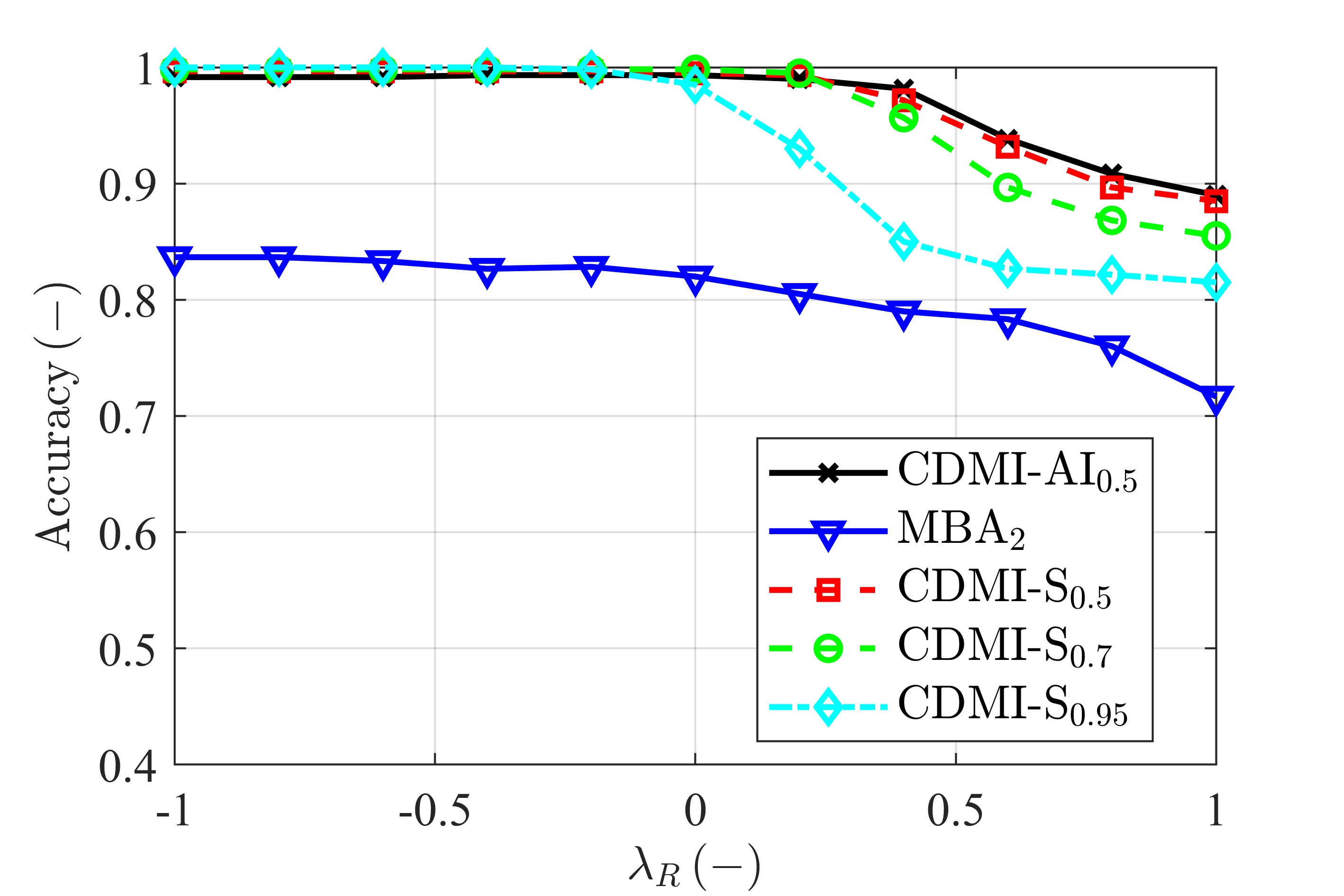}
	}
	\caption{Robustness analysis results for the multiple-epoch measurement cases. \subref{fig22a} $\Delta v$. \subref{fig22b} $\boldsymbol{P}_0$. \subref{fig22c} $\boldsymbol{R}$.}
	\label{fig22}
\end{figure}

Effects of the magnitude of the maneuver ($\Delta v$), initial estimated covariance (${\boldsymbol{P}_0} \leftarrow {10^{{\lambda _{{P_0}}}}}{\boldsymbol{P}_0}$), and measurement noise level ($\boldsymbol{R} \leftarrow {10^{{\lambda _R}}}\boldsymbol{R}$) on the maneuver detection accuracy are investigated in this subsection, with the corresponding results given in Figs.~\ref{fig21}-\ref{fig22}. One can see from Figs.~\ref{fig21}-\ref{fig22} that the proposed CDMI-AI$_{0.5}$ approach has the highest maneuver detection accuracy in most situations, except for two. One is the maneuver cases under single-epoch measurement (as shown in Fig.~\ref{fig21a}), where the competitive methods (\emph{i.e.}, the OCDM and MBA$_2$) work better. However, the OCDM and MBA$_2$ methods are more inclined to make misdetections in non-maneuver cases ($\Delta v=0$ in Fig.~\ref{fig21a}), leading to a lower overall accuracy. The other exception is when large initial uncertainties are quantified (see Fig.~\ref{fig21b}). In this situation, all these methods have seriously degraded maneuver detection accuracy, and the MBA$_2$ (red squares) performs marginally better than the proposed methods in terms of accuracy (at the cost of being much more computationally expensive, as shown in Table~\ref{tab7}). Finally, it is worth noting that the proposed CDMI-AI$_{0.5}$ is always the best when measurements at multiple epochs are available. 
As shown in Fig.~\ref{fig22}, CDMI-AI$_{0.5}$ (black lines) consistently outperforms CDMI-I$_{0.5}$ (red lines) in detection accuracy in different scenarios, despite using the same threshold of 0.5. This demonstrates the effectiveness of the integrated strategy employed by CDMI-AI.

\section{Conclusion} \label{sec:Conclusion}
A confidence-dominance maneuver indicator (CDMI) has been proposed, which relies on a recursive polynomial optimization (RPO) method to generate the maximum likelihood of the observed measurement given the confidence of the state estimate. 
The CDMI flags the potential maneuver by comparing the confidence levels associated with the state estimate and the observation likelihood. 
In addition, an integrated CDMI approach has been derived to avoid the need to select the state confidence level.
The effectiveness of the proposed approaches has been investigated through maneuver detection scenarios in cislunar space.
When applied to maneuver detection, the proposed RPO method outperforms the competitive optimization methods in optimality, convergence rate, and computational efficiency. 
Monte Carlo simulations demonstrate that the integrated CDMI approach achieves an overall maneuver detection accuracy of 89.33\% with a single angle measurement and 99.33\% with three angle measurements-at least 10\% higher than competing methods-while maintaining a significantly lower computational burden.
Furthermore, robustness analysis confirms that the proposed method maintains superior detection performance across varying levels of initial uncertainty and measurement noise.

\bibliographystyle{IEEEtran}
\bibliography{mybibfile}
\end{document}